\documentclass[a4paper,12pt,oneside]{article}  
\usepackage{geometry}
\usepackage{dsfont}
\usepackage[english]{babel}
\usepackage{microtype}
\usepackage{longtable}

\usepackage[round,authoryear]{natbib} 

\usepackage{rotating}
\usepackage{lscape}

\usepackage{import, multicol,lipsum} 
\usepackage{makecell,booktabs}

\usepackage[utf8]{inputenc}
\usepackage[T1]{fontenc}
\usepackage{caption}
\usepackage{subcaption}
\usepackage{pifont}
\usepackage{xcolor, blindtext}
\usepackage{url}
\usepackage{hyperref} 
\hypersetup{
colorlinks=true,
linkcolor=red,
citecolor=blue,     
urlcolor=blue
}

\usepackage{emptypage} 
\usepackage[export]{adjustbox} 
\usepackage{amsthm, amsmath, amssymb, amsfonts, mathtools} 
\usepackage[scr=boondox]{mathalpha} 
\usepackage{nicefrac}
\usepackage{enumitem}
\usepackage{graphicx} 
\usepackage{multirow, nicematrix, array} 
\graphicspath{{img/}}  
\usepackage{wrapfig}
\usepackage{float} 
\usepackage{tikz} 
\usepackage{verbatim}
\usepackage{bbm}
\RequirePackage{fix-cm}

\usepackage{multicol}
\usepackage[table]{xcolor}
\definecolor{lightgray}{gray}{0.95}

\usepackage{soul}

\usepackage[ruled,vlined]{algorithm2e} 
\DontPrintSemicolon
\SetAlgoLined
\SetKwInput{KwInput}{Input}
\SetKwInput{KwOutput}{Output}

\newtheorem{theorem}{Theorem}
\newtheorem{definition}{Definition}
\newtheorem{proposition}{Proposition} 
\newtheorem{lemma}{Lemma} 
\newtheorem{corollary}{Corollary}
\newtheorem{assumption}{Assumption}

\newcommand{\set}[1]{\left\{#1\right\}}

\newcommand{\ft}[1]{\breve{#1}}

\def \vveps {\boldsymbol{\varepsilon}}
\def \TTheta {\boldsymbol{\Theta}}
\def \DDelta {\boldsymbol{\Delta}}
\def \tr {\operatorname{tr}}

\def \dd {\mathbf{d}}
\def \SS {\mathbf{S}}
\def \XX {\mathbf{X}}
\def \YY {\mathbf{Y}}

\title{\bf Network Estimation for Stationary Time Series}
\author{
  Madeline A.\ Shelley \thanks{Department of Mathematics, University of York, UK}, 
  Chiara Boetti \thanks{Department of Mathematical Sciences, University of Bath, UK}, 
  Marina I.\ Knight\footnotemark[1], 
  Matthew A.\ Nunes\footnotemark[2] 
}
\date{\today}

\begin{document}
\maketitle
\thispagestyle{empty}

\begin{abstract}
 High-dimensional multivariate time series are common in many scientific and industrial applications, where the interest lies in identifying key dependence structures within the data for subsequent analysis tasks, such as forecasting. An important avenue to achieve this is through the estimation of the conditional independence graph via graphical models, although for time series data settings the underpinning temporal dependence can make this task challenging.   In this article, we propose a novel wavelet domain technique that allows the data-driven inference of the sparse conditional independence graph of a high-dimensional stationary multivariate time series. By adopting the locally stationary wavelet modelling framework, we repose the estimation problem as a well-principled wavelet domain graphical lasso formulation. Theoretical results establish that our associated estimation scheme enjoys good asymptotic properties when determining sparse dependence structure in input time series data. The performance of the proposed method is illustrated using extensive simulations and we demonstrate its applicability on a real-world dataset representing hospitalisations of COVID-19 patients.
\end{abstract}

\section{Introduction}\label{sec:intro}
In the statistical literature, graphical models are powerful tools to capture key dependence structures within multivariate data. Given a collection of random variables, one wishes to assess the relationship between two variables, conditioned on the remaining variables.  In the classical Gaussian graphical model literature, undirected graphs represent such dependencies among variables through the conditional independence graph ($\mathcal{CIG}$): each node is a variable, and the absence of an edge between two nodes encodes conditional independence between their corresponding variables given all other nodes (variables). For high-dimensional data, it is often desirable to impose sparsity in the graphical structure of the covariance \citep{friedman2008sparse}, as driven by applications in various settings, spanning genetics \citep{wille2006low}, biology \citep{jordan2004graphical, ni2022bayesian}, economics \citep{cerchiello2016conditional}, neurosciences \citep{dyrba2020gaussian, hinne2015bayesian}, and environmental science \citep{engelke2020graphical}.  For comprehensive statistical reviews of graphical models, see e.g., \cite{lauritzen1996graphical} and \cite{ maathuis2018handbook}.

\begin{figure}[H]
  \centering
  \begin{subfigure}[t]{0.4\linewidth}
      \includegraphics[width=\linewidth]{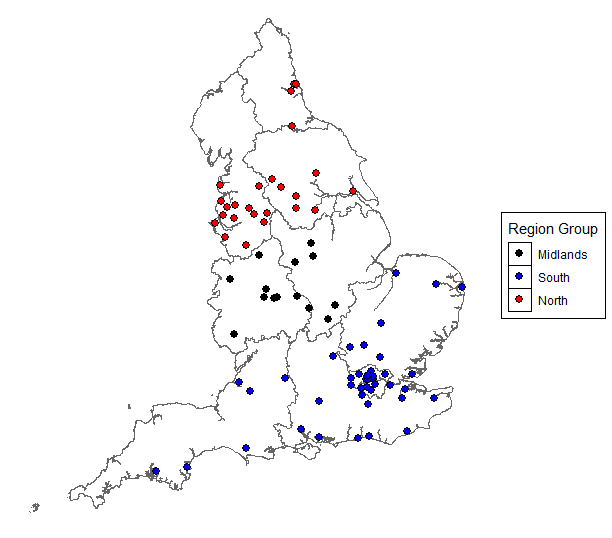}
      \caption{}
      \label{fig:UK-map-covid}
  \end{subfigure}
  \hfill
  \begin{subfigure}[t]{0.5\linewidth}
      \includegraphics[width=\linewidth]{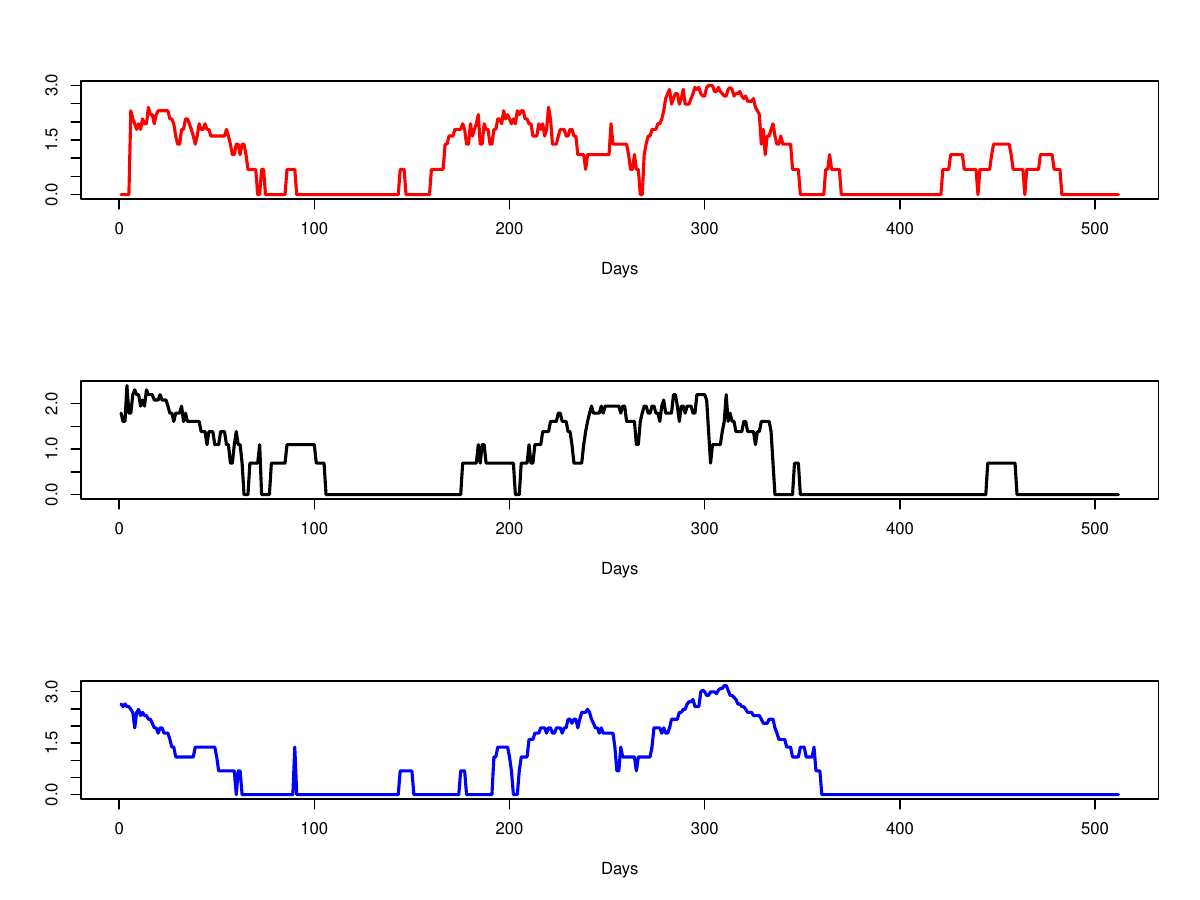}
      \caption{}
      \label{fig:covid-ts-realisations}
  \end{subfigure}
  \caption{(\subref{fig:UK-map-covid}) NHS Trust locations by region: North (red), Midlands (black), South (blue). (\subref{fig:covid-ts-realisations}) Example COVID-19 time series: Airedale (red), Wye Valley (black), Univ. Hospitals Plymouth (blue).} \label{fig:motivation}   
\end{figure}

Improvements in modern data collection methods have resulted in large multivariate time series being routinely collected across many fields.  Consequently, the estimation of conditional independence structures in such data has received growing attention in recent years \citep{brillinger1996remarks, gibberd2015estimating, jung2015graphical, dahlhaus2000graphical, dallakyan2022time}. The work in this article is in particular motivated by the analysis of spatially-observed hospitalisations, namely time series representing COVID-19 ventilation bed occupancy rates at 84 hospitals across the UK from April 2020 to July 2021. In this context, it is of interest to model and e.g., forecast the data in order to inform staffing resource needs. The data show clear spatial and temporal dependence (see Figure~\ref{fig:motivation}), but due to the dimension of the multivariate time series, analysis is challenging without the use of sparse modelling frameworks \citep{nason2023new}. Thus, being able to identify the key dependencies within the data in a principled manner is a vital initial aspect of analysis. 

Directly analogous to the classical setting, the dependencies in multivariate time series processes are mapped onto the Fourier spectral domain \citep{brillinger1996remarks, dahlhaus2000graphical}. Specifically, missing edges in the $\mathcal{CIG}$ correspond to zeros in the inverse spectral density (precision) matrix  across all frequencies.
Estimating the inverse spectral density matrix, however, poses significant challenges \citep{fiecas2019spectral}. A naive approach involves computing the smoothed periodogram—as it is a consistent estimator of the spectral density matrix—and subsequently inverting it. This strategy is suboptimal since the smoothed estimator often becomes ill-conditioned in high-dimensional settings or when the number of available observations is limited \citep{bohm2009shrinkage}. As a result, its inversion can severely amplify estimation errors, leading to inaccurate graphical structure inference.  

These difficulties have motivated the development of two broad classes of regularised estimators. The first class of methods involves shrinking the spectral density matrix, either before or after inversion \citep{fiecas2010functional, fiecas2011generalized, fiecas2014data, schneider2016partial}. While computationally straightforward, such techniques do not directly promote sparsity in the precision matrix, potentially obscuring the most informative conditional dependencies. The second class leverages the property that, for second-order stationary processes, the discrete Fourier coefficients at distinct frequencies are asymptotically independent and follow a complex-valued Gaussian distribution. This enables the extension of Gaussian graphical modelling frameworks to the frequency domain, yielding sparse and well-defined estimates of the precision matrix. We refer to this methodology as the \emph{time series graphical lasso}.
In this context, \cite{jung2015graphical} developed an estimation algorithm to learn the graphical dependence structure of time-dependent data. Later work extended this approach \citep{dallakyan2022time, tugnait2022sparse} and studied related settings, including cases where the data exhibit long memory \citep{baek2023local} or estimation of complex-valued Fourier spectral quantities \citep{deb2024regularized, deb2026inference}. 

While these approaches naturally extend graphical models to time series, they all rely on the Fourier representation of the process. In this article we propose a wavelet domain approach to conditional independence graph estimation, motivated by the observation that wavelet-based time series decompositions can provide improvements over Fourier-based methods in a range of analysis tasks, even in the stationary time series setting \citep{fay2009estimators, guerrier2022robust}. More precisely, we adopt the \emph{locally stationary wavelet} modelling framework introduced by \cite{nason2000wavelet} and extended to the multivariate setting by \cite{park2014estimating}, in which the wavelet spectral matrix captures process dependence through a time-scale representation of a multivariate time series.

Similar to Fourier-based methodology, the inverse of the wavelet spectrum matrices can be used for the inference of direct and indirect associations among the components of the process across both time and scale. Building on this, \cite{gibberd2015multiresolution} introduced an algorithmic sparse estimation method for the inverse local wavelet spectrum based on an estimator of the spectrum; however, their work does not consider the true distributional properties of the wavelet coefficients associated to the process when finding an optimal inverse spectrum, leading to a theoretically unjustified estimation scheme. We also note here that their work does not propose a way to recover the graphical dependence structure from the estimated inverse spectral quantities.

In this article we develop an alternative approach, the \emph{wavelet-based time series graphical lasso} framework for multivariate stationary time series, by considering the wavelet precision matrix and properly accounting for the distributional properties associated to the spectral matrix estimator. Our methodology allows sparse estimation of the inverse wavelet spectra and aids the recovery of a multi-resolution decomposition of the conditional independence graph structure.  Our proposed approach shows improved recovery of the true $\mathcal{CIG}$ compared to Fourier-based techniques in the literature, particularly when the underlying conditional dependence structure is sparse. This is of particular relevance in high-dimensional settings, where the identification of a relatively small number of genuine dependencies amongst a large number of potential connections is central to obtaining interpretable and parsimonious representations of the data.

In addition to estimation of the $\mathcal{CIG}$ for stationary time series, we also develop a tool for identifying the underlying graph, or network, in so-called network time series models.
Originally proposed by \cite{knight2016modelling}, Generalised Network Autoregressive (GNAR) processes model autoregressive dynamics while explicitly exploiting inter-dependencies encoded by a given graph, $\mathcal{N=(K,E^{N})}$, via nodal neighbourhoods (subsets of $\mathcal K$) as determined through the presence / absence of edges in $\mathcal{E^{N}}$. Focusing on relevant connections was shown to improve parameter estimation and forecasting accuracy, as well as to enhance computational efficiency \citep{knight2020generalized}.
However, when the underlying graph is unknown (i.e., unknown edge set $\mathcal{E^{N}}$), determining it from the data is a crucial first modelling task \citep{jimenez2025gnar}. One way to approach this issue is to fit network time series models on several candidate graphs and to select the one with the smallest prediction error \citep{nason2025forecasting}. This strategy, however, identifies a graph that aids prediction, which may differ from the graph that best represents the data-generating structure. In this work, we propose a data-driven procedure that infers the graph $\mathcal{N}$ directly from the dependence structure $\mathcal{CIG}$ estimated by our wavelet-based time series graphical lasso, by means of an edge-clustering approach. 

The remainder of this paper is organised as follows. Section~\ref{sec:Background} reviews the necessary background on graphical modelling for stationary time series and multivariate locally stationary wavelet processes. Section~\ref{sec:Wavelet-model-def} introduces the proposed wavelet-based time series graphical lasso methodology while Section~\ref{sec:Theoretical Properties} establishes the theoretical properties of the proposed estimator. Section~\ref{sec:Simulation Study} evaluates the methodology through a series of simulation studies and comparisons with existing approaches. Section~\ref{sec:graph-discovery} considers a related problem of recovering an underlying network from the estimated conditional dependence graph, when the graphs are different. Section~\ref{sec:Application} applies the proposed methodology to COVID-19 mechanical ventilation data across NHS Trusts in England. Finally, Section~\ref{sec:Conclusions} summarises our findings and outlines directions for future research.

\section{Background}\label{sec:Background}
In this section, we review the time series graphical lasso (TSglasso) for multivariate stationary time series as developed by \cite{jung2015graphical}, and give a brief overview of  multivariate locally stationary wavelet processes \citep{park2014estimating}. These two aspects provide the necessary background to formulate our proposed modelling framework in Section~\ref{sec:Wavelet-model-def}. 

In what follows, we consider a $P$-dimensional, zero-mean  stationary Gaussian process\\ $\XX=\{ \XX_t \}_{t=0}^{T-1}$, where $\XX_t = \left(X^{(1)}_t,\ldots, X^{(P)}_t\right)^{\top}$ is observed at time $t$ across $P$ channels. 

Our aim is to characterise the interactions among the $P$ components of the process $\XX$ by means of its underlying conditional independence graph $\mathcal{CIG} = \left( \mathcal{K}, \mathcal{E} \right)$, with node set $\mathcal{K}=\set{1,\ldots,P}$ and undirected edge set
$\mathcal{E} = \set{(p, q) \in \mathcal{K} \times \mathcal{K},\, p\neq q : p \leftrightsquigarrow q}$, where the edge $p \leftrightsquigarrow q$ indicates conditional dependence between channels $p$ and $q$. In particular, we denote by $\mathbf{G}$ the $P\times P$-dimensional adjacency matrix of the $\mathcal{CIG}$, with entries $\mathbf{G}_{p,q} = 1$ if $(p, q) \in \mathcal{E}$, and $\mathbf{G}_{p,q} = 0$ otherwise. By definition, $\mathbf{G}$ is symmetric with zeros on the diagonal. 

The $\mathcal{CIG}$ can be determined from the process spectral properties, as follows.
Let $\mathbf{f}(\omega)\in \mathbb{C}^{P\times P}$ be the spectral density matrix of $\XX$ at frequency $\omega \in[-\pi, \pi]$, and consider the precision matrix function $\mathbf{Q}(\omega):=\mathbf{f}^{-1}(\omega)$. \cite{dahlhaus2000graphical} shows that the channel $p, \, q$ processes, $\{X_{t}^{(p)}\}$ and  $\{X_{t}^{(q)}\}$, are independent conditional on all remaining components if and only if $\mathbf{Q}_{p,q}(\omega)=0, \forall \omega$, thus rendering a bijective characterisation of the edge set $\mathcal{E}$.

\subsection{Fourier-based Time Series Graphical Lasso}\label{sec:Fourier}

Denote the $P$-dimensional Fourier coefficients of $\XX$ by $\set{\dd(\omega_n)}_{n=0}^{T-1}$, obtained by applying the normalised discrete Fourier transform to $\XX$, where $\omega_n = \tfrac{n}{T}$. 
Under the assumption that~$\XX$ has an absolutely summable autocovariance function, \cite{brillinger2001time} showed that, as $T \to \infty$, $\dd(\omega_n) \dot\sim N^{\mathbb{C}}_{P}\left(\mathbf{0}, \mathbf{f}(\omega_n) \right)$ independently for $n = 1, \ldots, \lfloor \tfrac{T}{2} \rfloor - 1$. Moreover, $\dd(\omega_n) \dot\sim N_{P}\left(\mathbf{0}, \mathbf{f}(\omega_n) \right)$ for $n=0, \lfloor \tfrac{T}{2} \rfloor$, though these coefficients are typically ignored. Here, $\dot\sim N^{\mathbb{C}}_{P}$ and $\dot\sim N_{P}$ indicate that variables are approximately distributed as complex- and real-valued Gaussian, respectively. 

In standard Fourier spectral estimation, local periodogram smoothing is applied over~$M = \lfloor \tfrac{T/2 - m - 1}{L} \rfloor$ consecutive blocks, each of length $L = 2m + 1$ with half-window length $m$. Evaluated at the aggregated frequencies $\tilde{\omega}_{l} = \frac{(l-1)L+m+1}{T}$, $l=1,\ldots,M$, the smoothed sample spectral density is defined as
\begin{equation*}
    \tilde{\mathbf{f}}(\tilde{\omega}_{l})= \frac{1}{L} \sum_{k=-m}^{m} \dd(\tilde{\omega}_{l,k}) \dd^H(\tilde{\omega}_{l,k}),
\end{equation*}
with $\tilde{\omega}_{l,k}=\frac{(l-1)L+m+1+k}{T}$, $k=-m,\ldots,m$, representing the $L$ disjoint frequencies in the $l$-th window, and ${\cdot}^H$ indicating the Hermitian transpose.

As usual in the time series Fourier spectral domain literature, following from the (complex-valued) Gaussian distributions of the Fourier coefficients and the sample spectral density matrix above, the log-likelihood function for the precision matrix $\mathbf{Q}(\cdot)$ can be written as
\begin{equation}\label{eq:Qll}
    \mathcal{L}(\mathbf{Q}(\cdot)) =  \sum_{l=1}^{M} L\left[ \log\det\left(\mathbf{Q}(\tilde{\omega}_{l})\right) - \tr\left(\mathbf{\tilde{f}}(\tilde{\omega}_{l})\mathbf{Q}(\tilde{\omega}_{l})\right)\right],
\end{equation}
where $\tr(\cdot)$ denotes the trace operator. Thus, in the spirit of the graphical lasso, a sparse estimator $\hat{\mathbf{Q}}(\cdot)$ may be obtained by solving the regularised optimisation problem
\begin{equation}\label{eq:Fourier-TSglasso}
    \min_{\mathbf{Q}(\cdot)} \set{-\mathcal{L}(\mathbf{Q}(\cdot)) + \mathcal{P}(\mathbf{Q}(\cdot), \lambda)},
\end{equation}
where $\mathcal{P}(\mathbf{Q}(\cdot), \lambda)$ is a sparsity inducing penalty function. For example, \cite{jung2015graphical} use an $\ell_1$-penalty, \cite{tugnait2022sparse} defines a sparse group penalty, whilst \cite{dallakyan2022time} explore an $\ell_2$-penalty in a two-stage sparse estimation procedure for vector autoregressive processes. 
The optimisation problem \eqref{eq:Fourier-TSglasso} is typically solved using the alternating direction method of multipliers (ADMM) \citep{boyd2011distributed}, exploiting variable splitting and the separability of the objective function across frequencies $\{\tilde{\omega}_{l}\}_{l=1}^M$. Because $\mathbf{Q}(\cdot)$ is complex-valued, Wirtinger calculus is used to derive suitable ADMM update steps while incorporating the assumption of smooth variation across nearby frequencies \citep{dallakyan2022time}. This translates into choosing an appropriate smoothing span $L$ in \eqref{eq:Qll}.

A further step in the TSglasso framework involves inferring the adjacency matrix $\mathbf{G}$ from the estimated precision matrix functions, since solving \eqref{eq:Fourier-TSglasso} yields $M$ estimates, i.e., $\hat{\mathbf{Q}}(\tilde{\omega}_{l}) \in \mathbb{C}^{P \times P}$ for all $l=1,\ldots,M$. Techniques in the literature aggregate these estimates in different ways to form an estimated adjacency, $\hat{\mathbf{G}}$. For example, \cite{dallakyan2022time} set $\hat{\mathbf{G}}_{p,q} = 0$, i.e., no edge between nodes $p$ and $q$, if $\hat{\mathbf{Q}}_{p,q}(\tilde{\omega}_{l}) = 0$ for all $l$. However, this approach can reduce sparsity: small nonzero entries at some frequencies will still result in edges in the final graph. 
As an alternative, \cite{tugnait2022sparse} determines the presence of an edge between the $p$-th and $q$-th nodes in the $\mathcal{CIG}$ if the $\ell_{2}$-norm of the vector $\left(\hat{\mathbf{Q}}_{p,q}(\tilde{\omega}_{1}), \dots, \hat{\mathbf{Q}}_{p,q}(\tilde{\omega}_{M})\right)$ is different from zero. 

To preserve sparsity, in our TSglasso implementation in Section~\ref{sec:Simulation Study} we summarise the $M$ estimates using the entrywise $\ell_{1}$-norm for each $p,q=1,\ldots,P$, $\overline{\mathbf{Q}}_{p,q} = \frac{1}{M} \sum_{l=1}^M  |\hat{\mathbf{Q}}_{p,q}(\tilde{\omega}_{l})|$,
where $\overline{\mathbf{Q}} \in \mathbb{R}^{P \times P}$ is the resulting matrix. The estimate $\hat{\mathbf{G}}$ is then formed by setting the corresponding entries to 1 where $\overline{\mathbf{Q}}$ is non-zero. In practice, to account for numerical noise, $\hat{\mathbf{G}}_{p,q}$ is set to zero if $|\overline{\mathbf{Q}}_{p,q}| < c$, where $c=10^{-5}$ \citep{dallakyan2022time}.

\paragraph{Tuning Parameter Selection.}
The parameter $\lambda$ of the penalty term $\mathcal{P}(\mathbf{Q}(\cdot) , \lambda)$ in the TSglasso problem~\eqref{eq:Fourier-TSglasso} controls the sparsity of the estimated precision matrix $\mathbf{Q}(\cdot)$, and is chosen to minimise the extended Bayesian Criterion (eBIC) \citep{chen2008extended}. By denoting the frequency-specific terms $\mathcal{L}\big(\mathbf{Q}(\tilde{\omega}_{l})\big) := \log\det \left(\mathbf{Q}(\tilde{\omega}_{l})\right)-\text{tr}(\mathbf{\tilde{f}}(\tilde{\omega}_{l})\mathbf{Q}(\tilde{\omega}_{l}))$ in \eqref{eq:Fourier-TSglasso} for each $l$, \cite{tugnait2022sparse} and \cite{dallakyan2022time} define the eBIC as follows
\begin{equation*}
    \operatorname{eBIC}(\lambda) = -2L \sum_{l=1}^{M}\mathcal{L}(\mathbf{Q}\big(\tilde{\omega}_{l})\big) + \log(2 LM)\sum_{l=1}^M E_l + 4 \gamma\log(P)\sum_{l=1}^M E_l,
\end{equation*}
where $E_l$ is the number of non-zero elements in $\mathbf{Q}(\tilde{\omega}_{l})$ and $\gamma\in[0,1]$ is an additional hyper-parameter.

\subsection{Multivariate Locally Stationary Wavelet Processes} \label{sec:mvLSW-review}

Although the process under consideration in this work is assumed to be stationary, we shall employ the locally stationary wavelet modelling framework of \cite{nason2000wavelet} to allow for projections into the wavelet spectral domain. This framework has found a variety of applications across fields such as biology \citep{Hargreaves2018}, neuroscience \citep{embleton2022multiscale, knight2024adaptive}, healthcare \citep{del2025analysing}  and economics \citep{killick2020, killick25}. Departing from univariate data, \cite{park2014estimating} propose an extension, the \emph{multivariate locally stationary wavelet} (mvLSW) model for the $P$-dimensional process $\XX = \set{\XX_{t}}_{t=0}^{T-1}$, with $T=2^J$ and $J\in\mathbb{N}$, which allows for time-varying dependence within and across channels of the multivariate time series. The model expresses the process as a superposition of discrete non-decimated wavelets across scales $j$ and times $k$, $\psi_{j,k}(\cdot)$,  
\begin{equation*}
    \XX_{t} = \sum_{j=1}^\infty \sum_{k\in\mathbb{Z}} \mathbf{V}_j(\nicefrac{k}{T})\psi_{j,k}(t) \vveps_{j,k}, 
\end{equation*}
where $\set{\psi_{j,k}(t):=\psi_{j,k-t}}_{j,k}$ is a set of non-decimated wavelets at scale $j$ and location $k$, $\mathbf{V}_{j}\left( u \right)$ is the scale $j$ transfer function matrix at rescaled time $u=\nicefrac{k}{T} \in [0,1]$, and $\set{\vveps_{j,k}}_{j,k}$ are uncorrelated $P$-dimensional random vectors with zero mean and identity covariance matrix.  Each $\mathbf{V}_{j}\left( \cdot \right)$ is assumed to have a lower-triangular matrix form, whose entries are Lipschitz continuous functions with Lipschitz constant $L_j^{(p,q)}$ satisfying $\sum_{j=1}^{\infty} 2^j L_j^{(p,q)} < \infty$. For clarity, here we use the indexing convention that small $j$ corresponds to fine levels, and large $j$ to coarse scales.

The \emph{local wavelet spectral} (LWS) matrix function $\SS_{j}(u) \in \mathbb{R}^{P\times P}$ yields a time-scale decomposition of the process power at each scale $j$ and rescaled time $u$, and it is defined as $\SS_{j}(u) = \mathbf{V}_j(u) \mathbf{V}_j(u)^{\top}$.
More in general, the LWS function is connected to the (localised) cross-channel covariance between channels $p$ and $q$ of $\XX$ at rescaled time $u$ and lag $\tau\in \mathbb{Z}$, as 
\begin{equation}\label{eq:lacv}
\mathbb{C}\mathrm{ov}(X^{(p)}_{\lfloor uT \rfloor, T}, X^{(q)}_{\lfloor uT\rfloor+\tau, T})  \approx c^{(p,q)}(u,\tau)  = 
    \sum_{j=1}^\infty S^{(p,q)}_j(u)
    \Psi_j(\tau),
\end{equation}
where $\Psi_j(\tau)=\sum_k\psi_{j,k}\psi_{j,k-\tau}$ is the discrete autocorrelation wavelet \citep{nason2000wavelet}, and $\approx$ denotes equality up to an $\mathcal{O} (T^{-1})$ factor \citep{park2014estimating}. 

Note the LWS is a real-valued continuous function defined on the rescaled time domain $[0,1]$; since in practice, we only observe the process at discrete time points $k=\lfloor uT \rfloor$ with $k=0,\dots,T-1$, the notation $\SS_{j,k} = \SS_{j}(\nicefrac{k}{T})$ is often adopted \citep{park2014estimating}. In this vein, denote the precision matrix function as $\TTheta_{j,k}=\SS_{j,k}^{-1}$, where we only work with non-degenerate spectral structures and assume the spectrum is positive definite, and thus invertible.

Let $\dd_{j,k} = \big( d_{j,k}^{(1)},...,d_{j,k}^{(P)} \big)^{\top}$ be the empirical wavelet coefficient vector, defined as $\dd_{j,k} = \sum_{t=0}^{T-1}\XX_t\psi_{jk}(t)$, and let $\mathbf{I}_{j,k} = \dd_{j,k}\dd_{j,k}^{\top}$ be the wavelet matrix periodogram. Then, \cite{park2014estimating} define an asymptotically unbiased estimator of $\SS_{j,k}$ for each $j$ and each $k$ as 
\begin{equation}\label{eq:LSW LWS estimate}
    \hat{\SS}_{j,k} := \sum_{l=1}^J (\mathbf{C}^{-1})_{jl} \tilde{\mathbf{I}}_{l,k},
\end{equation}
where $J=\log_2(T)$, $\tilde{\mathbf{I}}_{l,k}$ is the wavelet periodogram, smoothed across time via a rectangular kernel to achieve consistency, and $\mathbf{C}$ is the discrete autocorrelation wavelet inner product matrix, shown to be positive definite with a bounded inverse, where $C_{jl} = \sum_{\tau} \Psi_{j}(\tau) \Psi_{l}(\tau)$ \citep{nason2000wavelet}.

Without connection to graph discovery, \cite{gibberd2015multiresolution} propose introducing sparsity in the inverse local wavelet spectrum, $\TTheta_{j,k}:= \SS_{j,k}^{-1}$, at each scale $j$ and location $k$  by means of the following $\ell_1$-regularised optimisation 
\begin{equation}\label{eq:wavelet-TSglasso-wrong}
  \hat{\TTheta}_{j,k} := \underset{\TTheta \succeq  0}{\arg\min} \set{ - \log\det(\TTheta) + \tr\bigr({\hat{\SS}_{j,k}} \,\TTheta \bigl) + \lambda_j \|\TTheta \|_1 }.
\end{equation}

Although it is reasonable to assume that the wavelet coefficients are zero-mean Gaussian and uncorrelated across locations $k$ within each scale $j$, as shown in \cite{park2014estimating}, the expectation of the raw wavelet periodogram $\mathbf{I}_{j,k}$ satisfies 
\begin{equation}\label{eq:beta-S relationship}
    \mathbb{E}(\mathbf{I}_{j,k}) \approx  \sum_{l=1}^J C_{jl}\SS_{l}(\nicefrac{k}{T}) = \boldsymbol{\beta}_j(\nicefrac{k}{T}) =: \boldsymbol{\beta}_{j,k}.
\end{equation}
Since $\mathbb{E}(\dd_{j,k})=\mathbf{0}$ due to the zero-mean definition of the random increments $\{\vveps_{j,k}\}_{j,k}$, then~\eqref{eq:beta-S relationship} implies $\dd_{j,k} \dot\sim N_P(\mathbf{0}, \boldsymbol{\beta}_{j,k})$, so that the actual inverse covariance matrix of the wavelet coefficients is $\boldsymbol{\beta}_{j,k}^{-1}$.  This is in contrast to the one employed in the optimisation problem \eqref{eq:wavelet-TSglasso-wrong}, namely $\SS_{j,k}^{-1}$, thus illustrating how this formulation does not correctly account for the true distribution of the wavelet coefficients. Under the process stationarity assumption, we next propose a principled way to bypass this problem.

\section{Proposed Wavelet-based Time Series Graphical Lasso}\label{sec:Wavelet-model-def}

Motivated by \citet[Proposition 3(a)]{nason2000wavelet}, we consider a stationary process $\XX$ admitting an mvLSW representation with Gaussian increments, where stationarity allows us to drop the time-localisation typical of locally stationary wavelet processes. 
In what follows we show that, under the assumption of process stationarity, we can characterise the adjacency matrix of the $\mathcal{CIG}$ in the \cite{dahlhaus2000graphical} sense described in Section \ref{sec:Background} by the corresponding adjacency matrix formed from the scale-dependent wavelet precision matrices, $\{\TTheta_j := \SS^{-1}_{j}\}_j$. This characterisation then becomes central to our proposal for wavelet-based $\mathcal{CIG}$ estimation.

\begin{proposition}\label{prop:wav_cig}
Let $\XX=\{ \XX_t \}_{t=0}^{T-1}$ be a $P$-dimensional, zero-mean stationary process with Gaussian increments and absolutely summable auto- and cross-covariance functions, such that $\sum_{\tau \in \mathbb{Z}} |c^{(p,q)}(\tau)|<\infty$ for any nodes $p, \, q =1,\ldots,P$, where $\XX_t = \big(X^{(1)}_t,\ldots, X^{(P)}_t\big)^{\top}$ is observed at time $t$ across $P$ channels (nodes). Let $\{\SS_{j}\}_j$ denote its wavelet spectral structure across the scales $j=1, \ldots, J=\log_2(T)$. 
Then, nodes $p$ and $q$ are conditionally independent given the remaining nodes if and only if $\left(\TTheta_j\right)_{p,q} = \big(\SS_{j}^{-1}\big)_{p,q} = 0$ for all $j=1,\ldots, J$.
\end{proposition}

The proof appears in Appendix~\ref{app:proofprop1} and makes use of the Fourier spectral domain characterisation typically used in graphical modelling \citep{dahlhaus2000graphical} applied to scale-dependent subprocesses defined akin to \cite{wu2023multi}. \newline

Stemming from the characterisation of Proposition~\ref{prop:wav_cig}, we are interested in sparsely estimating the wavelet precision matrices to detect the conditional independence structure of stationary processes. To this end, let $\XX$ be a $P$-dimensional stationary process with Gaussian increments. Following the arguments in Section~\ref{sec:mvLSW-review} and recalling  process Gaussianity and stationarity, at each scale $j$ the wavelet coefficients of $\XX$ are zero-mean $P$-dimensional Gaussian random vectors with covariance matrix $\boldsymbol{\beta}_j(\nicefrac{k}{T}) = \boldsymbol{\beta}_j$ for every time $k$ or, equivalently, $\dd_{j,k} \dot\sim N_P(\mathbf{0}, \boldsymbol{\beta}_j)$ for $k=0, \ldots, T-1$. 
As discussed in Section~\ref{sec:Fourier}, the distributional properties of the discrete Fourier coefficients were exploited to formulate a sparse estimation problem for the inverse spectral density matrix function. In contrast, the discrete wavelet coefficients yield the quantities $\boldsymbol{\beta}_{j}$---linear combinations of the $\SS$-spectra, as shown in equation~\eqref{eq:beta-S relationship}---and cannot be directly used instead of $\SS_{j}$. 

Explicitly, the probability distribution of a single wavelet coefficient $\dd_{j,k}$ at scale $j$ is 
\begin{equation*}
    p(\dd_{j,k}) = \frac{1}{(2\pi)^{P/2}}(\det(\boldsymbol{\beta}_j))^{-1/2} \exp\left\{-\frac{1}{2} \dd_{j,k}^{\top} \boldsymbol{\beta}_j^{-1} \dd_{j,k}\right\},
\end{equation*}
and hence the log-likelihood expressed in terms of $\boldsymbol{\beta}_j^{-1}$ up to an additive constant and multiplicative factor of $1/2$ can be derived as
\begin{equation}\label{eq:L-scale-time}
    \mathcal{L}_{j,k}(\boldsymbol{\beta}_j^{-1}) 
    = \log\det(\boldsymbol{\beta}_j^{-1}) - \tr(\mathbf{I}_{j,k} \boldsymbol{\beta}_j^{-1}),
\end{equation}
where $\mathbf{I}_{j,k}=\dd_{j,k}\dd_{j,k}^\top$ denotes the raw periodogram at scale $j$ and time $k$.

We construct a scale dependent likelihood-based criterion by aggregating the individual likelihood contributions across locations, and specifically at scale $j$ we define
\begin{equation}\label{eq:L-star}
    \mathcal{L}_j^*(\boldsymbol{\beta}_j^{-1})=\sum_{k=0}^{T-1} \mathcal{L}_{j,k}(\boldsymbol{\beta}_j^{-1})=T\big[\log\det(\boldsymbol{\beta}_j^{-1}) - \tr(\bar{\mathbf{I}}_{j} \boldsymbol{\beta}_j^{-1})\big] 
\end{equation}
with $\bar{\mathbf{I}}_j=\frac{1}{T}\sum_{k=0}^{T-1}\mathbf{I}_{j,k}$. Note here that smoothing over the entire time domain is used for the periodogram since $\XX$ is a stationary process. Hence, in the wavelet context we propose to solve the optimisation problem given by
\begin{equation} \label{eq:optB}  
    \mathbf{\hat{B}}_j = \underset{\mathbf{B}_j}{\arg\min} \set{ -T \left[ \log\det(\mathbf{B}_j) - \tr(\bar{\mathbf{I}}_j\mathbf{B}_j) \right] + \lambda_j\|\mathbf{B}_j \|_{1,\text{off}},},
\end{equation}
where $\mathbf{B}_j = \boldsymbol{\beta}_j^{-1}$ and $\|\mathbf{B}_j \|_{1,\text{off}}  = \sum_{p\neq q}|(\mathbf{B}_j)_{p,q}|$. 
Observe the evident multiresolution aspect, resulting in a separate optimisation problem for each scale $j$. 

Proposition~\ref{prop:Ij-consistent} demonstrates that $\bar{\mathbf{I}}_j$ is a consistent estimator of $\boldsymbol{\beta}_j$, making it suitable for  the above optimisation problem. Akin to the approach in \cite{cardinali2011costationarity}, we first introduce Assumption~\ref{ass:spectral-boundness} which imposes a decay condition on the marginal wavelet spectra across the scales. (Note this also provides a bound on the cross-spectral entries.)

\begin{assumption}[Uniform spectral boundedness]\label{ass:spectral-boundness}
    There exists a constant $D>0$, independent of $T,\, P,$ and nodes $p$ such that $S_j^{(p,p)}\leq D2^{-j}$, for $j=1,\ldots, J$ and $p=1,\ldots,P$.
\end{assumption}

\begin{proposition}\label{prop:Ij-consistent}
    Let $\XX=\{\XX_t\}_{t=0}^{T-1}$ be a process as in Proposition~\ref{prop:wav_cig} and suppose that Assumption~\ref{ass:spectral-boundness} hold.   
    Let $\boldsymbol{\beta}_j := \sum_{l=1}^J C_{jl}\SS_{l}$, with $C_{jl}$ denoting $(j,l)$-th entry of the discrete autocorrelation wavelet inner product matrix, and let $\bar{\mathbf{I}}_j=\frac{1}{T}\sum_{k=0}^{T-1}\mathbf{I}_{j,k}$ be the time-averaged raw periodogram at scale $j=1,\ldots,J=\log_{2}(T)$.        
    Then, for any fixed, fine enough scale $j$ such that $2^j=o(T)$, $\bar{\mathbf{I}}_j$ is asymptotically unbiased and a consistent estimator of $\boldsymbol{\beta}_j$, with $\mathbb{V}\mathrm{ar}\big(\bar{I}_j^{(p,q)}\big) = \mathcal{O}\big({2^j}{T^{-1}}\big), \, \forall (p,q).$
\end{proposition} 

Appendix~\ref{app:asymptotic-Ij} details the proof. \newline

Note that the penalty term in equation~\eqref{eq:optB} encourages sparsity by forcing zeros in the off-diagonal entries of the inverse $\boldsymbol{\beta}$-spectrum. This differs from the formulation of \cite{gibberd2015multiresolution}, which imposes sparsity directly on the inverse wavelet spectrum $\SS_j^{-1}$. In general, however, $\boldsymbol{\beta}_{j} = \sum_{j=1}^JC_{jl} \SS_l \neq \SS_j$ (see e.g. \cite{knight2024adaptive}), and consequently the zero pattern of $\boldsymbol{\beta}_{j}^{-1}$ does not necessarily coincide with that of $\SS_j^{-1}$. By Proposition~\ref{prop:wav_cig}, it is the latter that characterises the conditional independence graph. We therefore require a construction that addresses this mismatch.

\subsection{Scale-based Wavelet Precision Estimation}\label{sec:sparsity}
We begin by considering an approach that allows us to enforce sparsity directly in the precision matrix, $\SS_{j}^{-1}$, while still using the distributional properties of the wavelet coefficients. 
We propose {\em to define a new process} $\YY = \{\YY_{t}\}_{t=0}^{T-1}$, with corresponding wavelet coefficients $\{\dd_{j,k}^{(Y)}\}_{j,k}$, whose covariance matrices $\{\boldsymbol{\beta}_{j}^{(Y)} \}_{j}$ are designed to match $\{\SS_{j}^{(X)}\}_{j}$. In fact, this would allow us to reformulate the optimisation problem~\eqref{eq:optB} in terms of the precision matrix $\TTheta^{(X)}_{j} := \big(\SS_{j}^{(X)}\big)^{-1}$, and therefore to estimate the $\mathcal{CIG}$ of the process $\XX$ on a {\em scale-by-scale basis}. Note  we now use the superscript $^{(\cdot)}$ to explicitly indicate the process. 

\begin{assumption}[Auxiliary spectral validity]\label{ass:spectral-validity}
    For each scale $j=1,\ldots,J$, the transformed spectral structure $\sum_{l=1}^J (\mathbf{C}^{-1})_{jl} \SS_l^{(X)}$ is positive definite.
\end{assumption}

Assumption~\ref{ass:spectral-validity} ensures that an auxiliary process $\YY$ can be constructed with a well-defined spectral structure $\{\SS_{j}^{(Y)}\}_{j}$ such that $\SS_{j}^{(Y)} := \sum\limits_{l=1}^J (\mathbf{C}^{-1})_{jl} \SS_{l}^{(X)}$, $\forall j$. 
By equation~\eqref{eq:beta-S relationship}, the covariance of the wavelet coefficients associated with $\YY$ then satisfies $\boldsymbol{\beta}_{j}^{(Y)} = \SS_{j}^{(X)}$ at each scale $j$ leading to $\dd_{j,k}^{(Y)} \dot\sim N_P(\mathbf{0}, \SS_{j}^{(X)})$, $\forall k$, as desired. 
 
Our proposal uses the {\em new process} wavelet coefficients, $\{\dd_{j,k}^{(Y)}\}_{k}$, to enforce sparsity in the original precision $\TTheta^{(X)}_{j}$ within each scale $j$.  
Specifically, we formulate {\em at each scale} $j$ an optimisation problem akin to equation~\eqref{eq:optB},
\begin{equation}\label{eq:stay optimisation Y}
    \hat{\TTheta}^{(X)}_{j} = \underset{\TTheta^{(X)}_{j}}{\arg\min}\left\{-T\Big[\log\det\big(\TTheta^{(X)}_{j}\big) - \tr\big(\bar{\mathbf{I}}^{(Y)}_{j}\TTheta^{(X)}_{j} \big)\Big] + \lambda_j\big\|\TTheta^{(X)}_{j}\big\|_{1,\text{off}}\right\}.
\end{equation}
Here, $\bar{\mathbf{I}}_{j}^{(Y)} = \frac{1}{T}\sum_{k=0}^{T-1}\mathbf{I}^{(Y)}_{j,k}$ is the time-averaged periodogram of the process $\YY$, where its stationarity allows smoothing to be carried out over the entire temporal domain. Notice~$\bar{\mathbf{I}}_{j}^{(Y)}$ is an asymptotically unbiased and consistent estimator of $\SS_{j}^{(X)}$ since, by Proposition~\ref{prop:Ij-consistent}, it is an asymptotically unbiased and consistent estimator of $\boldsymbol{\beta}_j^{(Y)}$.

\subsubsection[Process Y Simulation]{Process $\YY$ Simulation}
Since we do not have access to the true spectral structure of the original process $\XX$, we replace it in the above development with its bias-corrected, smoothed estimator $\{\hat{\SS}_{j}^{(X)}\}_j$ in~\eqref{eq:LSW LWS estimate}, with the smoothing being taken over the entire time domain due to the process stationarity. In other words, the new process spectral structure is given by 
\begin{equation}\label{eq:hat-S-j-Y}
    \hat{\SS}_{j}^{(Y)}=\sum\limits_{l=1}^J (\mathbf{C}^{-1})_{jl}\hat{\SS}_{l}^{(X)}, \, \forall j,
\end{equation}
which consistently estimates the target quantity $\SS_{j}^{(Y)} = \sum\limits_{l=1}^J (\mathbf{C}^{-1})_{jl} \SS_{l}^{(X)}, \forall j$ \citep{park2014estimating}.  

\subsubsection[Mitigating Sensitivity to the Simulated Realisation of Y]{Mitigating Sensitivity to the Simulated Realisation of $\YY$} In order to ensure robustness of our proposed procedure to $\YY=\{\YY_t\}_{t=0}^{T-1}$, we use a bootstrap-like procedure. Specifically, we simulate $R$ realisations $\{\YY^{(r)}\}_{r=1}^R$, all sharing the spectral structure $\{\hat{\SS}_{j}^{(X)}\}_{j}$, and obtain the associated smoothed periodograms $\bar{\mathbf{I}}_j^{(\YY^{(r)})}$ at each scale $j$,  for each replicate $r = 1, \ldots,R$. 

At each scale $j$, the average periodogram across the $R$ replicates, 
\begin{equation}\label{eq:S-hat-hat}
    \hat{\hat{\SS}}_j=\frac{1}{R}\sum_{r=1}^R \bar{{\mathbf{I}}}_j^{(\YY^{(r)})},
\end{equation} is a consistent estimator of the original spectral structure $\SS_{j}^{(X)}$, hence we propose its use in solving~\eqref{eq:stay optimisation Y} and yielding $\hat{\TTheta}_j^{(X)}$ as
\begin{equation}\label{eq:S-inverse_optimisation}
    \hat{\TTheta}_j^{(X)} = \underset{\TTheta_j^{(X)}}{\arg\min}  \left\{-T\bigg[\log\det\big(\TTheta_j^{(X)}\big) - \tr\Big(\hat{\hat{\SS}}_j\TTheta_j^{(X)}\Big)\bigg] + \lambda_j\big\|\TTheta^{(X)}_{j}\big\|_{1,\text{off}}\right\}.
 \end{equation}
\bigskip

\subsubsection{Model Selection}\label{sec:model-selection}
We now focus on the tuning parameter $\lambda_j$ of the penalty term in \eqref{eq:S-inverse_optimisation} that controls the sparsity in the precision matrix. For the sake of simplicity, we drop the precision matrix process superscript and refer to $\TTheta^{(X)}_{j}$ as $\TTheta_{j}$.

Motivated by the use of eBIC for tuning parameter selection in the Fourier-domain TSglasso literature, we introduce an analogous criterion for the proposed estimator. At each scale $j$ and location $k$, we define
\begin{equation*}
    \operatorname{eBIC}_{j,k}(\lambda) = -2\mathcal{L}_{j,k}(\TTheta_j)+\log(T)E_j + 4\gamma\log(P)E_j,
\end{equation*}
with $\mathcal{L}_{j,k}$ as in~\eqref{eq:L-scale-time}, $E_j$ is the number of non-zero elements in $\TTheta_j$, and $\gamma \in [0,1]$ is an hyperparameter.
As the likelihood-based quantity~\eqref{eq:L-star} aggregates the location-specific contributions over $k$, we similarly construct the scale-specific eBIC-inspired criterion
\begin{equation*}
    \operatorname{eBIC}^{*}_{j}(\lambda) 
    = \sum_{k=0}^{T-1} \operatorname{eBIC}_{j,k}(\lambda)
    = -2\mathcal{L}_j^*(\TTheta_j) +T\log(T)E_j +4T\gamma\log(P)E_j
\end{equation*}
Since optimisation problem \eqref{eq:S-inverse_optimisation} is defined separately for each scale $j$, the information criteria above is scale-dependent and the tuning parameter $\lambda$ will also vary with $j$.

\subsubsection{\texttt{WavTSglasso} Algorithm}
Algorithm~\ref{alg:WavTSglasso} summarises our proposed wavelet-based time series graphical lasso method, which we refer to as \texttt{WavTSglasso}. At each scale, the optimisation problem \eqref{eq:S-inverse_optimisation} is solved numerically via the \texttt{glasso} algorithm \citep{glasso}. This approach leverages a block coordinate descent procedure, and applies the soft-thresholding operator to induce exact zeros in the resulting precision estimates $\hat{\Theta}_{j}$.

\begin{algorithm}[H]
\caption{\texttt{WavTSglasso} for Sparse Precision Estimation}
\label{alg:WavTSglasso}
\KwInput{$\mathbf{x}=\{\mathbf{x}_t\}_{t=0}^{T-1}$, a sample of the time series $\XX$ of length $T=2^{J}$ for some $J\in\mathbb{N}$; $R$, a number of bootstrap replicates.}
\KwOutput{$\{\hat{\boldsymbol{\Theta}}_j\}_{j=1}^J$, the estimated scale-dependent, sparse precision matrices.}
\BlankLine
\BlankLine
Obtain $\{\hat{\SS}^{(X)}_{j}\}_{j}$ from $\mathbf{x}$ as in \eqref{eq:LSW LWS estimate} smoothed over the entire time-domain.\;
Compute $\{\hat{\SS}^{(Y)}_{j}\}_{j}$ as in \eqref{eq:hat-S-j-Y}.\;

\For{$r=1,\dots,R$}{
  Simulate $\mathbf{y}^{(r)}$ with $\{\hat{\SS}^{(Y)}_{j}\}_{j}$.\;
  Compute smoothed periodogram $\{\bar{\mathbf{I}}^{(\YY^{(r)})}_{j}\}_{j}$.\;
}

Compute $\big\{\hat{\hat{\SS}}_j\big\}_{j}$ as in \eqref{eq:S-hat-hat}.

\BlankLine
\For{$j=1,\dots,J$}{
  Select $\lambda_j$ by minimising the chosen information criterion.\;
  Obtain $\hat{\TTheta}_{j}$ via solving \eqref{eq:S-inverse_optimisation} by means of the \texttt{glasso} algorithm.\;
}

\end{algorithm}

\subsection[\texttt{WavTSglasso} for CIG Estimation]{\texttt{WavTSglasso} for $\mathcal{CIG}$ Estimation}\label{sec:scale-selection} 

While Fourier-based TSglasso aggregates estimated precision matrices across frequencies (Section~\ref{sec:Fourier}), our approach estimates the precision matrix at each scale. Consequently, these must be combined across scales to infer the $\mathcal{CIG}$ of the process $\XX$.
As shown in Proposition~\ref{prop:wav_cig}, the scale-dependent conditional independence graphs coincide across all scales. Thus, we propose to use the following \textit{similarity measure} for each pair of scales $j$ and $j'$,
\begin{equation}\label{eq:sim measure non scaled}
    \Delta(j,j') = \sum_{p=1}^P\sum_{q\neq p} \mathds{1} \big\{ \big( \TTheta_{j} \big)_{p,q}\neq0, \big( \TTheta_{j'} \big)_{p,q}\neq 0 \big\}, 
\end{equation}
from which we can compute the \textit{scaled similarity measure}
\begin{equation}\label{eq:sim measure}
    \Delta_{S}(j,j')=\frac{\Delta(j,j')}{\sqrt{E_j}\sqrt{E_{j'}}}. 
\end{equation}
The similarity measure in equation~\eqref{eq:sim measure non scaled} calculates the number of common predicted edges in scales $j$ and $j'$, or put simply, it identifies the common dependencies across scales. Equation~\eqref{eq:sim measure} then scales $\Delta(j,j')$ by the number of edges corresponding to each of the scales $j$ and $j'$ individually, so that $\Delta_S(j,j')=1$ if the two the scales $j$ and $j'$ yield exactly the same edge(s). We compute $\Delta_S(j,j')$ for every pair $(j,j')$ and choose the scale pair $({j}_1, {j}_2)$ which maximises $\Delta_S(j,j')$. If there are no edges in either of the scales, then $\Delta(j,j')=0$, and thus $\Delta_S(j,j')=0$. In line with both our theoretical results and with previous literature \citep{sanderson2010nonstationary, mcgonigle2020wavelet}, we use scales $j=1, \ldots, J^{\star}=\lfloor 0.7J \rfloor$.

Our proposal for the estimation of the conditional graph dependencies is to  {\em combine the estimated precision matrices corresponding to the pair of scales with highest scaled similarity}, $\hat{\TTheta}_{j_1}$ and $\hat{\TTheta}_{j_2}$, in order to construct the estimated $\mathcal{CIG}$ adjacency, $\hat{\mathbf{G}}$. Specifically, we define $\hat{\mathbf{G}}_{p,q} = 1$ if $(\hat{\TTheta}_{j_1})_{p,q} \neq 0$ or $(\hat{\TTheta}_{j_2})_{p,q} \neq 0$, and 0 otherwise.
This approach aims to balance the trade-off between sparsity and accuracy. Indeed, when computing the similarity, we focus on edges that are commonly predicted across scales, while the final adjacency incorporates all edges identified at either of the selected scales.

We anticipate that Theorems~\ref{thm1}--\ref{thm3} in Section~\ref{sec:Theoretical Properties} establish fixed wavelet scale results; the data-driven selection of the pair $(j_1,j_2)$ introduced above is an additional aggregation step and is not covered by these results. We therefore regard this scale-selection procedure as a practical aggregation strategy similar to that needed in the Fourier TSglasso setting, whose performance is investigated empirically through simulation studies in Section~\ref{sec:Simulation Study}.

Our proposed $\mathcal{CIG}$ estimation method is summarised in Algorithm~\ref{alg:WavTSglasso-CIG}.

\paragraph{Similarity measures using more scales.}
While we propose to compare two scales at a time, equations~\eqref{eq:sim measure non scaled} and~\eqref{eq:sim measure} can be easily extended to include more scales if necessary. For example, the corresponding similarity measure formulae with three scales are
\begin{eqnarray*}
    \Delta(j,j',j'') &=& \sum_{p=1}^P\sum_{q\neq p} \mathds{1} \big\{ \big( \TTheta_{j} \big)_{p,q}\neq0, \big( \TTheta_{j'} \big)_{p,q}\neq 0, \big( \TTheta_{j''} \big)_{p,q}\neq 0 \big\}, \\
    \Delta_S(j,j',j'')&=&\frac{\Delta(j,j',j'')}{\sqrt[3]{E_j}\sqrt[3]{E_{j'}}\sqrt[3]{E_{j''}}} . 
\end{eqnarray*}

\begin{algorithm}[H]
\caption{\texttt{WavTSglasso} for $\mathcal{CIG}$ Estimation}
\label{alg:WavTSglasso-CIG}
\KwInput{$\{\hat{\boldsymbol{\Theta}}_j\}_{j=1}^J$, estimated sparse precision matrices obtained using Algorithm~\ref{alg:WavTSglasso}.}
\KwOutput{$\hat{\mathbf{G}}$, the estimated $\mathcal{CIG}$  adjacency.}
\BlankLine
\BlankLine

\For{$j=1,\dots,J^{\star}$}{
    \For{$j'=j+1,\ldots,J^{\star}$}{
        Compute scaled similarity $\hat\Delta_S(j,j')$ as in \eqref{eq:sim measure}.
  }
}

Obtain $(j_{1},j_{2}) = \arg\max_{(j,j')} \hat{\Delta}_{S}(j,j')$.\;
Form $\hat{\mathbf{G}}$ by combining $\hat{\TTheta}_{j_{1}}$ and $\hat{\TTheta}_{j_{2}}$.\;

\end{algorithm}


\section{Theoretical Properties}\label{sec:Theoretical Properties}

In this section, we establish theoretical properties of the estimator in our proposed \texttt{WavTSglasso} algorithm. Throughout this section we consider growing network-dimension with time hence we denote $P=P_T$ allowed to increase as $T \rightarrow \infty$ subject to rate conditions specified in the following results.

We start by establishing a high-probability Frobenius norm bound of the proposed precision estimator obtained via the optimisation problem~\eqref{eq:S-inverse_optimisation}  at a fixed wavelet scale $j$. Theorem~\ref{thm1} below provides an explicit convergence rate on the estimation error under the following regularity conditions. We first establish notation, and make use of the following three additional assumptions. 

\begin{assumption}[Uniform covariance summability]\label{ass:covariance-summability}
    The $P_T$-dimensional process $\XX$ is zero-mean, stationary and Gaussian, with 
    \begin{equation*}
        \max_{1\leq p,q\leq P_T} \sum_{\tau \in \mathbb{Z}} |c_X^{(p,q)}(\tau)| \leq M_X <\infty, 
    \end{equation*} 
for some constant $M_X$ independent of $T$ and $P_T$. 
\end{assumption}

\begin{assumption}[Uniform mvLSW regularity]\label{ass:mvLSW-regularity}
    For the process $\XX$, the entries $V_j^{(p,q)}(u)$ of the mvLSW transfer function matrix are Lipschitz continuous in $u \in [0,1]$, with Lipschitz constants $L_{j}^{(p,q)}$ satisfying $$\max_{1 \leq p,q\leq P_T}\sum_{j=1}^J 2^j L_{j}^{(p,q)}<\infty.$$ 
\end{assumption}

\begin{assumption}[Uniform eigenvalue bounds]\label{ass:eigenvalues}
    For the process $\XX$, the eigenvalues of $\TTheta_{j}$ lie in the interval $[\phi_{\min}^{j}, \phi_{\max}^{j}]$, where the minimum and maximum eigenvalues satisfy $0<\phi_{\min}^{j}<\phi_{\max}^{j}<\infty$, for each scale $j$. 
\end{assumption}

\paragraph{Practical interpretation for Assumptions~\ref{ass:covariance-summability}-\ref{ass:eigenvalues}.} Assumption~\ref{ass:covariance-summability} requires the auto- and cross- covariances to be absolutely summable, so temporal dependence must decay sufficiently quickly and not become increasingly strong as the dimension grows. Assumption~\ref{ass:mvLSW-regularity} ensures that the mvLSW regularity and approximation conditions hold uniformly across component pairs as the dimension increases. Assumption~\ref{ass:eigenvalues} requires the eigenvalues of the true precision matrix to remain bounded away from both $0$ and~$\infty$, preventing the problem from becoming ill-conditioned. Collectively, these assumptions ensure that the spectral estimator remains sufficiently well-behaved as $P_T$ increases and that the underlying graphical model remains well-conditioned. Naturally, assumptions made on the $\XX$ are mirrored by its replicates.

\begin{definition}[Scale-specific graph support]\label{def1}
    Define the support of the non-zero off-diagonal entries of the precision matrix at a fixed scale $j$ to be $\mathcal{E}_{j}=\{(p,q):(\TTheta_{j})_{p,q}\neq 0,\, p\neq q\}$, where $\TTheta_{j}$ denotes the true inverse spectrum at scale $j=1, \ldots, J=\log_2(T)$. Denote its complement as $\mathcal{E}_j^c=\{(p,q):(\TTheta_{j})_{p,q}= 0,\, p\neq q\}$. Let $E_j:=|\mathcal{E}_j|$ denote the number of non-zero off-diagonal entries of $\TTheta_j$.  In particular, $0\leq E_{j}\leq P_T(P_T-1)$, where $E_j$ may depend on $P_T$.
\end{definition}

\begin{theorem}[High-probability Frobenius norm bound of the penalised precision estimator]\label{thm1}
    Let $\hat{\TTheta}_j$ denote the penalised estimator of $\TTheta_j$ at a fixed wavelet scale $j$, obtained from solving optimisation problem \eqref{eq:S-inverse_optimisation} by means of Algorithm~\ref{alg:WavTSglasso}. Suppose for $\nu>2$, $JP_T^\nu/T \rightarrow 0$ as $T \rightarrow \infty$, where $P_T$ is allowed to increase with $T$. Let $R\geq 1$ be fixed, and define
    \begin{equation}\label{eq:ajT}
        a_{j,T}=\frac{\tilde{\mu}_j}{\sqrt{T}}+\sqrt{\frac{J\kappa^*_jP_T^\nu}{T}},
    \end{equation}
    and $r_{j,T}=P_Ta_{j,T}$, where $\tilde{\mu}_j,\kappa_j^*>0$ are finite constants, independent of $T$ and $P_T$, such that
    \begin{equation*}
        \max_{1\leq p,q\leq P_T}\bigg|\mathbb{E}\Big(\hat{\hat{S}}_j^{(p,q)}\Big)-S_j^{(p,q)}\bigg|\leq \frac{\tilde{\mu}_j}{\sqrt{T}}
        \quad \text{ and }\quad
        \max_{1 \leq p,q \leq P_T} \mathbb{V}\mathrm{ar}\Big(\hat{\hat{S}}_j^{(p,q)}\Big) \leq \frac{J\kappa_j^*}{T},
    \end{equation*}
    where we recall that $\hat{\hat{\SS}}_j$ denotes the replicate and time averaged periodogram in equation~\eqref{eq:S-hat-hat}. Suppose Assumptions \ref{ass:spectral-boundness}-\ref{ass:eigenvalues} hold and $r_{j,T}=o(1)$. Let $\lambda_j'=\lambda_j/T$ satisfy $\lambda_j'\leq C_2^j a_{j,T}$ for some finite constant $C_2^j>0$, and define $C_1^j=\frac{1}{2(\phi_{\max}^j+\eta)^2}$. Then, for any $\tilde{C}_j>\frac{1+C_2^j}{C_1^j}$,
    \begin{equation*}
        \mathbb{P}\bigg(\Big\|\hat{\TTheta}_j-\TTheta_j\Big\|_F> \tilde{C}_jr_{j,T}\bigg)\leq P_T^{2-\nu}
    \end{equation*}
    for all sufficiently large $T$, where $\|\mathbf{M}\|_F=\sqrt{\tr(\mathbf{M}^{\top} \mathbf{M})}$ for a generic square matrix $\mathbf{M}$.   
    Consequently, as $T \rightarrow \infty$ and $r_{j,T}=o(1)$,
    \begin{equation*}
        \Big\|\hat{\TTheta}_j-\TTheta_j\Big\|_F=\mathcal{O}_P(r_{j,T}) \,\,\,\,\,\, \mbox{ and } \,\,\,\,\,\,\,\Big\|\hat{\TTheta}_j-\TTheta_j\Big\|_F=o_P(1) 
    \end{equation*}
    hold with probability at least $1-P_T^{2-\nu}$.
\end{theorem}
\medskip
The proof of Theorem~\ref{thm1} appears in Appendix~\ref{app:proofthm1}, and is heavily reliant on three lemmas, introduced therein. 

We now consider stronger element-wise and graph recovery properties of the estimator. In order to do so we impose two additional conditions controlling the interaction between entries corresponding to the true graph and those outside its support, and to restrict the growth of the graph complexity as $P_T$ increases.

\begin{assumption}[Mutual incoherence]\label{ass:mutual-incoherence}
    Let $\mathcal{D}_T=\big\{(p,p): p=1,\ldots, P_T\big\}$ denote the diagonal indices and define the augmented support $\mathcal{S}_j=\mathcal{E}_j \cup \mathcal{D}_T$, where $\mathcal{E}_j$ is the off-diagonal support of $\TTheta_j$. Let $\boldsymbol{\Gamma}_j=\SS_j \otimes \SS_j$, where $\otimes$ denotes the usual Kronecker product. Assume that there exists a constant $\alpha_j\in (0,1]$, independent of $T$ and $P_T$ such that $$\max_{e \in \mathcal{S}_j^c} \bigg \|\big(\boldsymbol{\Gamma}_j\big)_{e\mathcal{S}_j}\Big[\big(\boldsymbol{\Gamma}_j\big)_{\mathcal{S}_j\mathcal{S}_j}\Big]^{-1}\bigg\|_1 \leq 1-\alpha_j,$$
    where $\|\mathbf{m}\|_1=\sum_{i=1}^K |m_i|$ for a general vector $\mathbf{m}=(m_1,\ldots, m_K)$.
\end{assumption}

\begin{assumption}[High-dimensional regularity]\label{ass:high-dim-regularity}
    For each fixed scale $j$, define $\text{deg}_{j,T}$ as the maximum number of non-zero entries in any row of the precision matrix $\TTheta_j$. Let $\kappa_{S,j,T}=\|\SS_j\|_{\infty,\text{op}}$ and $\kappa_{\Gamma,j,T}=\Big\|\big[(\boldsymbol{\Gamma}_j)_{\mathcal{S}_j\mathcal{S}_j}\big]^{-1}\Big\|_{\infty,\text{op}}$ where $\mathcal{S}_j$ denotes the augmented support defined in Assumption~\ref{ass:mutual-incoherence} and $\|\mathbf{M}\|_{\infty,\text{op}}=\max_p\sum_q |M_{pq}|$ for a generic square matrix $\mathbf{M}$. Let $a_{j,T}$ be as defined in equation~\eqref{eq:ajT} and assume that as $T \rightarrow \infty$, 
    \begin{equation}\label{eq:assumption7}
        \text{deg}_{j,T}(1+8/\alpha_j)\max\{\kappa_{S,j,T}\kappa_{\Gamma,j,T},\kappa_{S,j,T}^3\kappa_{\Gamma,j,T}^2\} a_{j,T} \rightarrow 0.
    \end{equation}
\end{assumption}

\paragraph{Practical interpretations of Assumptions~\ref{ass:mutual-incoherence} and~\ref{ass:high-dim-regularity}.} Assumption~\ref{ass:mutual-incoherence} requires the coordinates of the Hessian corresponding to absent edges not to be too strongly coupled to those in the true augmented support. This condition allows the graphical lasso to distinguish between true and absent edges, helping to prevent false-positive edge selection. Assumption~\ref{ass:high-dim-regularity} controls the graph complexity and the relevant matrix norms relative to the spectral estimation error, ensuring the remainder terms in the primal-dual witness argument are sufficiently small.

Under these additional conditions, Theorem~\ref{thm2} combines the concentration result for the spectral estimator and the primal-dual witness argument of \cite{Ravikumar2011consistency} to obtain a high-probability element-wise error bound for the penalised precision estimator.

\begin{theorem}[High-probability element-wise bound for the penalised precision estimator]\label{thm2}
    Suppose Assumptions \ref{ass:spectral-boundness}--\ref{ass:high-dim-regularity} hold. Let $T \rightarrow \infty$ and $P_T$ allowed to increase with $T$, such that $JP_T^\nu/T \rightarrow 0$ for $\nu > 2$. Then, for each fixed scale $j$, with $\lambda_j'=\lambda_j/T=\frac{8}{\alpha_j} a_{j,T}$, where $a_{j,T}$ is defined in equation~\eqref{eq:ajT}, the graphical lasso estimator satisfies $$\mathbb{P}\Bigg(\Big\| \hat{\TTheta}_j-\TTheta_j\Big\|_\infty > 2\kappa_{\Gamma,j,T}\Bigg(1+\frac{8}{\alpha_j}\Bigg)a_{j,T}\Bigg)\leq P_T^{2-\nu},$$ 
    where $\|\mathbf{M}\|_\infty=\max_{p,q}|M_{p,q}|$ for a generic square matrix $\mathbf{M}$. 

    Hence, as $T \rightarrow \infty$, $a_{j,T}\rightarrow 0$, and since $\kappa_{\Gamma,j,T}$ and $\alpha_j$ are constants,
    \begin{equation*}
        \Big\| \hat{\TTheta}_j-\TTheta_j\Big\|_\infty=\mathcal{O}_P\Big(a_{j,T}\Big) \,\,\,\,\,\,\mbox{ and }\,\,\,\,\,\,\,\Big\| \hat{\TTheta}_j-\TTheta_j\Big\|_\infty=o_P(1)
    \end{equation*}
\end{theorem} 

The proof of Theorem~\ref{thm2} appears in Appendix~\ref{app:proof-of-graph-recovery}.

While the above result controls the element-wise estimation error, this alone does not guarantee exact graph recovery, since a true non-zero entry may be too small relative to the error bound. To ensure exact recovery with high probability, we additionally require the following assumption.  non-zero entries of the true precision matrix to remain sufficiently separated from zero.

\begin{assumption}[Minimal signal strength]\label{ass:theta-min}
    For each fixed scale $j$ such that $\mathcal{E}_j \neq \emptyset$, define $\theta_{\min,j,T} := \min_{(p,q)\in \mathcal{E}_j} |(\Theta_j)_{p,q}|$, and let $a_{j,T}$ be as in equation~\eqref{eq:ajT}. For sufficiently large $T$, assume $$\theta_{\min,j,T}>2\kappa_{\Gamma,j,T}(1+8/\alpha_j)a_{j,T}.$$
\end{assumption}

\paragraph{Practical interpretation of Assumption~\ref{ass:theta-min}.} Assumption~\ref{ass:theta-min} requires the smallest non-zero entry of the true precision matrix to remain sufficiently large relative to the element-wise precision matrix estimation bound. This ensures that the true edges do not become indistinguishable from $0$ as $P_T$ increases, helping prevent false-negative edge selection.

Theorem~\ref{thm3} next combines the element-wise bound of Theorem~\ref{thm2} with the additional minimum signal condition to establish exact graph recovery of the true edge set with high probability. The argument parallels the model selection consistency result of \citet[Theorem 2]{Ravikumar2011consistency}.

\begin{theorem}[High-probability graph support recovery]\label{thm3}
    Define the estimated edge set at scale $j$ by $\hat{\mathcal{E}}_j=\Big\{(p,q): p \neq q, \big(\hat{\TTheta}_j\big)_{p,q}\neq 0\Big\}$. Under the Theorem~\ref{thm2} setup and notation, and under the additional Assumption~\ref{ass:theta-min},  as $T \rightarrow \infty$, we have for each fixed scale $j$, $\mathbb{P}\Big(\hat{\mathcal{E}}_j = \mathcal{E}_j\Big) \geq 1-P_T^{2-\nu}$.
\end{theorem}

The proof of Theorem~\ref{thm3} appears in Appendix~\ref{app:proof-of-graph-recovery}.

Theorems \ref{thm1}--\ref{thm3} concern the fixed-scale results for a tuning parameter satisfying the stated rate condition. The results established therefore consider an admissible tuning parameter rather than theoretical guarantees of the data-driven eBIC selection procedure as introduced in Section~\ref{sec:model-selection}. The eBIC-inspired criterion is used as a practical method for selecting the tuning parameter and is evaluated empirically in Section~\ref{sec:Simulation Study}.



\section{Simulation Study}\label{sec:Simulation Study}
In this section, we present the results from several simulation studies conducted to evaluate the performance of our proposed \texttt{WavTSglasso} methodology under different scenarios. In particular, we assess how the process dimension $P$, the length of the time series $T$, and the variability in the sparsity levels (the number of edges relative to their potential total, i.e., $s:=|\mathcal{E}|/\binom{P}{2}$) and in the connection intensities affect $\mathcal{CIG}$ estimation. 

As performance metrics, we compute the following edge detection rates: true positive rate (TPR), false positive rate (FPR), true discovery rate (TDR) and $F_1$-score based on the estimated adjacency matrix $\hat{\mathbf{G}}$ compared to the true adjacency matrix $\mathbf{G}$ for the conditional independence graph. The $F_1$-score provides a summary measure of graph recovery performance by balancing the proportion of true edges recovered against the reliability of the identified edges. Let TP and FP denote the number of true positives and false positives, respectively, and let TN and FN denote the number of true negatives and false negatives. Then,
\begin{equation*}
\text{TPR} = \frac{\text{TP}}{\text{TP} + \text{FN}}, \quad \text{FPR} = \frac{\text{FP}}{\text{FP} + \text{TN}}, \quad \text{TDR} = \frac{\text{TP}}{\text{TP} + \text{FP}}, \quad F_1=\frac{2\times\mathrm{TPR}\times \mathrm{TDR}}{\mathrm{TPR}+\mathrm{TDR}}.
\end{equation*}

Section \ref{sec:bootstrap-study} considers an empirical evaluation of {\texttt{WavTSglasso}} when varying the number of bootstrap samples of $\YY$, $R$. 
We then compare our proposed wavelet-based TSglasso algorithm with Fourier-based TSglasso methods, with either $\ell_1$ or $\ell_2$ penalisation.
All methods are implemented in \texttt{R} \citep{Rcore}.
We use the \texttt{glasso} package \citep{glasso} to solve the optimisation problem \eqref{eq:S-inverse_optimisation}, yielding sparse estimates of the wavelet precision matrix $\TTheta_{j}$ at each scale $j$ via a block coordinate descent algorithm.
For the Fourier-$\ell_1$ approach, we implement the ADMM algorithm to optimise \eqref{eq:Fourier-TSglasso}, as proposed by \cite{jung2015graphical}. In addition, we exploit separability of \eqref{eq:Fourier-TSglasso} and solve it at each frequency via the \texttt{glasso} package.
For the Fourier-$\ell_2$ approach, we adopt the ADMM implementation in the \texttt{tsglasso} package by \cite{dallakyan2022time}. Although their method is a two-stage procedure for sparse VAR estimation, its first stage estimates the precision matrix function using likelihood \eqref{eq:Fourier-TSglasso} with an $\ell_2$-penalty.

\paragraph{Data Generating Processes.}
Each simulation study comprises $K=50$ replicates of a stationary process with known $\mathcal{CIG}$, broadly pertaining to a vector autoregressive structure. 

We generate data based on 10-, 20-, and 25-dimensional processes with varying graphs and time-dependent dynamics. In particular, some of these processes are simulated using the generalised network autoregressive (GNAR) model \citep{knight2020generalized, GNAR}, a multivariate autoregressive process whose univariate series are associated with the nodes of a given graph, $\mathcal{N}$. 
We use the $r$-stage neighbour definition of \cite{knight2020generalized} which establishes two nodes as $r$-stage neighbours if and only if the shortest path between them comprises $r$ edges of $\mathcal{E^N}$. Then, in the GNAR framework, each nodal time series depends on its own past and on the past of its neighbours. Mathematically, a $P$-dimensional zero-mean Gaussian process $\XX$ is GNAR($p,[\mathbf{s}]$) on the given $\mathcal{N}$ if
\begin{equation*}
    \XX_{t} = \sum_{l=1}^{p} \mathbf{A}_l \XX_{t-l} + \vveps_{t}, \, \text{ where } \, \mathbf{A}_l = \operatorname{diag}(\alpha_{l,1},\ldots,\alpha_{l,P}) + \sum_{r=1}^{s_l} \beta_{l,r} \mathbf{W}^{(r)},
\end{equation*}
and $\mathbf{W}^{(r)}$ denotes the (weighted) neighbour structure at stage $r$. 
Here, $p\in\mathbb{N}$ is the maximal autoregressive lag, $[\mathbf{s}]=(s_{1},\ldots,s_{p})$ with $s_l\in\mathbb{N}$ the maximal stage dependence at lag $l$, $\alpha_{l,i}$ is the $i$-th nodal autoregressive coefficient at lag $l$, and $\beta_{l,r}$ is the network autoregressive coefficient for the $r$-th stage neighbour at the $l$-th lag.

A subtle but important point to make here is that the given network structure $\mathcal{N}$ differs from the conditional independence graph $\mathcal{CIG}$. As noted by \cite{nason2023new}, inter-variable relationships depend on nodes up to twice the largest stage dependence. For example, in a GNAR(1,[1]) model, the adjacency matrix $\mathbf{G}$ includes both first- and second-stage neighbours in $\mathcal{N}$. The numerical experiments for the $\mathcal{CIG}$ estimation across GNAR processes are presented in Section~\ref{sec:ER-models} and \ref{sec:Ring-Models}, and for VAR and VARMA processes in Section~\ref{sec:VAR Models}. We postpone the recovery of the underlying graph $\mathcal{N}$ of GNAR models to Section~\ref{sec:graph-discovery}. Table~\ref{tbl:Sparsity} (Appendix~\ref{app:Sparsity-tbl}) reports the sparsity level of the $\mathcal{CIG}$ for each simulated model in Sections \ref{sec:ER-models}--\ref{sec:VAR Models}. 

\paragraph{Computational Considerations.} The matrix multiplications that yield the input spectral proxy, $\{\hat{\SS}_j^{(X)}\}_j$ and the auxiliary spectral estimates $\{\hat{\SS}_j^{(Y)}\}_j$ occasionally may result in matrices that are not positively definite. As \cite{knight2024adaptive} document, such issues only occur when using wavelets with short support, e.g. Haar wavelets, and regularisation may be required. In practice, we regularise the estimates using the method of \cite{schnabel}, as also implemented by \cite{park2014estimating}.

Furthermore, for each scale $j$, we restrict the search for $\lambda$ to a range $(\lambda_j^l, \lambda_j^u)$, following a similar approach to that in \cite{tugnait2022sparse}. First, we compute $\lambda_j^{sm}$, defined as the smallest value of $\lambda$ for scale $j$ for which the estimated network contains no edges, i.e., $\hat{\TTheta}_j$ is a diagonal matrix. We then set $\lambda_j^u = \lambda_j^{sm}/3$ and $\lambda_j^l = \lambda_j^u/10$. This choice of $\lambda$ values is dependent on the scale representation of the observed process, and excludes both extremely dense and extremely sparse graphs, allowing us to focus the search on a scale-specific range that yields more interpretable and practically relevant models.

\subsection{Empirical Evaluation of Bootstrapping}\label{sec:bootstrap-study}

We begin by empirically assessing the sensitivity of our proposed methodology to the number of bootstrap replicates $(R)$, with the aim of identifying a suitable choice for the subsequent simulation study. Recall that our proposed \texttt{WavTSglasso} methodology averages the time-smoothed wavelet periodograms across $R$ realisations of the auxiliary process $\YY$ (see equation~\eqref{eq:S-inverse_optimisation}), to reduce variability associated with any individual simulated realisation. We consider $R \in\{1,10,50\}$ and examine the extent to which increasing $R$ improves the accuracy of the estimated conditional independence graph.

\begin{figure}[!ht]
    \centering
    \begin{subfigure}[H]{0.3\linewidth}
    \includegraphics[width=\linewidth]{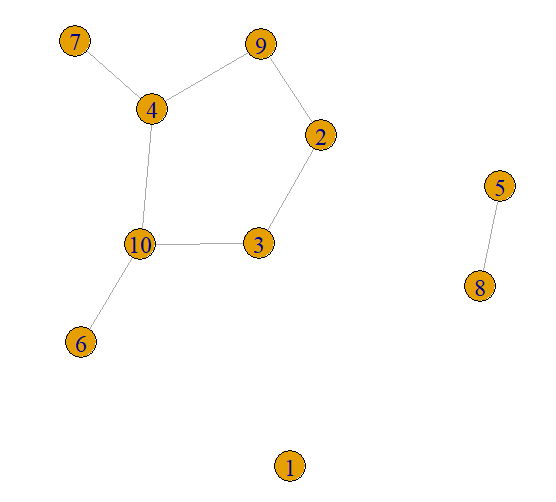}\bigskip
    \caption{ Sparse ER}
    \end{subfigure}
    \hspace{20mm}
    \begin{subfigure}[H]{0.3\linewidth}
    \includegraphics[width=\linewidth]{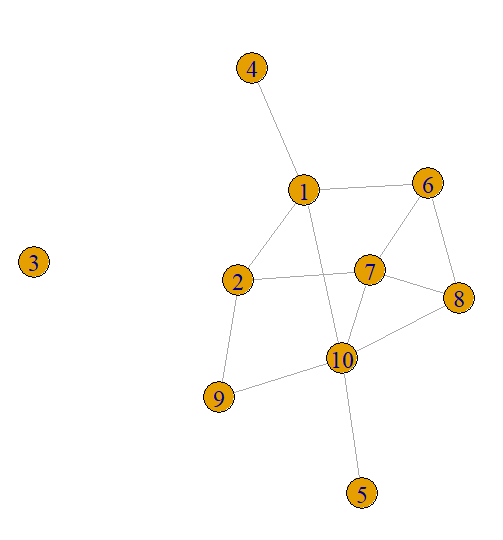}\bigskip
    \caption{ Dense ER}
    \end{subfigure}
    
    \caption{Two representative 10-dimensional Erd\H{o}s-Rényi graphs with edge probability $\rho=0.1$ for the Sparse ER and $\rho=0.25$ for the Dense ER.}
  \label{fig:10D-ER Models}
\end{figure}

To do this, we generate 10-dimensional Erd\H{o}s-Rényi (ER) graphs \citep{erdos1959random}, where edges are formed between pairs of nodes with a given probability $\rho$. Hence, the level of sparsity of the graph can be controlled by varying $\rho$. We consider two probability levels, $\rho \in\{0.1, 0.25\}$, which we refer to throughout as the {\em Sparse ER} and {\em Dense ER} settings, respectively, and generate 10 random graphs for each setting. Note that our methodology accommodates general graph topologies, including those with disconnected nodes, as shown in the illustrative examples of ER graphs in Figure~\ref{fig:10D-ER Models}. For every generated graph, we then simulate $K=50$ replicates of a GNAR(1,[1]) model with $\beta_{1,1} = 0.85$ and with lengths $T \in \{256,1024,2048\}$. We then apply the proposed methodology for each $R \in \{1,10,50\}$. 

Table~\ref{tbl:ER Model Boot T=1024} (below) reports the averaged TPR, FPR, TDR and $F_1$ scores for $T=1024$, whilst Tables~\ref{tbl:ER Model Boot T=256} and \ref{tbl:ER Model Boot T=2048} (Appendix \ref{app:add_sims}) provide the corresponding rates for $T=256$ and $T=2048$.
Across all time series lengths, increasing the number of bootstrap simulations improves performance, with improvements already observed when increasing $R$ from $1$ to $10$, particularly for the denser graphs. Given the generally marginal improvements in graph recovery beyond $R=10$ and the additional computational cost associated with larger values of $R$, we set $R=10$ for the remainder of the simulation study. For both sparsity settings, increasing the time series length leads to a reduction in the FPR and a corresponding increase in the TDR, whilst the TPR remains relatively stable for larger $T$ - indicating increasing $T$ improves the reliability of the identified dependencies. 

\begin{table}[!ht]
    \centering
    \begin{tabular}{l*{8}{c}}
    \toprule
    & \multicolumn{4}{c}{Sparse ER} & \multicolumn{4}{c}{Dense ER} \\
    \cmidrule(lr){2-5}\cmidrule(lr){6-9}
    & TPR & FPR & TDR & $F_1$ & TPR & FPR & TDR & $F_1$ \\
    \midrule

$R=1$
& 0.9190 & 0.0174 & 0.9522 & 0.9353
& 0.5968 & 0.1078 & 0.9072 & 0.7200\\
\midrule

$R=10$
& 0.9189 & 0.0156 & 0.9584 & 0.9382
& 0.6729 & 0.1101 & 0.9105 & 0.7739\\
\midrule

$R=50$
& 0.9132 & 0.0131 & 0.9607 & 0.9363
& 0.6806 & 0.1140 & 0.9087 & 0.7783 \\

\end{tabular}
\caption{10-dimensional GNAR(1,[1]) models over Erd\H{o}s-Rényi graphs for $T=1024$.}
\label{tbl:ER Model Boot T=1024}
\end{table}

\subsection{Effect of Graph Sparsity}\label{sec:ER-models}
We now investigate the performance of our proposed methodology in recovering ER graphs under varying levels of sparsity. The corresponding edge detection rates over the $K=50$ series realisations for all methods with $T=1024$ are reported  in Table~\ref{tbl:ER Model Results GNAR(1,[1]),T=1024} below.

Our \texttt{WavTSglasso} estimation procedure performs particularly well in sparse settings, consistently achieving lower FPR and higher TDR values than the competitors. Although the Fourier methods achieve higher TPR values, these come at the expense of identifying a substantially larger number of spurious connections. In contrast, the proposed wavelet approach maintains high TPR values whilst achieving accurate identification of absent edges.

In denser settings, the {\texttt{WavTSglasso}} and Fourier methods become more comparable. In particular, Fourier-$\ell_1$ (glasso) achieves a higher TPR at the expense of a slightly higher FPR and consequently lower TDR, with both methods providing a favourable balance between true edge recovery and sparsity preservation.

\begin{table}[!ht]
\begin{center}
\setlength{\tabcolsep}{4pt}
\begin{tabular}{l*{8}{c}}
\toprule
      & \multicolumn{4}{c}{Sparse ER} & \multicolumn{4}{c}{Dense ER} \\
\cmidrule(lr){2-5}\cmidrule(lr){6-9}
  & TPR & FPR & TDR & $F_1$ & TPR & FPR & TDR & $F_1$ \\
\midrule

W - $\ell_1$ 
& \textbf{0.9189} & \textbf{0.0156} & \textbf{0.9584} &\textbf{0.9382}
& \textbf{0.6729} & \textbf{0.1101} & \textbf{0.9105} & \textbf{0.7739} \\
\midrule

F - $\ell_1$ (glasso) 
& 0.9781 & 0.0870 & 0.6851 & 0.8058
& \textbf{0.8226} & \textbf{0.1403} & \textbf{0.8535} & \textbf{0.8378}\\ 
\midrule

F - $\ell_1$ (ADMM) 
&  0.9716 & 0.3047 & 0.3826 & 0.5490
& 0.8440  & 0.2549 & 0.7518 & 0.7952 \\
\midrule

F - $\ell_2$ (ADMM) 
&  0.9783 & 0.1074 & 0.7403 & 0.8428
&  0.8932 & 0.3448 & 0.7777 & 0.8315\\

\end{tabular}
\caption{10-dimensional GNAR(1,[1]) model over Erd\H{o}s-Rényi graphs for $T=1024$. Values in bold identify the best method(s) for $\mathcal{CIG}$ recovery across the four aggregated measures.}
\label{tbl:ER Model Results GNAR(1,[1]),T=1024}
\end{center}
\end{table}

\subsection{Effect of Strength of Neighbourhood Dependence}\label{sec:Ring-Models}
We now investigate the accuracy of the estimation methods when the strength of dependence on the neighbours in the graph varies. In the GNAR framework, this is encoded in the network parameter $\beta$, where a high $\beta$-value (close to $1$) refers to a strong connection between neighbours and low $\beta$-value (close to $0$) refers to a weak connection.
In these experiments, we simulate processes of length $T=1024$ from a GNAR(1,[1]) model over either a 10- or 25-dimensional Ring, as displayed in Figure~\ref{fig:Ring Networks} (Appendix~\ref{app:add_sims}). 

Tables~\ref{tbl:25D Ring Results}  (below) and~\ref{tbl:10D Ring Results} (Appendix~\ref{app:add_sims}) report the average edge recovery rates across $K=50$ replicates, comparing the performance of our proposed wavelet-based methodology with the Fourier methods. For each graph dimension, we consider two GNAR models corresponding to $\beta_{1,1}\in\{ 0.85, \, 0.55\}$, with the best results for each model highlighted in bold.
In particular, our proposed wavelet-based method outperforms all Fourier-based methods for the $25$-dimensional Ring structure, achieving higher TPR values. Similar findings  are observed for the $10$-dimensional Ring, where {\texttt{WavTSglasso}} generally outperforms all Fourier methods-- when $\beta_{1,1}=0.55$ the Fourier-$\ell_2$ achieves a slightly lower FPR balanced by a lower TPR when compared to {\texttt{WavTSglasso}}.

Also note here that the TDR becomes more sensitive to false positives when moving from the $10$-dimensional Ring to the $25$-dimensional Ring as the number of possible false positives increases from $25$ to $250$. As a result, even if the FPR remains roughly constant, the absolute number of false positives is substantially higher, leading to a decrease in the TDR.

The averaged estimated adjacency matrices for the $10$- and $25$-dimensional models in Figures~\ref{fig:10D-Ring-beta65-beta35} and~\ref{fig:25D-Ring-beta65-beta35} (Appendix~\ref{app:Ring Models}) visually illustrate the benefits of our proposed method.

\begin{table}[!ht]
\begin{center}
\setlength{\tabcolsep}{4pt}
\begin{tabular}{l*{8}{c}}
\toprule
& \multicolumn{4}{c}{$\beta=0.85$} & \multicolumn{4}{c}{$\beta=0.55$} \\
\cmidrule(lr){2-5}\cmidrule(lr){6-9}
  & TPR & FPR & TDR & $F_1$ & TPR & FPR & TDR & $F_1$ \\
\midrule

W - $\ell_1$ 
& \textbf{0.9892} & \textbf{0.0114} & \textbf{0.9442} & \textbf{0.9662}
& \textbf{0.7384} & \textbf{0.0708} & \textbf{0.7477} & \textbf{0.7430} \\
\midrule

F - $\ell_1$ (glasso) 
& {0.7776} & {0.1832} & {0.4539} & 0.5732
& {0.6524} & {0.0878} & {0.6136} & 0.6324 \\
\midrule

F - $\ell_1$ (ADMM) 
& {0.8704} &{0.0851} &{0.6836} & 0.7658
&{0.2764} &{0}  &{1} & 0.4331\\
\midrule

F - $\ell_2$ (ADMM) 
&{0.5652} &0.0105 &{0.9172} & 0.6994
&{0.6468} & 0.0622 &{0.6978} & 0.6713\\

\end{tabular}
\caption{25-dimensional GNAR(1,[1]) models for $T=1024$ over Ring graph. Values in bold identify the best method for $\mathcal{CIG}$ recovery across the four aggregated measures.}
\label{tbl:25D Ring Results}
\end{center}
\end{table}

\subsection{Effect of process dynamics}\label{sec:VAR Models}
We turn our attention to how different time series processes affect the recovery of $\mathbf{G}$.
To this end, we consider a 10-dimensional VAR(1) process, $\XX_{t} = \mathbf{A}\XX_{t-1} + \vveps_{t}$, where $\mathbf{A} = (A_{p,q})\in \mathbb{R}^{10\times 10}$ is the coefficient matrix and $\{\vveps_{t}\}_t$ is a zero-mean Gaussian noise process with $\boldsymbol{\Sigma}_{\vveps} = \mathbf{I}_{10}$ for all $t$.
Define
\begin{equation}\label{eq:VAR Model1}
A_{p,q} =
\begin{cases}
    \beta_p, & \text{if } |p - q| \in \{1, 9\}, \\
    0, & \text{otherwise},
\end{cases}
\qquad \text{for each }p,q \in \{1, \ldots, 10\},
\end{equation}

so that variability in the coefficient matrix is introduced via the  coefficients $\{\beta_p\}_{p=1}^{10}$. Specifically, we consider two intervals of differing widths, namely $I_{1}=(0.6,0.7)$,  $I_{2}=(0.4,0.9)$, and we randomly draw $\beta_p$, $p=1,\ldots,10$, from each interval in turn. 
The dependence graph of these two VAR(1) processes matches the 10-dimensional GNAR(1,[1]) models with Ring graph of Section~\ref{sec:Ring-Models}. In particular, the special case $\beta_p=\beta_{1,1}$ for all $p=1,\ldots,10$ coincides with the GNAR(1,[1]) Ring model with network parameter $\beta_{1,1}$. All simulation studies used $K=50$ realisations.

As shown by the edge detection rate results in Table~\ref{tbl:VAR Model1} (below), increasing variability in the coefficient matrix $\mathbf{A}$ generally reduces the ability of estimation methods to recover true edges, with the exception of Fourier-$\ell_1$ glasso, for which the TPR remains relatively stable. In particular, the TPR of our proposed wavelet-based method decreases as variability increases, accompanied by a reduction in the FPR and an improvement in the TDR, indicating a sparser estimated structure. The wavelet method achieves a favourable balance between true edge recovery and the preservation of sparsity across both settings. In comparison, the Fourier methods employing ADMM tend to produce sparser estimates with lower TPR values, whilst Fourier-$\ell_1$ (glasso) yields undesirably high FPR values in both settings. Figure~\ref{fig:VAR-I1-I2} (Appendix~\ref{app:VAR Models}) visually reinforces the superiority of our proposed method in capturing the true data dependence patterns.

\begin{table}[ht!]
\begin{center}
\setlength{\tabcolsep}{4pt} 
\begin{tabular}{l*{8}{c}}
\toprule
& \multicolumn{4}{c}{$I_{1}$} & \multicolumn{4}{c}{$I_{2}$} \\
\cmidrule(lr){2-5}\cmidrule(lr){6-9}
& TPR & FPR & TDR & $F_1$ & TPR & FPR & TDR & $F_1$ \\
\midrule
W - $\ell_1$
& \textbf{0.7910}
& \textbf{0.0947}
& \textbf{0.8628}
& \textbf{0.8253}
& \textbf{0.6170}
& \textbf{0.0460}
& \textbf{0.9200}
& \textbf{0.7386} \\
\midrule
F - $\ell_1$ (glasso)
& 0.7720
& 0.2307
& 0.7005
& 0.7345
& 0.7460
& 0.1987
& 0.7222
& 0.7339 \\
\midrule
F - $\ell_1$ (ADMM)
& 0.4950
& 0.0020
& 0.9906
& 0.6601
& 0.4440
& 0.0127
& 0.9731
& 0.6098 \\
\midrule
F - $\ell_2$ (ADMM)
& 0.6530
& 0.0333
& 0.9398
& 0.7706
& \textbf{0.5620}
& \textbf{0.0080}
& \textbf{0.9809} 
& \textbf{0.7146} \\
\end{tabular}
\caption{VAR models with coefficient matrix entries \eqref{eq:VAR Model1} randomly picked in their corresponding intervals and $T=1024$.  Values in bold identify the best method(s) for $\mathcal{CIG}$ recovery across the four aggregated measures.}
\label{tbl:VAR Model1}
\end{center}
\end{table}

We also consider the case when the dependence graph is not determined by the coefficient matrix $\mathbf{A}$. 
Similar to \cite{fiecas2019spectral}, we construct a 10-dimensional VAR(1) process in which the true sparse structure of $\mathbf{G}$ matches that of the precision matrix of the noise process. This is achieved by using a diagonal coefficient matrix, here we set $\mathbf{A} = 0.5 \, \mathbf{I}_{10}$. Then, the off-diagonal entries of $\boldsymbol{\Sigma}_{\vveps}^{-1}$ are randomly set to $0$ or $0.5$ with probability $0.5$, and the diagonal entries are adjusted to ensure positive definiteness. 

Further, we also consider the 20-dimensional VARMA(2,2) model in \cite{wilms2023sparse} 
\begin{equation}\label{eq:VARMA Model}
    \XX_{t} = \mathbf{A}_{1}\XX_{t-1} + \mathbf{A}_{2}\XX_{t-2} + \vveps_{t} + \mathbf{B}_{1}\vveps_{t-1} + \mathbf{B}_{2}\vveps_{t-2},
\end{equation}
where $\{\vveps_{t}\}_{t}$ is a zero-mean Gaussian noise process with identity covariance matrix. 
The autoregressive coefficient matrices are diagonal: $\mathbf{A}_1 = 0.4 \, \mathbf{I}_{20}$ and $\mathbf{A}_2 = 0.2 \, \mathbf{I}_{20}$. 
The moving-average matrices are block-diagonal, with $5\times 5$-dimensional blocks $1.5 \, (\mathbf{I}_{5} + \mathbf{J}_{5})$ for $\mathbf{B}_1$, and $0.75 \, (\mathbf{I}_{5} + \mathbf{J}_{5})$ for $\mathbf{B}_2$, where $\mathbf{J}_{5}$ denotes the $5\times 5$-dimensional matrix whose entries are all $1$. Consequently, the dependence graph of \eqref{eq:VARMA Model} inherits this block-diagonal structure.

Table~\ref{tbl:VAR Model2} reports the estimation performance of the four methods. For the VAR model, all approaches struggle to accurately recover the adjacency matrix $\mathbf{G}$, point also immediately salient from Figure~\ref{fig:Fiecas19} in Appendix~\ref{app:VAR Models}. Both Fourier-$\ell_1$ methods yield extremely sparse estimates with low TPR values, whilst {\texttt{WavTSglasso}} recovers a greater proportion of true edges, albeit at the expense of higher FPR. The Fourier-$\ell_2$ method offers a more balanced trade-off between recovery of true and absent edges, although its overall recovery performance remains limited.
For the VARMA model \eqref{eq:VARMA Model}, our proposed \texttt{WavTSglasso} achieves near-perfect recovery, whereas the Fourier methods perform poorly due to high FPRs. Inspecting the aggregated recovery heatmaps in Figure~\ref{fig:Deb24} (Appendix~\ref{app:VAR Models}), the block-diagonal structure is still discernible in the Fourier-based estimates. However, these methods also identify a considerable number of spurious connections between blocks, whilst our proposed approach successfully extracts the underlying block-diagonal structure.

\begin{table}[!ht]
\begin{center}
\begin{tabular}{l*{8}{c}}
\toprule
& \multicolumn{4}{c}{VAR(1) Model} & \multicolumn{4}{c}{VARMA(2,2) Model} \\
\cmidrule(lr){2-5}\cmidrule(lr){6-9}
& TPR & FPR & TDR & $F_1$ & TPR & FPR & TDR & $F_1$ \\
\midrule
W - $\ell_1$
& {{0.5073}}
& {{0.3971}}
& {{0.4984}}
& 0.5028
& {\textbf{0.9850}}
& {\textbf{0.0077}}
& {\textbf{0.9704}} 
&\textbf{0.9776} \\
\midrule
F - $\ell_1$ (glasso)
& {0.0018}
& {0}
& {1}
& 0.0036	
& {1}
& {0.8639}
& {0.2250} 
& 0.3673 \\
\midrule
F - $\ell_1$ (ADMM)
& {0.0609}
& {0.0257}
& {0.7217}
& 0.1123
& {1}
& {0.5386}
& {0.3180}
& 0.4825 \\
\midrule
F - $\ell_2$ (ADMM)
& {0.3964}
& {0.2279}
& {0.6250}
& 0.4851
& {1}
& {0.8840}
& {0.2205} 
& 0.3613 \\
\end{tabular}
\caption{VAR and VARMA models whose dependence graph are not determined by the autoregressive component and $T=1024$. Values in bold identify the best method for $\mathcal{CIG}$ recovery across the four aggregated measures.}
\label{tbl:VAR Model2}
\end{center}
\end{table}

Across all simulation studies, our proposed {\texttt{WavTSglasso}} methodology is particularly advantageous over the existing Fourier-based methods in settings characterised by a sparse true conditional independence graph, with the wavelet-based approach achieving high levels of true edge recovery whilst successfully limiting the identification of spurious connections. 

In such settings, with the number of absent edges being much larger than the number of true edges,  even relatively low false positive rates can result in a substantial proportion of spurious connections in the estimated graph, and the ability of {\texttt{WavTSglasso}} to maintain high true positive rates whilst limiting false discoveries is therefore particularly desirable.


\section{Underlying Graph Discovery}\label{sec:graph-discovery}
So far, we have focussed on inferring the conditional dependence structure, $\mathcal{CIG}=(\mathcal{K},\mathcal{E})$ underlying the observed data. In this section, we focus on finding the underlying network (graph) $\mathcal{N}=(\mathcal{K},\mathcal{E}^{\mathcal{N}})$, when such a network is indeed present in the data, such as for the GNAR models in Section~\ref{sec:Simulation Study}. Note that these two tasks are crucially not identical.

If the data follows a GNAR($p,[\mathbf{s}]$) process or, more generally, if we assume the data has an underlying graph structure $\mathcal{N}$, then estimating the edge set $\mathcal{E}^{\mathcal{N}}$ of such a graph is crucial for inference. We denote the adjacency matrix of the underlying graph by $\mathbf{G}^{\mathcal{N}}$, and recall we denote the adjacency of the conditional independence graph, $\mathcal{CIG}$, by $\mathbf{G}$. Since $\mathcal{E}^{\mathcal{N}}$ is contained within $\mathcal{E}$ \citep[Theorem 2]{nason2023new}, the task is to determine which edges in~$\mathbf{G}$ correspond to those of $\mathbf{G}^{\mathcal{N}}$. 

From the proposed \texttt{WavTSglasso} approach described in Section~\ref{sec:Wavelet-model-def}, we obtain the sparse estimated precision matrices $\hat{\boldsymbol{\Theta}}_j$ for each scale $j=1,...,J$, and define the corresponding scale-dependent edge and precision value sets as
\begin{eqnarray*}
      \mathcal{E}^{(\mathbf{G})}_j &=& \{(p,q)\in \mathcal{K}\times\mathcal{K},\, p \neq q: \big(\hat{\boldsymbol{\Theta}}_j\big)_{p,q} \neq 0\},\\
    {\vartheta}^{(\mathbf{G})}_j &=&\{\big(\hat{\boldsymbol{\Theta}}_j\big)_{p,q}: (p,q) \in \mathcal{E}^{(\mathbf{G})}_j\}, \mbox{ respectively}.
\end{eqnarray*}

We propose applying $k$-means clustering to the precision values ${\vartheta}_j^{(\mathbf{G})}$ to separate them into $k=2$ clusters: those that correspond to edges in $\mathbf{G}^{\mathcal{N}}$ and the additional edges in $\mathbf{G}$ (a `noise' cluster). The $k$-means clustering procedure yields a cluster, $\mathscr{C}_{1j}$, containing edges with low values of $\big(\hat{\boldsymbol{\Theta}}_j\big)_{p,q}$, or in other words, those edges between node pairs $(p,q)$ that are the least significant; this is the cluster whose centre is closest to $0$, hence the `noise' cluster. The second cluster, $\mathscr{C}_{2j}$, contains the remaining edges in $\mathbf{G}$, thus we let $(\hat{\mathbf{G}}_j^{\mathcal{N}})_{p,q}=1$ when $(p,q)\in \mathscr{C}_{2j}$, and zero otherwise, in order to obtain a {\em scale $j$-dependent} sparse estimator for the underlying graph adjacency $\mathbf{G}^{\mathcal{N}}$.

Our strategy for obtaining $\hat{\mathbf{G}}^{\mathcal{N}}$ exploits the information available at the finest two scales. The restriction to finer scales is motivated by Proposition~\ref{prop:Ij-consistent}, which highlights the scale-dependent variability of spectral estimation. Since the distribution of precision matrix entries may vary across scales, clustering at a single scale may result in the omission of genuine network connections. We therefore combine the scale-specific network estimates to obtain $\hat{\mathbf{G}}^\mathcal{N}_{p,q} = 1$ if $(\hat{\mathbf{G}}^\mathcal{N}_{1})_{p,q} =1$ or $(\hat{\mathbf{G}}^\mathcal{N}_{2})_{p,q} =1$, and 0 otherwise.

This approach aims to to exploit complementary information whilst limiting the accumulation of spurious edges that may arise when incorporating a large number of scales.

The effects of dimension, sparsity and length of the time series on the network estimation are assessed in what follows via simulation, akin to our \texttt{WavTSglasso} experiments. We compute TPR, FPR, TDR, and F1-scores for each simulation, comparing the estimated adjacency matrix $\hat{\mathbf{G}}^{\mathcal{N}}$ to the true adjacency matrix $\mathbf{G}^{\mathcal{N}}$.
Since, to our knowledge, there are no other algorithms which attempt to identify the underlying network, $\mathcal{N}$, we only report results using our clustering methodology. 
\paragraph{Use of other clustering techniques.}
We choose $k$-means for edge-selection due to its simplicity and speed but other clustering algorithms can be used.  We have found that experimental results change only minimally when other clustering techniques such as tree-based procedures \citep{Johnson_1967} or $k$-medoids \citep{kaufman1987medoids} are used.

\subsection{Simulation Results for GNAR Models defined on Erd\H{o}s-Rényi Graphs}\label{sec:ER-clustering}
We report results for the Erd\H{o}s-Rényi (ER) network study introduced in Section~\ref{sec:ER-models}.
Recall that the baseline setting consists of ten randomly generated 10-dimensional ER graphs, with sparsity controlled via the edge probability $\rho \in \{0.1, 0.25\}$. For each ER graph, we simulate time series of length $T \in \{256,1024,2048\}$ for a GNAR(1,[1]) model with $\beta_{1,1}=0.85$. 

Table~\ref{tbl:ER Model Clustering-Results GNAR(1,[1])} reports the four measure results, TPR, FPR, TDR and $F_1$, for the estimated underlying networks. For both sparsity settings, the TPR remains relatively stable as the length of the time series increases, while the FPR decreases considerably. As the edge sparsity decreases, stability and overall performance unsurprisingly deteriorate, as consistent with the growing difficulty of recovering the underlying dependence structure. Nevertheless, the clustering procedure maintains relatively low FPRs, particularly for longer time series.

\begin{table}[!ht]
\begin{center}
\setlength{\tabcolsep}{5pt}
\begin{tabular}{l*{8}{c}}
\toprule
& \multicolumn{4}{c}{Sparse ER} & \multicolumn{4}{c}{Dense ER} \\
\cmidrule(lr){2-5}\cmidrule(lr){6-9}
& TPR & FPR & TDR & $F_1$ & TPR & FPR & TDR & $F_1$ \\
\midrule
$T=256$
 & 0.7719 & 0.0367 & 0.7562 & 0.7640
 & 0.4168 & 0.0608 & 0.7733 & 0.5417\\
$T=1024$
 & 0.8208 & 0.0092 & 0.9351 & 0.8742
 & 0.4192 & 0.0156 & 0.9395& 0.5797\\
$T=2048$
 & 0.8163 & 0.0021 & 0.9821 & 0.8916
 & 0.4290 & 0.0066 & 0.9711 & 0.5951 \\
\end{tabular}
\caption{10-dimensional GNAR(1,[1]) models with different $T$ over Erd\H{o}s-Rényi graphs.}
\label{tbl:ER Model Clustering-Results GNAR(1,[1])}
\end{center}
\end{table}

\subsection{Simulation Results for GNAR Models defined on Ring Graphs}\label{sec:Ring-clustering}
We now assess how the estimation performance varies as the strength of dependence on the neighbours varies. We can vary the dependence on the neighbours in the GNAR models by varying the network parameter $\beta$, with a strong connection depicted by a $\beta$-value close to $1$ and a weak connection depicted by a $\beta$ value close to $0$. We simulate GNAR(1,[1]) models over a 10- or 25-dimensional Ring with length $T=1024$ (recall Figure~\ref{fig:Ring Networks}), and as in Section~\ref{sec:Ring-Models}, we evaluate two models with $\beta\in \{0.85,0.55\}$ defined over both Ring graphs. 

The results in Table~\ref{tbl:Clustering-Results Rings} are consistent with those reported in Section~\ref{sec:Ring-Models}. Edge-discovery rates decline as the neighbourhood strength weakens in both dimensions, whilst the ability to identify absent edges remains very high across all models. These points are readily visualised using the average estimated adjacency matrices that appear in Figure~\ref{fig:10D-beta65-beta35-Ring-clustering} (Appendix~\ref{app:graph-discovery}).

\begin{table}[H]
\begin{center}
\setlength{\tabcolsep}{5pt}
\begin{tabular}{l*{8}{c}}
\toprule
&\multicolumn{4}{c}{$\beta=0.85$} & \multicolumn{4}{c}{$\beta=0.55$}\\
\cmidrule(lr){2-5}\cmidrule(lr){6-9}
  & TPR & FPR & TDR & $F_1$ & TPR & FPR & TDR & $F_1$ \\
\midrule
$P=10$
 & 1 & 0 & 1 & 1
 & 0.8960 & 0.0035 & 0.9870 & 0.9393\\
$P=25$
 & 0.9992 & 0 & 1 & 0.9996
 & 0.9000 & 0.0008 & 0.9914 & 0.9435\\
\end{tabular}
\caption{10- and 25-dimensional GNAR(1,[1]) models over Ring graphs with $T=1024$.}
\label{tbl:Clustering-Results Rings}
\end{center}
\end{table}

\section{Application to COVID-19 Data}\label{sec:Application}
We now apply our network estimation methodology to the motivating epidemiology dataset
introduced in Section~\ref{sec:intro}. We consider the daily recorded number of COVID-19 patients transferred to mechanical ventilation beds across various NHS Trusts in England, covering the period from April 2020 to July 2021. Specifically, the dataset consists of count data for 140 NHS Trusts, each with $T = 452$ observations. A logarithmic transformation is applied to stabilise the variance and bring the data closer to normality. The dataset is available in the \texttt{GNAR} package \citep{GNAR} and was first analysed by \cite{nason2023new}. Since the true graph structure of dependencies is unknown, our aim is to estimate and visualise a sparse dependence graph pertaining to the data. 

We begin by extending each time series to length $T = 512$  via symmetric padding to match $J = 9$ wavelet scales. We apply a stationarity test \citep{nason2013test} and retain only those series that satisfy the stationarity assumption. Based on this test, we exclude 50 NHS Trusts. Additionally, we remove 6 trusts whose codes do not appear in the NHS reference spreadsheet, resulting in an observed multivariate process over $P = 84$ nodes.

As in \cite{nason2023new} who used geographic distance as a basis for defining network structure, we organise the nodes by NHS region to examine how nearby Trusts may influence one another. We group NHS Trusts into regional categories in order to explore how dependency patterns vary geographically from North to South. Specifically, nodes 1-25 correspond to Trusts in the North, nodes 26-42 to the Midlands, and nodes 43-84 to the South. Table~\ref{tbl:covid-tbl} (Appendix~\ref{app:covid}) breaks down the node allocation for each NHS England Region.

\begin{figure}[H]
\centering
\begin{subfigure}{0.3\linewidth}
    \includegraphics[width=\linewidth]{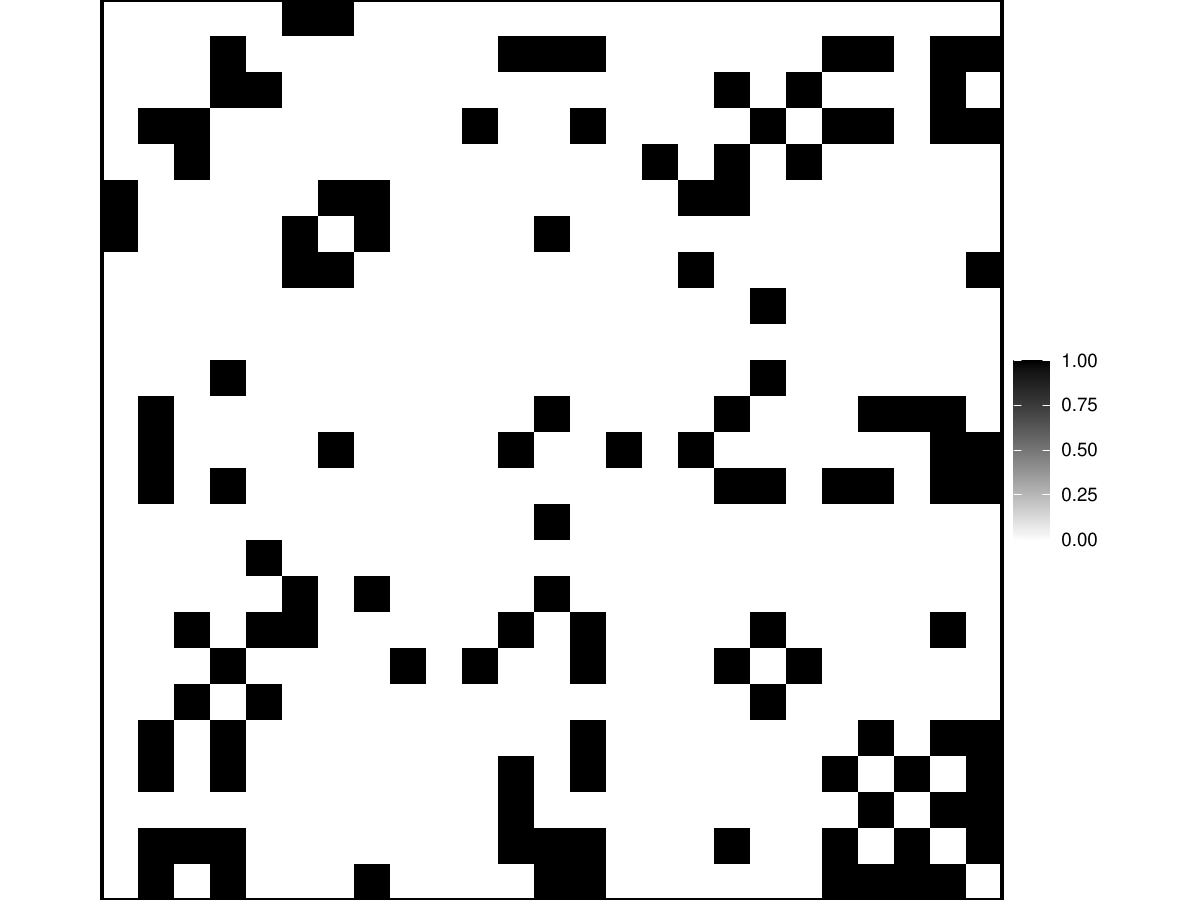}
    \caption{North.}\label{fig:CIG-North}
\end{subfigure}
\hfill
\begin{subfigure}{0.3\linewidth}
    \includegraphics[width=\linewidth]{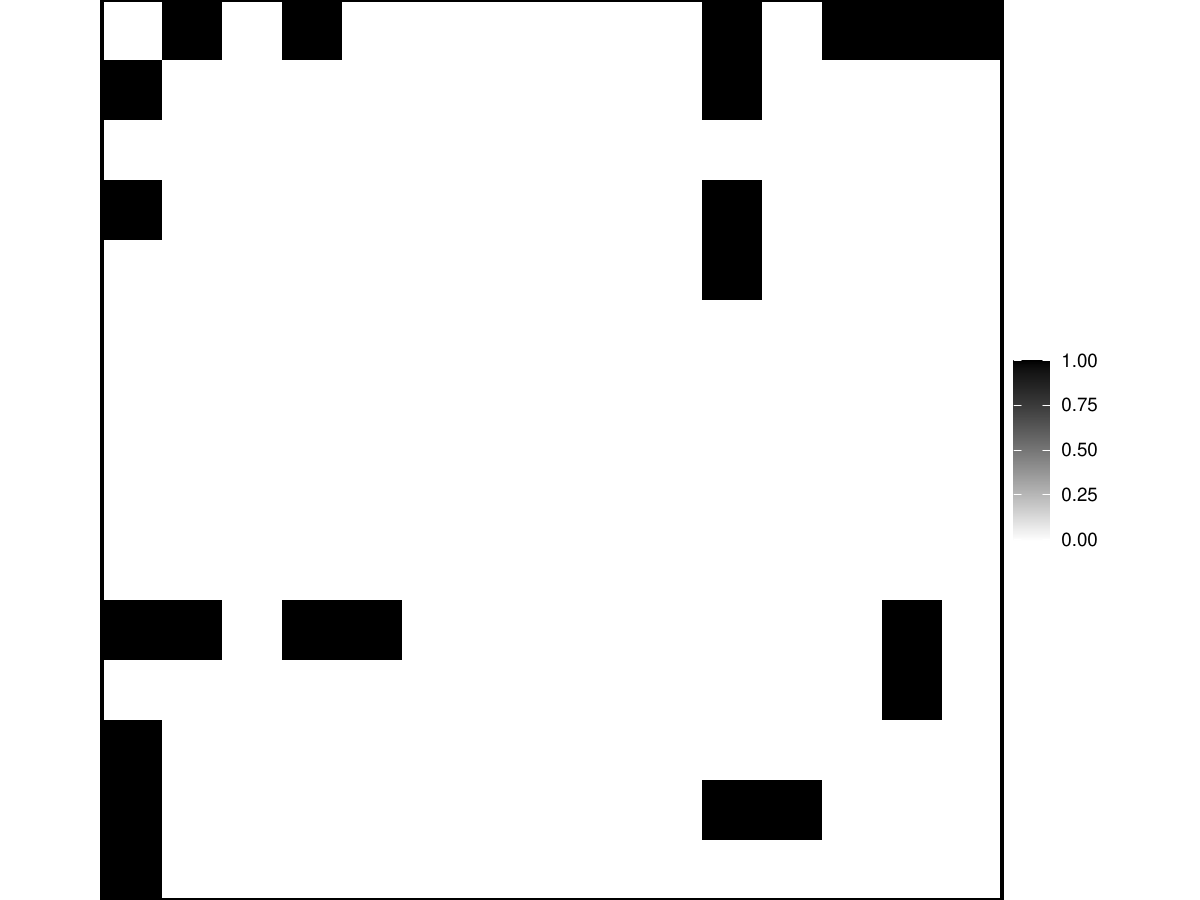}
    \caption{Midlands.}\label{fig:CIG-Mid}
\end{subfigure}
\hfill
\begin{subfigure}{0.3\linewidth}
    \includegraphics[width=\linewidth]{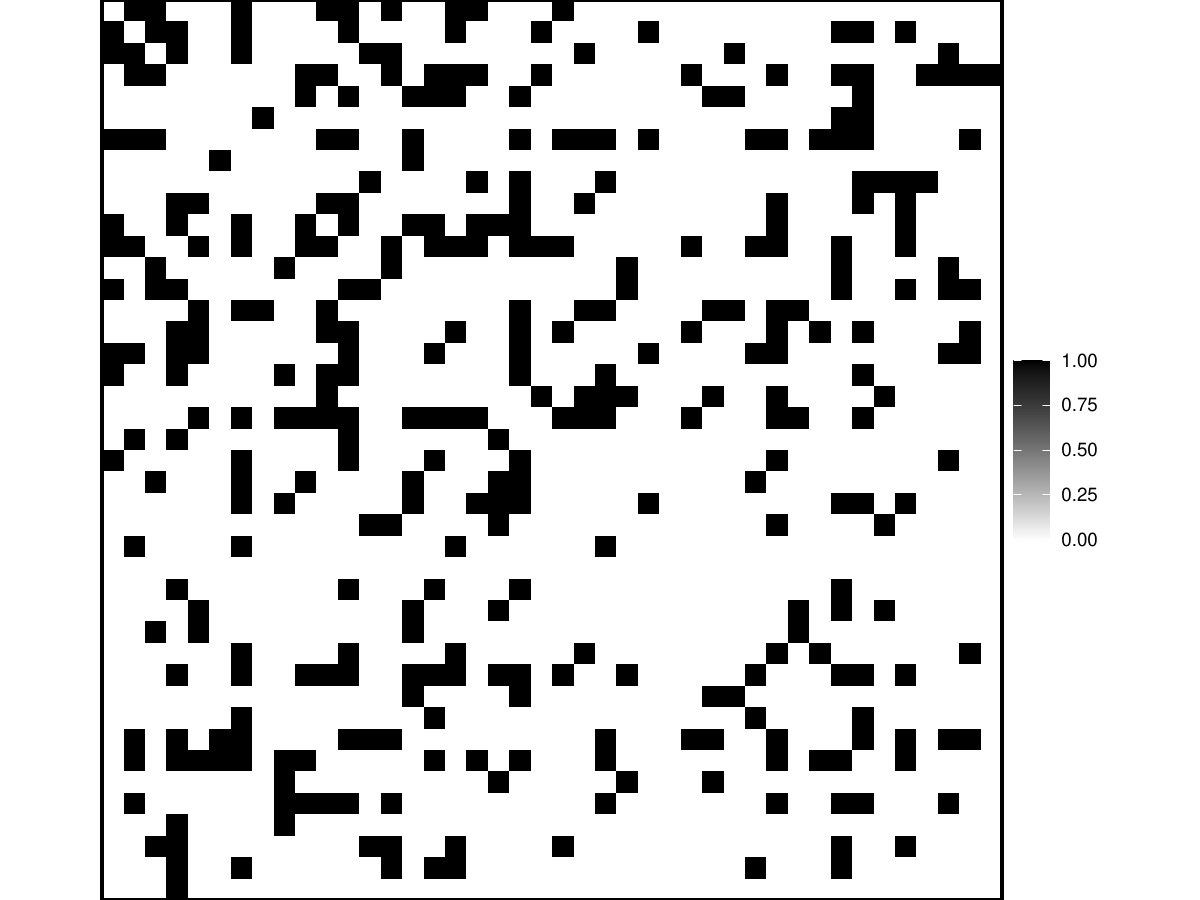}
    \caption{South.}\label{fig:CIG-South}
\end{subfigure}
\caption{Estimated adjacency matrices for the $\mathcal{CIG}$ for the (a) North, (b) Midlands, and (c) South of England. For each plot, NHS trusts are plotted from West to East.} 
\label{fig:covid-region-CIG}
\end{figure}

\paragraph{$\mathcal{CIG}$ estimation using the \texttt{WavTSglasso} proposal from Section~\ref{sec:Wavelet-model-def}.}
Firstly, we estimate and visualise the dependence structure in the data. We apply our proposed wavelet-based method, as described in Algorithm \ref{alg:WavTSglasso}, with $R=10$ bootstraps as aligned to our simulation study findings in Section~\ref{sec:bootstrap-study}. 
The similarity measure in \eqref{eq:sim measure} selects the two finest scales, $(j_1,j_2):=(1,2)$, for the estimation of the adjacency matrix $\mathbf{G}$.  
The estimated $\mathcal{CIG}$ exhibits a relatively dispersed dependence structure, with connections occurring throughout the network rather than being confined to geographically neighbouring Trusts. Figure~\ref{fig:covid-region-CIG} shows the estimated adjacencies restricted to Trusts within the North (left), Midlands (middle) and South (right), thus illustrating different conditional dependence structures within geographical regions. A black (white) square corresponds to a (no) connection between trusts. 

Within the North (Figure~\ref{fig:CIG-North}), the estimated $\mathcal{CIG}$ exhibits a broadly connected structure with several prominent Trusts rather than a single dominant centre. St Helens and Knowsley Teaching Hospitals (node 24) is particularly well connected, with dependencies extending beyond Cheshire and Merseyside to Trusts in the North East, Lancashire and Greater Manchester. Newcastle upon Tyne Hospitals (node 4) similarly exhibits connections spanning several constituent northern regions, including Yorkshire and Humber, Lancashire, Greater Manchester and Cheshire and Merseyside. This suggests that the dependence is not simply localised within NHS administrative areas, with some Trusts appearing to bridge otherwise geographically distinct parts of the North. Newcastle is also one if the largest specialist teaching Trusts in the country, whereas St Helens and Knowsley is principally an acute teaching Trust centred on Whiston Hospital. Their joint prominence suggests that connectivity in the estimated $\mathcal{CIG}$ does not appear to be explained solely by hospital size or specialist status.

A markedly different structure is observed in the Midlands (Figure~\ref{fig:CIG-Mid}). The estimated $\mathcal{CIG}$ is substantially sparser, with several Trusts exhibiting no connections within the Midlands subnetwork and the remaining dependence concentrated around a small connected core. Nottingham University Hospitals (node 26) occupies a particularly prominent position, with connections spanning each of the North, West and Central Midlands. As one of the largest acute teaching Trusts in England and a major provider of specialist services, its position may partly reflects its wider regional role. However, the broader contrast with the North and South indicates that the prevalence of large hospitals cannot explain the structure: despite containing several major providers, the Midlands $\mathcal{CIG}$ remains comparatively sparse. The estimated network therefore identifies a small number of Trusts that may act as informational bridges between otherwise weakly connected parts of the regional system.

The South (Figure~\ref{fig:CIG-South}) exhibits a much more extensively connected and multi-centered structure. Dependencies are distributed throughout the subnetwork, with substantial connectivity between the East and London and between London and Kent, Surrey and Sussex. An informative example is the Princess Alexandra Hospital (node 49), which has connections extending from the East to Trusts in London, the South East and further South. Unlike the large specialist Trusts highlighted above, the Princess Alexandra Hospital is a district general hospital. Its prominent position therefore provides further evidence that connectivity in the $\mathcal{CIG}$ is not simply a proxy for hospital size, teaching status or the provision of highly specialised services.

Together, these findings illustrate the benefit of estimating the network directly from the data. A network constructed solely from geographical proximity necessarily assumes that dependence is primarily spatial and applies the same geographical rule throughout the country. By contrast, the estimated $\mathcal{CIG}$ reveals substantial heterogeneity: the North contains several bridging Trusts, the Midlands is organised around a much smaller connected core, and the South exhibits a broader multi-centered dependence structure. Moreover, prominent nodes include both large specialist teaching Trusts and district general hospitals, suggesting that neither geographical proximity, administrative region nor hospital type alone is sufficient to characterise the dependence structure. The estimated $\mathcal{CIG}$ provides conditional statistical relationships that may reflect common epidemic pressure, population movements, referral patterns or general hospital demand similarities rather than evidence of direct patient transfers or causal relationships. Therefore the estimated $\mathcal{CIG}$ provides a data-driven means of identifying Trusts and relationships that may contain information about wider system pressure which would be omitted by a purely geographically specified network, while providing a more parsimonious network representation for subsequent modelling or forecasting.

\paragraph{Network $\mathcal{N}$ estimation using the strategy proposed in Section~\ref{sec:graph-discovery}.}
We now estimate the underlying network for GNAR modelling using the clustering method in Section~\ref{sec:graph-discovery}. \cite{nason2023new} show that the data can be appropriately modelled by a GNAR(1,[1]) model, and in-line with our previous findings we use the finest two scales ($j=1,2$) to apply the clustering procedure to the non-zero entries of the estimated precision matrices.

\begin{figure}[!ht]
\centering
\begin{subfigure}{0.30\linewidth}
  \includegraphics[width=\linewidth]{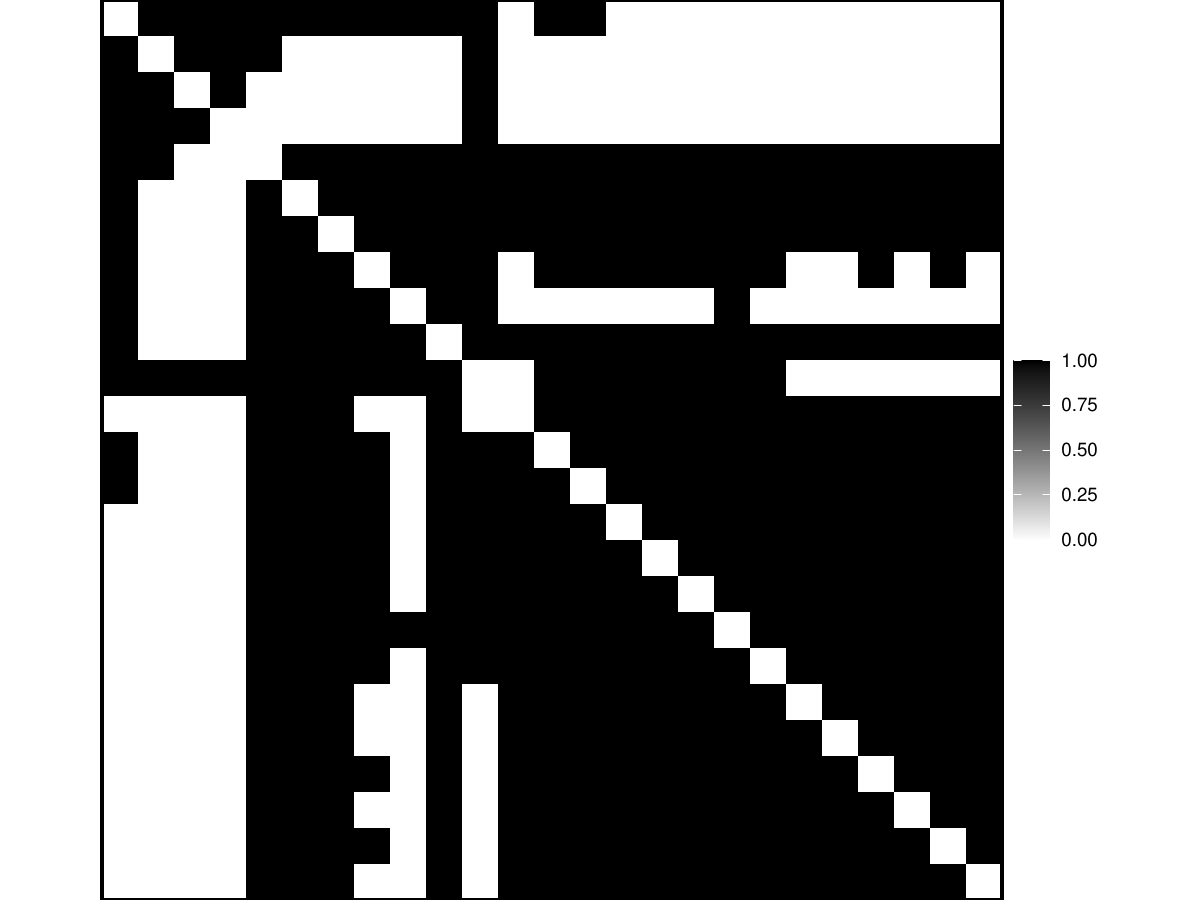}
  \caption{North}
\end{subfigure}\hfill
\begin{subfigure}{0.30\linewidth}
  \includegraphics[width=\linewidth]{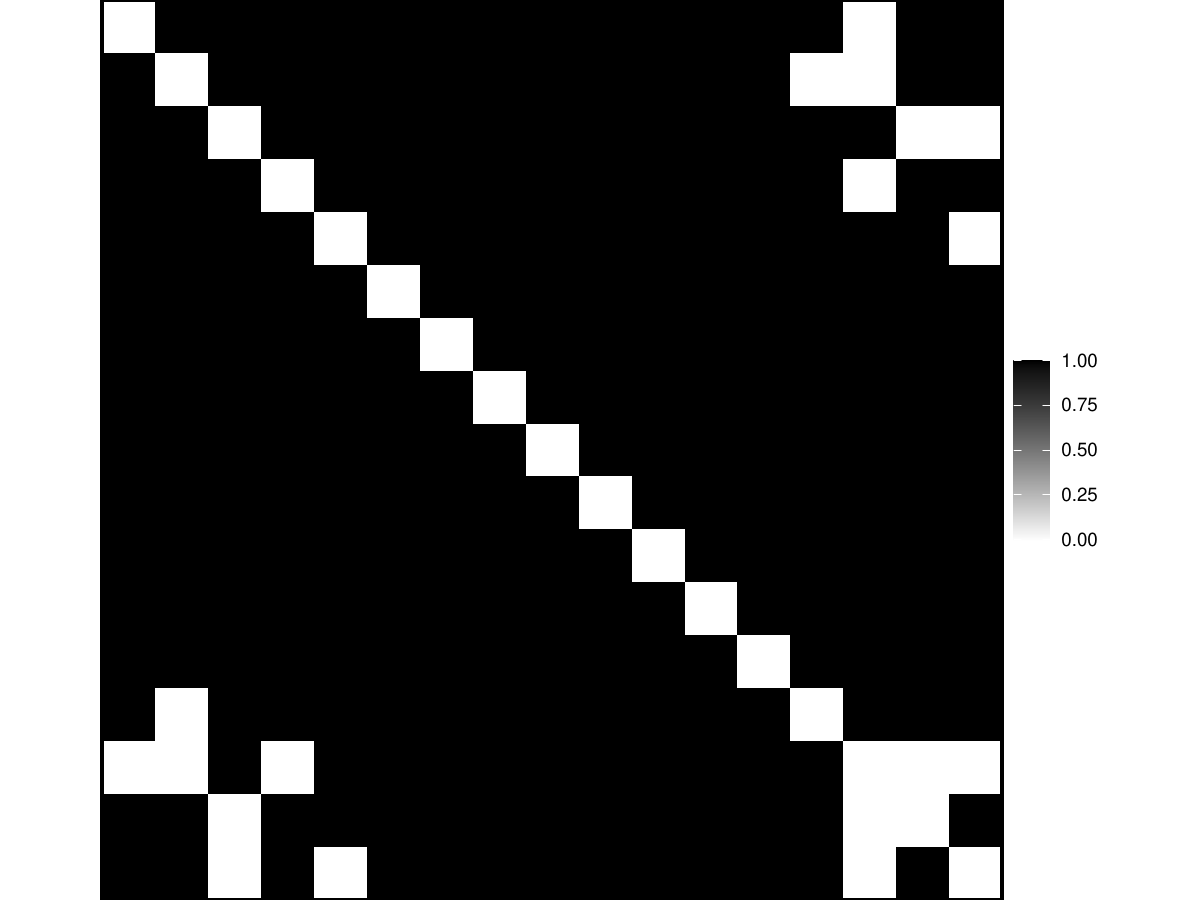}
  \caption{Midlands}
\end{subfigure}\hfill
\begin{subfigure}{0.30\linewidth}
  \includegraphics[width=\linewidth]{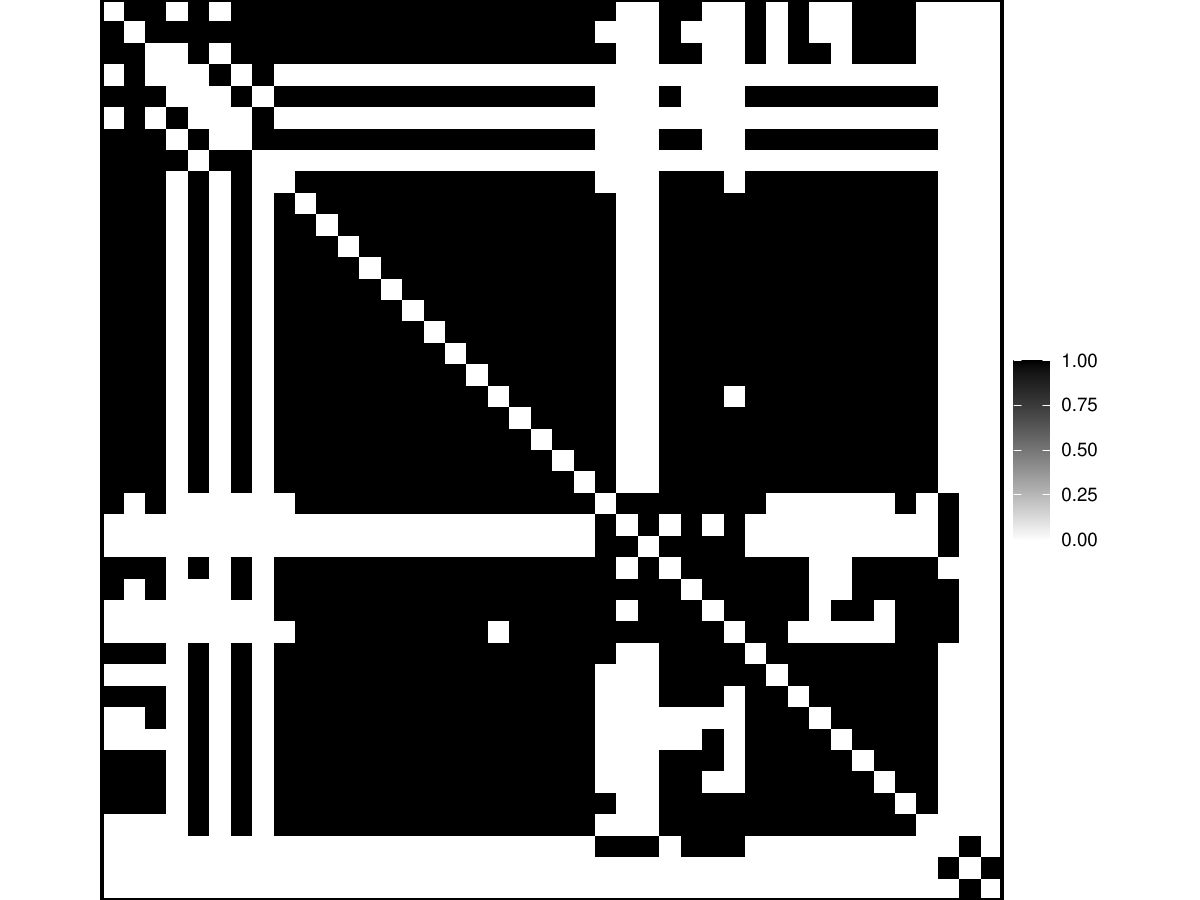}
  \caption{South}
\end{subfigure}

\vspace{4mm}

\begin{subfigure}{0.30\linewidth}
  \includegraphics[width=\linewidth]{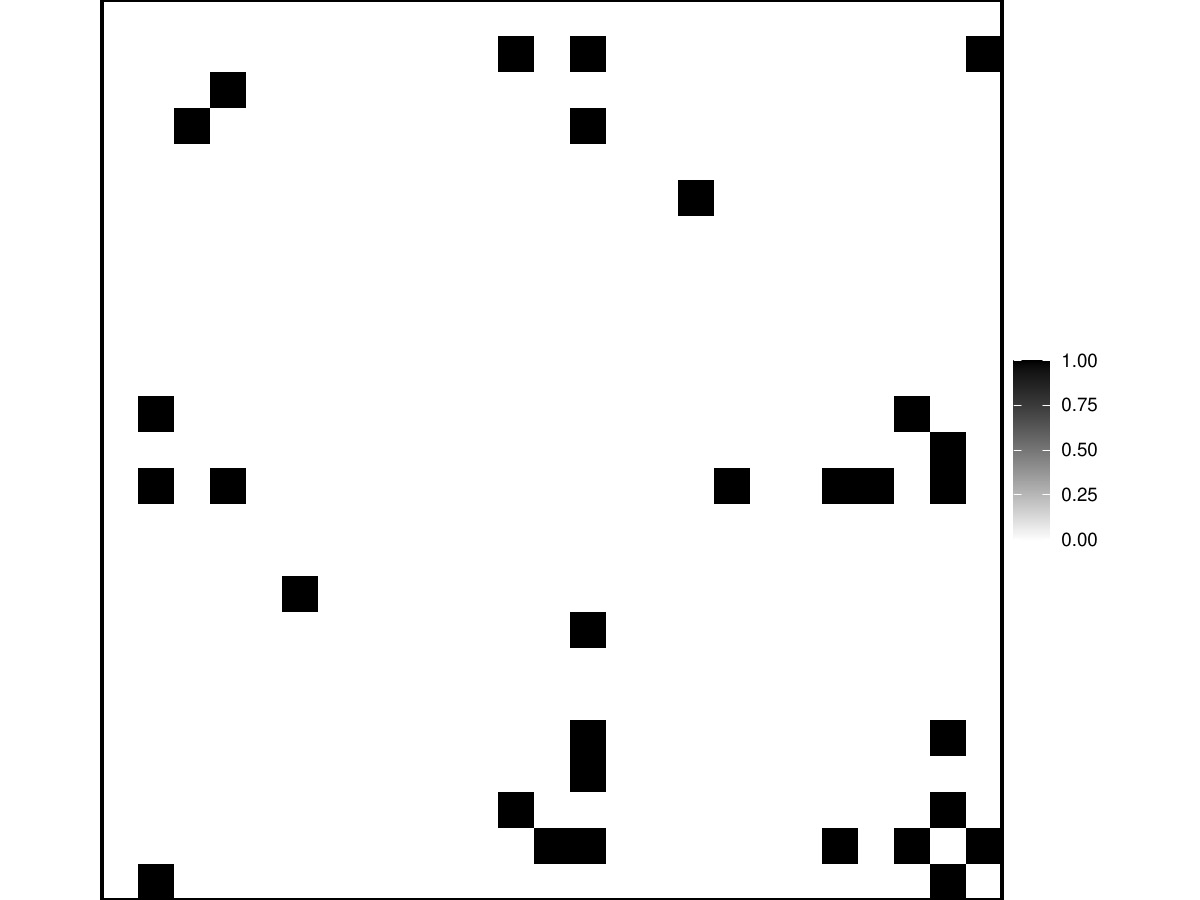}
  \caption{North}
\end{subfigure}\hfill
\begin{subfigure}{0.30\linewidth}
  \includegraphics[width=\linewidth]{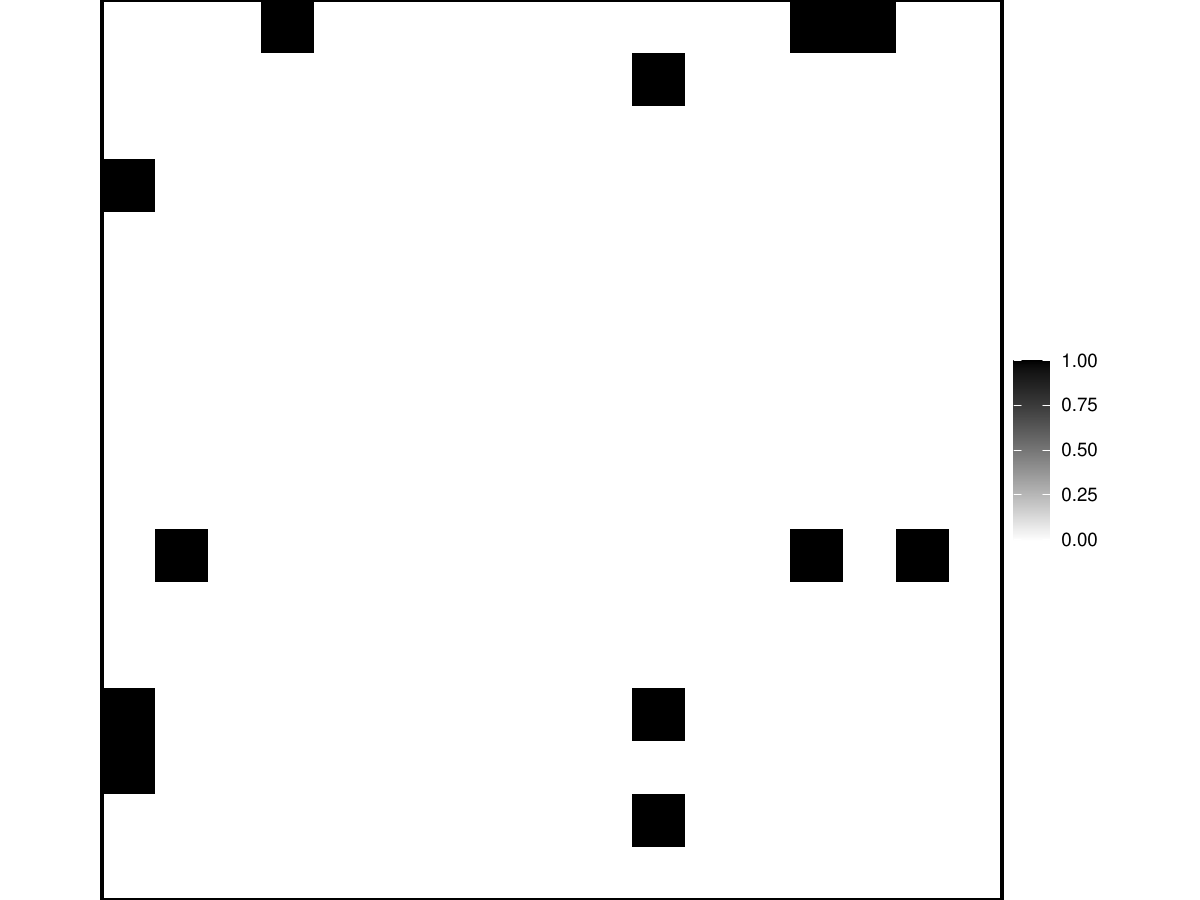}
  \caption{Midlands} 
\end{subfigure}\hfill
\begin{subfigure}{0.30\linewidth}
  \includegraphics[width=\linewidth]{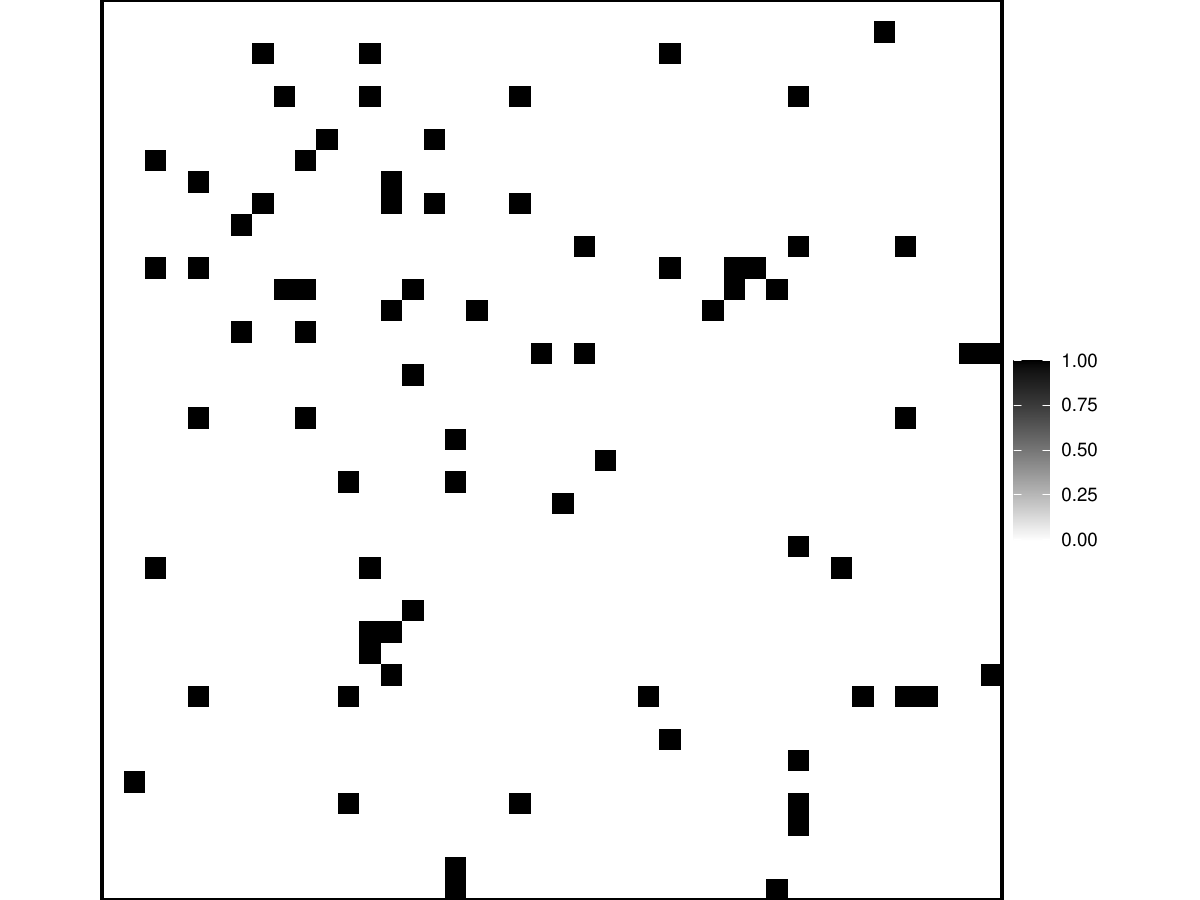}
  \caption{South}
\end{subfigure}

\caption{Underlying network adjacency matrices for the COVID-19 ventilation beds data. Columns correspond to North, Midlands, and South; top row: adjacencies used in \cite{nason2023new}; bottom row: data-driven estimated adjacencies using the clustering method.}
\label{fig:covid-agg-underlying}
\end{figure}

The top panels of Figure~\ref{fig:covid-agg-underlying} show the network used by \citet{nason2023new} at a regional level, whereas our estimated network is reported in the bottom panels. Recall that \cite{nason2023new} employ a user-defined distance-metric to obtain the network, namely by connecting any two trusts which are within 120km of each other. Although this threshold was specifically chosen to minimise the predictive error of the resulting GNAR model, it produces a more densely connected network than the data-driven one estimated by our clustering method.

\paragraph{Benefit of data-driven network estimation for forecasting.} We now evaluate our data-inferred network versus a user-based construction, in the context of prediction---one of the key targets in \cite{nason2023new}, although not a task in itself for our development here.  Table~\ref{tbl:forecasting-comparison} reports two-step-ahead mean squared prediction errors (MSPEs) for forecasts based on the adjacency matrix proposed by \cite{nason2023new}  and on the {\tt{WavTSglasso}}-derived network, where we adopted a GNAR(1,[1]) model as suggested by \cite{nason2023new}. 
Despite containing considerably fewer connections and the geographic network used in \cite{nason2023new} being designed to optimise prediction, our estimated network achieves competitive forecasting performance, with the {\tt{WavTSglasso}}-derived network achieving lower MSPEs at both forecasting horizons, with reductions of $6.4\%$ and $8.1\%$, respectively. 

These findings suggest that a substantially sparser, data-driven network can deliver crucial dependence information, potentially useful for forecasting and without relying on user-defined connections-- however, this topic is left for future research.

\begin{table}[!ht]
  \centering
  \begin{tabular}{c|ccc}
  Step ahead & MSPE (\cite{nason2023new}) & MSPE (\texttt{WavTSglasso})\\
  \midrule
    1   &  32.6 & 30.5 \\ 
    2   &  69.1 & 63.5 \\ 
  \end{tabular}
  \caption{Mean Squared Prediction Errors ($\times 10^{3}$) for one- and two-step-ahead forecasts from a GNAR(1,[1]) model using the geographical network suggested by \cite{nason2023new} and the network learned using our proposed wavelet-based approach.}    \label{tbl:forecasting-comparison}
\end{table}

\section{Conclusions and Further Work}\label{sec:Conclusions}
In this article, we proposed a novel wavelet domain method for estimating the conditional independence graph ($\mathcal{CIG}$) for (weakly) stationary time series. By considering the distributional properties of the wavelet coefficients associated to the process, we formulate a principled penalised optimisation problem to achieve sparse estimation of the inverse wavelet spectra of the time series, upon which our $\mathcal{CIG}$ estimation procedure is built. We establish theoretical results showing that our $\mathcal{CIG}$ estimator achieves desirable asymptotic properties, and demonstrate that the underlying conditional independence structure of the process can be successfully recovered.

Through a range of numerical experiments, we evaluated the performance of our proposed method under different scenarios, including varying data-generating processes, graph structures, and data dimensions. Our findings demonstrate that the proposed wavelet-based approach is particularly effective in recovering sparse conditional independence structures, where it achieves a favourable balance between the identification of genuine dependencies and the avoidance of spurious connections compared with existing Fourier-based techniques. We also proposed an agnostic procedure to recover the underlying graph of GNAR models from the estimated conditional independence graph. Finally, we applied the network estimation method to a dataset of COVID-19 patients in England and found a sparse dependence structure that differs substantially from the geographical network produced by \cite{nason2023new}. The estimated underlying network also demonstrated competitive forecasting performance in comparison to the user-defined adjacency built by \cite{nason2023new}. 

Several avenues for further methodological development arise from this work. Firstly, further investigation into the selection and aggregation of wavelet scales would be of interest, including alternative data-driven selection of the number of scales to aggregate. Secondly, the proposed clustering-based approach for recovering the underlying network could be further developed, particularly with regard to scale selection. A natural extension would be to adapt the wavelet-based TSglasso to nonstationary data settings. This extension is non-trivial, as it requires careful definitions of network locally stationary processes and of their associated conditional independence graphs, as well as determining their efficient aggregation.

\bibliographystyle{agsm}
\bibliography{references.bib}
\newpage

\begin{appendix}
\section{Proofs} \label{app:proofs}
In this appendix, we provide details of the proofs of the results established in the main text.

\subsection{Proof of Proposition~\ref{prop:wav_cig}}\label{app:proofprop1}
\begin{proof} 

In order to relate the individual wavelet spectral matrices $\SS_j$ to the frequency-domain characteristic of conditional independence, we first isolate the contribution of each wavelet scale to the original process. This allows the dependence structure associated with a single $\SS_j$ to be studied separately from the remaining scales. Akin to \cite{wu2023multi}, let us define a scale-dependent subprocess of the original process $\XX$ to be $\XX_{j,t}=\left(X_{j,t}^{(1)},\ldots,X_{j,t}^{(P)}\right)^\top$, with the following representation,
\begin{align} \label{eqa:subprocess}
    \XX_{j,t}= \sum_{k \in \mathbb{Z}} \mathbf{V}_j(\nicefrac{k}{T}) \psi_{j,k}(t) \vveps_{j,k}
\end{align} 
where the terms $\mathbf{V}_j(\nicefrac{k}{T}), \psi_{j,k}(t)$ and $\vveps_{j,k}$ are as in the multivariate LSW definition (see Section \ref{sec:mvLSW-review} for details). We can conceptualise the time series $\{X_{j,t}^{(p)}\}_t$ as giving the scale-$j$ representation of the original node $p$ process, $\{X_{t}^{(p)}\}_t$.
From the zero-mean assumption for the innovations $\{\vveps_{j,k}\}_{j,k}$, we immediately obtain that the subprocess is also zero-mean, i.e. $\mathbb{E}(\XX_{j,t})=\mathbf{0}$, at all times $t$.
Coupling the zero-mean property with representation~\eqref{eqa:subprocess} yields
\begin{align*}
    &\mathbb{C}\mathrm{ov}(X_{j,\lfloor uT \rfloor}^{(p)},X_{j,\lfloor uT \rfloor+\tau}^{(q)}) = \mathbb{E} \left[ X_{j,\lfloor uT \rfloor}^{(p)} X_{j,\lfloor uT \rfloor+\tau}^{(q)}\right] \notag \\
    &=\mathbb{E} \left[  \left(\sum\limits_{k} \mathbf{V}_j^{(p)}(\nicefrac{k}{T}) \psi_{j,k}(\lfloor uT \rfloor) \vveps_{j,k}\right)
     \times  \left(\sum\limits_{k'} \mathbf{V}_{j}^{(q)}(\nicefrac{k'}{T})\psi_{j,k'}(\lfloor uT \rfloor+\tau)\vveps_{j,k'} \right)^{\top}  \right] \notag \\
    & = \sum\limits_{k}\sum\limits_{k'} \mathbf{V}_j^{(p)}(\nicefrac{k}{T}) \psi_{j,k}(\lfloor uT \rfloor) \mathbb{E}(\vveps_{j,k}\vveps^{\top}_{j,k'})\psi_{j,k'}(\lfloor uT \rfloor+\tau)\left(\mathbf{V}_{j}^{(q)}(\nicefrac{k'}{T})\right)^{\top},
\end{align*}
where $\mathbf{V}_j^{(p)}(u)$ denotes the $p$th row of $\mathbf{V}_j(u)$.
From the assumptions on the innovation process, namely $\mathbb{E}( \vveps_{j,k}\vveps^{\top}_{j,k'})=\mathbb{C}\mathrm{ov}(\vveps_{j,k},\vveps_{j,k'} ) = \delta_{j,j'}\delta_{k,k'}\mathbf{I}_P$, with $\mathbf{I}_P$ denoting the $P$-dimensional identity matrix, and using the definition of the LWS matrix, $S_{j}^{(p,q)}(u)= \mathbf{V}_{j}^{(p)}(u)\left(\mathbf{V}_{j}^{(q)}(u)\right)^{\top}$, by letting $m=k-\lfloor uT \rfloor$ we obtain
\begin{align*}
    \mathbb{C}\mathrm{ov}(X_{j,\lfloor uT \rfloor}^{(p)},X_{j,\lfloor uT \rfloor+\tau}^{(q)})= \sum\limits_{m}S^{(p,q)}_j\left(\nicefrac{(\lfloor uT \rfloor+m)}{T}\right)\psi_{j,m}(0)\psi_{j,m}(\tau).
\end{align*}
Motivated by equation~\eqref{eq:lacv}, define $c_j^{(p,q)}(u,\tau)=S^{(p,q)}_j(u) \Psi_j(\tau)$ and,
using the Lipschitz continuity of $S_j^{(p,q)}(u)$,  we bound
\begin{align*}
   & \left| \mathbb{C}\mathrm{ov}(X_{j,\lfloor uT \rfloor}^{(p)},X_{j,\lfloor uT \rfloor+\tau}^{(q)})-c_j^{(p,q)}(u,\tau) \right| \\
    &=\left| \sum\limits_{m}S^{(p,q)}_j\left(\nicefrac{(\lfloor uT \rfloor+m)}{T}\right)\psi_{j,m}(0)\psi_{j,m}(\tau) - c_j^{(p,q)}(u,\tau)\right| \\
    &\leq T^{-1}\sum\limits_{m}|m|L^{(p,q)}_{j}|\psi_{j,m}(0)\psi_{j,m}(\tau)|=\mathcal{O}(T^{-1}),
\end{align*}
where we used the scale-$j$ autocorrelation wavelet property of having a compact support of length $2^j$, coupled with $\Psi_{j}(\tau)=\mathcal{O}(1)$ and the property of the Lipschitz constants. 

Thus the process covariance is $\mathbb{C}\mathrm{ov}(X_{j,[uT]}^{(p)},X_{j,[uT]+\tau}^{(q)}) \approx c_j^{(p,q)}(u,\tau)$, which shows that under the assumption of stationarity of the original process $\XX$, the covariance function of the scale-$j$ subprocess is 
$c_j^{(p,q)}(\tau)=S^{(p,q)}_j \Psi_j(\tau), \, \forall \tau$. Also note that the covariance of $\XX$ is given by $c^{(p,q)}(\tau)=\sum_j S^{(p,q)}_j \Psi_j(\tau)$ \citep{park2014estimating}, which can be reformulated using the subprocess notation as 
\begin{equation}\label{eq:ubc}
c^{(p,q)}(\tau)=\sum_j c_j^{(p,q)}(\tau).
\end{equation}

Continuing under the process $\XX$ stationarity assumption, since the mean and covariance structures for the scale-$j$ subprocess are not time-dependent, as we have seen above, the scale-dependent subprocesses $\{\XX_{j,t}\}_t$ are themselves (weakly) stationary for each scale $j$.

In this setting, we can establish a relationship between the (Fourier) spectral density matrix and the stationary wavelet spectrum for each {\em subprocess}. Specifically, let $\ft{g}\left(\omega\right)$ denote the Fourier transform of $g(\cdot)$ at frequency $\omega$. Then for a node pair $(p,q)$ and scale-$j$ subprocess $\{\XX_{j,t}\}_t$, using the Fourier transform linearity, we have the following relationship between the associated Fourier and wavelet spectra

\begin{equation}\label{eq:ftow}
f_j^{(p,q)}(\omega)= \ft{c}_j^{(p,q)}\left(\omega\right)=S^{(p,q)}_j\ft{\Psi}_j(\omega), 
\end{equation}
where we denote by $\mathbf{f}_j(\cdot)$ and $\SS_{j}$ the $P \times P$-dimensional Fourier and wavelet spectral matrices of the scale-$j$ subprocess, respectively. 
In what follows, we work with frequencies $\omega$ such that $\mathbf{f}_j(\omega)$ is of full rank as stipulated by \citet[Theorem 2.4]{dahlhaus2000graphical}. 
 
We next show that the conditional independence of two nodes given the remaining ones is equivalent to the conditional independence of their subprocesses across all scales $j$. Let us denote by $\{e_t^{(p)}\}_t$ the channel $p$ residual process obtained by removing the linear effects of the nodes other than $p, \, q$ from the original process $\XX$ and $\{e_{j,t}^{(p)}\}$ denote the corresponding residual subprocess at scale-$j$; similarly, define the residual process at node $q$ and its associated subprocesses. Following the arguments in \cite{dahlhaus2000graphical}, the process $\{(e_t^{(p)},e_t^{(q)})^{\top}\}_t$ is bivariate stationary, and the Fourier and wavelet cross-spectra of their scale $j$-subprocess are connected akin to equation~\eqref{eq:ftow}, namely
\begin{equation}\label{eq:residf}
f_j^{e;(p,q)}(\cdot)=S^{e;(p,q)}_j\ft{\Psi}_j(\cdot).
\end{equation}
Additionally, following~\eqref{eq:ubc} for the residual process, we may express its covariance using the covariances of the residual subprocesses as
\begin{equation}\label{eq:residc}
c^{e;(p,q)}(\tau)=\sum_j c_j^{e;(p,q)}(\tau)=\sum_j S^{e;(p,q)}_j \Psi_j(\tau), \, \forall \tau.
\end{equation}

To establish the conditional independence equivalence, first assume the absence of an edge $(p,q)$ in the conditional independence graph of $\XX$, and recall this implies that $c^{e;(p,q)}(\tau)=0,$ for all $\tau$, following \cite{dahlhaus2000graphical}. Combining equation~\eqref{eq:residc} with the linear independence of the autocorrelation wavelets $\{\Psi_j\}_j$ \citep{nason2000wavelet}, this is equivalent to  $S^{e;(p,q)}_j=0, \, \forall j$. Using equation~\eqref{eq:residf}, we obtain that $f_j^{e;(p,q)}(\cdot)\equiv 0, \, \forall j$, and hence the nodes $p$ and $q$ are conditionally independent in each scale-dependent subprocess. 

Conversely, if nodes $p$ and $q$ are conditionally independent in every scale-dependent subprocess, then the corresponding residual cross-covariances satisfy $c^{e;(p,q)}_j(\tau)=0, \, \forall j, \, \forall \tau$ \citep{dahlhaus2000graphical}, hence so do their Fourier transforms $f_j^{e;(p,q)}(\cdot)\equiv 0, \, \forall j$. Summing over scales in equation~\eqref{eq:residc} yields $c^{e;(p,q)}(\tau)=0, \forall \tau$ and therefore by \cite{dahlhaus2000graphical} nodes $p$ and $q$ are conditionally independent in the original process $\XX$. 
Further, using the graph characterisation of each subprocess in Theorem 2.4 of \cite{dahlhaus2000graphical}, the absence of the $(p,q)$ edge equivalently translates into $\big(\mathbf{f}_j^{-1}(\cdot)\big)_{p,q}\equiv 0, \forall j$.

Since $\SS_{j}$ is nonsingular, the scale-$j$ spectral density matrix is also invertible and from \eqref{eq:ftow}
\begin{equation*} 
    \mathbf{f}^{-1}_j(\cdot) = \SS_{j}^{-1}\ft{\Psi}^{-1}_j(\cdot),
\end{equation*}

which implies that
\begin{equation*}
   (p,q) \notin \mathcal{E} \iff \big(\SS_{j}^{-1}\big)_{p,q} = 0, \forall j.
\end{equation*}

\end{proof}

\subsection{Proof of Asymptotic Properties of Time Averaged Periodogram}\label{app:asymptotic-Ij}
Before showing that the time-averaged raw periodogram $\bar{\mathbf{I}}_j$ is an unbiased and consistent estimator of $\boldsymbol{\beta}_j$ at each scale $j$, we first characterise the cross-covariance structure of the wavelet coefficients at different locations for a general multivariate LSW process.

\begin{lemma}\label{lem:d_jk-expec}
    Let $\{\XX_t\}_{t=0}^{T-1}$ be a $P$-dimensional mvLSW process with Gaussian innovations $\{\boldsymbol{\varepsilon}_{j,k}\}_{j,k}$ and wavelet coefficients $\dd_{j,k}=\big(d_{j,k}^{(1)},\ldots, d_{j,k}^{(P)}\big)^\top$ at scale $j$ and location $k$. For each scale $j$, locations $k,k'$ and nodes $p,q=1,\ldots, P$ it holds
    \begin{equation*}
        \mathbb{E}\big(d_{j,k}^{(p)}d_{j,k'}^{(q)}\big)=\sum_\tau c^{(p,q)}\Big(\frac{k}{T}, \tau\Big)\Psi_j(\tau+k'-k) + \mathcal{O}(T^{-1}), 
    \end{equation*}
    where $\Psi_j(\cdot)$ is the autocorrelation wavelet and $\mathbf{c}(u,\tau)=\sum_{l=1}^\infty \SS_{l}(u)\Psi_l(\tau)$ is the localised wavelet cross-covariance matrix defined in equation~\eqref{eq:lacv}. 
\end{lemma}

\begin{proof}
    Recall from Section~\ref{sec:mvLSW-review} that the $p$-th component of an mvLSW process is a $\XX^{(p)}_t=\sum_{j=1}^J\sum_{k \in \mathbb{Z}}\sum_{q=1}^P V_j^{(p,q)}(k/T)\psi_{j,k}(t)\varepsilon_{j,k}^{(q)}$, where the increments $\{\boldsymbol{\varepsilon}_{j,k}\}_{j,k}$ are such that $\mathbb{C}\mathrm{ov}\big(\varepsilon_{j,k}^{(p)}\varepsilon_{l,h}^{(q)}\big)=\delta_{p,q}\delta_{j,l}\delta_{k,h}$. Then, 
    \begin{align*}
        \mathbb{E}\big(d_{j,k}^{(p)}d_{j,h}^{(q)}\big)&=\mathbb{E}\Bigg(\bigg(\sum_{t_1} X_{t_1}^{(p)}\psi_{j,k}(t_1)\bigg)\bigg(\sum_{t_2} X_{t_2}^{(q)}\psi_{j,h}(t_2)\bigg)\Bigg) \\ 
        &= \mathbb{E}\Bigg(\bigg(\sum_{t_1}\sum_{l_1=1}^J\sum_{k'_1}\sum_{r_1=1}^P V_{l_1}^{(p,r_1)}(k'_1/T) \psi_{l_1,k'_1}(t_1)\varepsilon_{l_1,k'_1}^{(r_1)}\psi_{j,k}(t_1)\bigg) \\
        & \, \, \, \, \, \, \, \, \, \, \, \, \, \, \, \times \bigg(\sum_{t_2}\sum_{l_2=1}^J\sum_{k'_2}\sum_{r_2=1}^P V_{l_2}^{(q,r_2)}(k'_2/T) \psi_{l_2,k'_2}(t_2)\varepsilon_{l_2,k'_2}^{(r_2)}\psi_{j,h}(t_2)\bigg)\Bigg) \\
        &= \sum_{l=1}^J\sum_{k'}\sum_{r=1}^P V_l^{(p,r)}(k'/T)V_l^{(q,r)}(k'/T)\bigg(\sum_{t_1} \psi_{l,k'}(t_1)\psi_{j,k}(t_1)\bigg)\bigg(\sum_{t_2} \psi_{l,k'}(t_2)\psi_{j,h}(t_2)\bigg)
    \end{align*}

    Since $\mathbf{V}_l(\cdot)$ is lower triangular by definition, we have  for each $k' \in \mathbb{Z}$
    \begin{equation*}
        \sum_{r=1}^PV_l^{(p,r)}(k'/T)V_l^{(q,r)}(k'/T)=\sum_{r=1}^{\min\{p,q\}}V_l^{(p,r)}(k'/T)V_l^{(q,r)}(k'/T)=S_l^{(p,q)}(k'/T),
    \end{equation*} leading to 
    \begin{align}        \mathbb{E}\big(d_{j,k}^{(p)}d_{j,h}^{(q)}\big)&=\sum_{l=1}^J\sum_{k'}S_l^{(p,q)}(k'/T)\bigg(\sum_{t_1}\sum_{t_2}\psi_{l,k'}(t_1)\psi_{l,k'}(t_2)\psi_{j,k}(t_1)\psi_{j,h}(t_2)\bigg) \nonumber\\
        &=\sum_{l=1}^J\sum_{i} S_l^{(p,q)}\bigg(\frac{i+k}{T}\bigg)\bigg(\sum_{t_1}\sum_{t_2} \psi_{l,i+k-t_1}\psi_{l,i+k-t_2}\psi_{j,k-t_1}\psi_{j,h-t_2}\bigg), \label{eq:crossexp}
    \end{align}
    where we used $k'=i+k$ in the last line. Recalling that $S_l^{(p,q)}(\cdot)$ is a Lipschitz continuous function with Lipschitz constant $L_l^{(p,q)}$ at each scale $l$, we obtain at each fixed $i$, 
    \begin{equation}\label{eq:lipschitz}
        \bigg|S_l^{(p,q)}\bigg(\frac{i+k}{T}\bigg)-S_l^{(p,q)}\bigg(\frac{k}{T}\bigg)\bigg| \leq L_l^{(p,q)}\frac{|i|}{T}.
    \end{equation}
    Coupling indicies $\tau:=t_1-t_2$, $a:=i+k-t_2$ and $b:=h-t_2$ in equation~\eqref{eq:crossexp} with \eqref{eq:lipschitz}, yields
    \begin{align*}
        \mathbb{E}\big(d_{j,k}^{(p)}d_{j,h}^{(q)}\big)&= \sum_\tau\sum_{l=1}^J S_l^{(p,q)}(k/T)\sum_a\psi_{l,a}\psi_{l,a-\tau}\sum_b\psi_{j,b}\psi_{j,b-(\tau+h-k)} +\mathcal{O}(T^{-1}) \\
        &= \sum_\tau\sum_{l=1}^J S_l^{(p,q)}(k/T)
        \Psi_l(\tau)\Psi_j(\tau+h-k)+\mathcal{O}(T^{-1}) \\
        &=\sum_\tau c^{(p,q)}(k/T)
        \Psi_j(\tau+h-k)+\mathcal{O}(T^{-1}).
    \end{align*}
\end{proof}

\subsubsection*{Proof of Proposition~\ref{prop:Ij-consistent}.}
\begin{proof}
    We proceed by proving the result component-wise; thus we restrict our attention to an arbitrary $(p,q)$-th entry of $\bar{\mathbf{I}}_j$. We recall $\beta_j^{(p,q)}:=\sum_{l=1}^JC_{jl}S_l^{(p,q)}$, where $C_{jl}$ denotes the $(j,l)$-th entry of the autocorrelation wavelet inner product matrix, and $\bar{I}_j^{(p,q)}=\frac{1}{T}\sum_{k=0}^{T-1}I_{j,k}^{(p,q)}$ is the time-averaged periodogram.

    Proposition 4 of \cite{park2014estimating} establishes that $\mathbb{E}\big(I_{j,k}^{(p,q)}\big)=\sum_{l=1}^J C_{jl}S_l^{(p,q)}(k/T)+\mathcal{O}(T^{-1})$. Therefore, for a stationary process, $\mathbb{E}\big(I_{j,k}^{(p,q)}\big)=\sum_{l=1}^J C_{jl}S_l^{(p,q)}+\mathcal{O}(T^{-1})$, for all $k$. Hence, $\bar{I}_j^{(p,q)}$ is asymptotically unbiased for $\beta^{(p,q)}_j$ at each scale $j$. 

    We now show that $\mathbb{V}\mathrm{ar}\big(\bar{I}_{j}^{(p,q)}\big)=\mathcal{O}\big(2^jT^{-1}\big)$. We have, 
    \begin{equation*}
        \mathbb{V}\mathrm{ar}\big(\bar{I}_j^{(p,q)}\big)=\frac{1}{T^2}\sum_{k=0}^{T-1}\mathbb{V}\mathrm{ar}\big(I_{j,k}^{(p,q)}\big)+\frac{2}{T^2}\sum_{k=0}^{T-1}\sum_{h=k+1}^{T-1}\mathbb{C}\mathrm{ov}\big(I_{j,k}^{(p,q)}, I_{j,h}^{(p,q)}\big).
    \end{equation*}
    By Proposition 4 of \cite{park2014estimating}, for scale $j$ and location $k$, we can write
    \begin{align*}        \mathbb{V}\mathrm{ar}\big(I_{j,k}^{(p,q)}\big)=&\Bigg(\sum_{l=1}^JC_{jl}S_l^{(p,p)}(k/T)+\mathcal{O}(T^{-1})\Bigg)\Bigg(\sum_{l=1}^JC_{jl}S_l^{(q,q)}(k/T)+\mathcal{O}(T^{-1})\Bigg) \\ &+\bigg(\sum_{l=1}^JC_{jl}S_l^{(p,q)}(k/T)+\mathcal{O}(T^{-1})\bigg)^2.
    \end{align*}
    Since $\{\XX_t\}_{t=0}^{T-1}$ is stationary, $S_l^{(p,q)}$ does not depend on time for any cross-nodal pair $(p,q)$ and scale $l$. Moreover, recall that $\Psi_j(\tau)$ is compactly supported with scale-dependent support and satisfies $|\Psi_j(\tau)|\leq\Psi_j(0)=1$ for all $\tau$ \citep{nason2000wavelet}. We then exploit the compact support of $\Psi_j(\cdot)$, say $[a,b]$, and obtain $|C_{jl}|=\big|\sum_\tau \Psi_j(\tau)\Psi_l(\tau)\big|\leq \sum_{\tau \in [a,b]}|\Psi_j(\tau)||\Psi_l(\tau)|=\mathcal{O}(\min\{2^j,2^l\})$. Coupled with Assumption~\ref{ass:spectral-boundness}, we obtain for any fixed scale $j$ 
    \begin{equation*}
        \Bigg|\sum_{l=1}^J C_{jl}S_l^{(p,q)}\Bigg| \leq \sum_{l=1}^J |C_{jl}|\Big|S_l^{(p,q)}\Big| \leq C_SD\Bigg(\sum_{l=1}^j 2^l2^{-l}+\sum_{l=j+1}^J2^j2^{-l}\Bigg) \leq C_SD(j+1)=\mathcal{O}(1),
    \end{equation*}
    where $C_S$ is a constant. Hence the variance of the raw periodogram for each scale $j$ and location $k$ becomes
    \begin{equation*}        \mathbb{V}\mathrm{ar}\big(I_{j,k}^{(p,q)}\big)=\Bigg(\sum_{l=1}^JC_{jl}S_l^{(p,p)}\Bigg)\Bigg(\sum_{l=1}^JC_{jl}S_l^{(q,q)}\Bigg)+\Bigg(\sum_{l=1}^JC_{jl}S_l^{(p,q)}\Bigg)^2+\mathcal{O}\left({2^j}{T}^{-1}\right).
    \end{equation*}
    Since $j$ is a fixed, fine enough scale with $2^j=o(T)$, we obtain,
    \begin{equation}\label{eq:Var-Ij}
        \frac{1}{T^2}\sum_{k=0}^{T-1}\mathbb{V}\mathrm{ar}\big(I_{j,k}^{(p,q)}\big)=\frac{1}{T^2}\sum_{k=0}^{T-1}\left(\mathcal{O}(1)+
        \mathcal{O}\left({2^j}{T}^{-1}\right)\right)=\mathcal{O}\left({T}^{-1}\right).
    \end{equation}
    On the other hand, similar calculations to \cite{park2014estimating} yield, 
    \begin{equation*}
        \mathbb{E}\big(I_{j,k}^{(p,q)}I_{j,h}^{(p,q)}\big) = \mathbb{E}\big(I_{j,k}^{(p,q)}\big)\mathbb{E}\big(I_{j,h}^{(p,q)}\big)+\mathbb{E}\big(d_{j,k}^{(p)}d_{j,h}^{(p)}\big)\mathbb{E}\big(d_{j,k}^{(q)}d_{j,h}^{(q)}\big)+\mathbb{E}\big(d_{j,k}^{(p)}d_{j,h}^{(q)}\big)\mathbb{E}\big(d_{j,k}^{(q)}d_{j,h}^{(p)}\big).
    \end{equation*}
    Together with Lemma \ref{lem:d_jk-expec}, for each $p,q,j,k,h$ the covariance term becomes,
    \begin{align*}        \mathbb{C}\mathrm{ov}\big(I_{j,k}^{(p,q)},I_{j,h}^{(p,q)}\big)&=\mathbb{E}\big(I_{j,k}^{(p,q)}I_{j,h}^{(p,q)}\big)-\mathbb{E}\big(I_{j,k}^{(p,q)}\big)\mathbb{E}\big(I_{j,h}^{(p,q)}\big) \\        &=\mathbb{E}\big(d_{j,k}^{(p)}d_{j,h}^{(p)}\big)\mathbb{E}\big(d_{j,k}^{(q)}d_{j,h}^{(q)}\big)+\mathbb{E}\big(d_{j,k}^{(p)}d_{j,h}^{(q)}\big)\mathbb{E}\big(d_{j,k}^{(q)}d_{j,h}^{(p)}\big)\\
        &= \bigg(\sum_\tau c^{(p,p)}\bigg(\frac{k}{T},\tau\bigg)\Psi_j(\tau+h-k)+\mathcal{O}(T^{-1})\bigg) \\
        &\times \bigg(\sum_\tau c^{(q,q)}\bigg(\frac{k}{T},\tau\bigg)\Psi_j(\tau+h-k)+\mathcal{O}(T^{-1})\bigg) \\
        &+\bigg(\sum_\tau c^{(p,q)}\bigg(\frac{k}{T},\tau\bigg)\Psi_j(\tau+h-k)+\mathcal{O}(T^{-1})\bigg)^2
    \end{align*}
    As before, we rely on the properties of the autocorrelation wavelet \citep{nason2000wavelet}
    \begin{align*}
        \mathbb{C}\mathrm{ov}\big(I_{j,k}^{(p,q)}, I_{j,h}^{(p,q)}\big)=&\bigg(\sum_\tau c^{(p,p)}\bigg(\frac{k}{T},\tau\bigg)\Psi_j(\tau+h-k)\bigg) \bigg(\sum_\tau c^{(q,q)}\bigg(\frac{k}{T},\tau\bigg)\Psi_j(\tau+h-k)\bigg)\\ &+\bigg(\sum_\tau c^{(p,q)}\bigg(\frac{k}{T},\tau\bigg)\Psi_j(\tau+h-k)\bigg)^2+\mathcal{O}\left({2^j}{T}^{-1}\right),
    \end{align*}
    and we leverage the compact support $[a,b]$ of $\Psi_j(\cdot)$ to bound the product sums as in \cite{cardinali2011costationarity}, namely
    \begin{align*}
        \Bigg(\sum_{\tau=a-(h-k)}^{b-(h-k)}c^{(p,p)}\bigg(\frac{k}{T}, \tau\bigg)\Psi_j(\tau+h-k)\Bigg)\Bigg(\sum_{\tau=a-(h-k)}^{b-(h-k)}c^{(q,q)}\bigg(\frac{k}{T}, \tau\bigg)\Psi_j(\tau+h-k)\Bigg) \\
        \leq \bigg(\sum_{r=a}^b \bigg|c^{(p,p)}\bigg(\frac{k}{T}, r-(h-k)\bigg)\bigg|\bigg)\bigg(\sum_{r=a}^b\bigg|c^{(q,q)}\bigg(\frac{k}{T}, r-(h-k)\bigg)\bigg|\bigg),
    \end{align*}
    and similarly for the second term.

    Using the process stationarity and $\sum_\tau|c^{(p,q)}(\tau)|<\infty$ for each pair $(p,q)$, together with the compact support length $\mathcal{O}(2^j)$ of $\Psi_j(\cdot)$, we denote $A_h^{(p,q)}=\sum_{r=a}^b|c^{(p,q)}(r-(h-k))|$. Then $\sup_hA_h^{(p,q)}<\infty$ and $\sum_h A_h^{(p,q)} = \mathcal{O}(2^j)$, so $\sum_h A_h ^{(p,p)}A_h^{(q,q)} = \mathcal{O}(2^j)$ and $\sum_h \big(A_h^{(p,q)}\big)^2=\mathcal{O}(2^j)$. Hence,
    \begin{align}\label{eq:Cov-Ij}
        \frac{2}{T^2}\sum_{k=0}^{T-1}\sum_{h=k+1}^{T-1}\mathbb{C}\mathrm{ov}\big(I_{j,k}^{(p,q)}, I_{j,h}^{(p,q)}\big)&=\frac{2}{T^2}\sum_{k=0}^{T-1} \mathcal{O}\big(2^j\big) + \frac{2}{T^2}\sum_{k=0}^{T-1}\sum_{h=k+1}^{T-1}\mathcal{O}\left({2^j}{T}^{-1}\right) = \mathcal{O}\left({2^j}{T}^{-1}\right)
    \end{align}

    Combining equations \eqref{eq:Var-Ij} and \eqref{eq:Cov-Ij}, we find that $\mathbb{V}\mathrm{ar}\big(\bar{I}_j^{(p,q)}\big)=\mathcal{O}\left({2^j}{T}^{-1}\right)$. Together with asymptotic unbiasedness, $\mathbb{V}\mathrm{ar}\big(\bar{I}_j^{(p,q)}\big)\rightarrow 0$ as $T \rightarrow \infty$ implies mean-square consistency.
\end{proof}

\subsection{Proof for Precision Estimator Convergence in Frobenius norm (Theorem~\ref{thm1})}\label{app:proofthm1}

Our proof of Theorem~\ref{thm1} follows the general strategy of \cite{Ravikumar2011consistency} and \cite{tugnait2022sparse} but is adapted to the wavelet-domain likelihood and scale-wise formulation. We first extend Proposition~\ref{prop:Ij-consistent} to the growing network-dimension setting, and use this result to establish a high-probability bound for the bias-corrected spectral estimator in Lemma \ref{lemma1}. We then establish a high-probability bound for the replicate-averaged spectral estimator used in the optimisation problem \eqref{eq:S-inverse_optimisation} in Lemma \ref{lemma2}. Next we develop a Taylor expansion for the log-determinant with explicit integral remainder in Lemmas \ref{lemma3} and \ref{lemma4}, and finally combine these to obtain a bound on the Frobenius norm of $\hat\TTheta_j^{(X)}-\TTheta_{j}^{(X)}$. 

\paragraph{Notational Simplifications.} From here on, we will drop the explicit notation  showing to which process standard quantities pertain: going forward $\TTheta_j$ denotes $\TTheta_j^{(X)}$; $\SS_j$ denotes $\SS_j^{(X)}$, and we will also write $\dd_{j,k}$ to denote $\dd_{j,k}^{(Y)}$; $\bar{\mathbf{I}}_j$ to denote $\bar{\mathbf{I}}_j^{(Y)}$, unless otherwise stated.

\subsubsection{Preliminary Spectral Concentration Results}
Proposition~\ref{prop:Ij-consistent} establishes component-wise consistency of the time-averaged wavelet periodogram when the process dimension $P$ is fixed. For the growing network-dimension analysis required for Theorem~\ref{thm1}, we require these bounds to hold uniformly over the $P_T^2$ component pairs. Corollary \ref{prop:high-dimensional-Ij-consistency} below provides this uniform extension, from which a high-probability bound for the bias-corrected spectral estimator follows.

\begin{corollary}[Uniform extension of Proposition~\ref{prop:Ij-consistent}]\label{prop:high-dimensional-Ij-consistency}
    Let $\XX=\{\XX_t\}_{t=0}^{T-1}$ be a $P_T$-dimensional zero-mean stationary Gaussian process satisfying Assumptions \ref{ass:covariance-summability} and \ref{ass:mvLSW-regularity}. Recall $\boldsymbol{\beta}_j=\sum_{l=1}^J C_{jl}\SS_l$, and let $\bar{\mathbf{I}}_j=\frac{1}{T}\sum_{k=0}^{T-1} \mathbf{I}_{j,k}$ be the time-smoothed periodogram at scale $j$. 
    
    Then, for each fixed scale $j$ such that $2^j=o(T)$, there exist finite constants $\mu_j$ (independent of $T, P_T, p$ and $q$), and $\kappa$ (additionally independent of $j$), such that 
    \begin{equation*}
        \max_{1\leq p,q\leq P_T}\bigg|\mathbb{E}\Big(\bar{I}_j^{(p,q)}\Big)-\beta_j^{(p,q)}\bigg| \leq \frac{\mu_j}{T},
    \end{equation*}
    and
    \begin{equation*}
        \max_{1\leq p,q \leq P_T} \mathbb{V}\mathrm{ar}\Big(\bar{I}_j^{(p,q)}\Big) \leq \frac{2^j \kappa}{T}.
    \end{equation*}
    Consequently, for any such fixed scale $j$,
    \begin{equation*}
        \max_{1\leq p,q\leq P_T} \mathbb{E}\bigg[\Big(\bar{I}_j^{(p,q)}-\beta_j^{(p,q)}\Big)^2\bigg]=\mathcal{O}(2^jT^{-1}).
    \end{equation*}
\end{corollary}

\begin{proof}
    Under the additional conditions of Assumptions \ref{ass:covariance-summability} and \ref{ass:mvLSW-regularity} along with Assumption~\ref{ass:spectral-boundness} used in Proposition~\ref{prop:Ij-consistent}, the covariance summability, spectral and mvLSW regularity bounds used in the proof of Proposition~\ref{prop:Ij-consistent} hold uniformly over $p,q=1,\ldots,P_T$, with constants independent of $T$ and $P_T$. Consequently, the constants appearing in the component-wise bias and variance bounds of Proposition~\ref{prop:Ij-consistent} may be chosen uniformly over the $P_T^2$ component pairs. Moreover, the scale-uniform bounds in Assumptions \ref{ass:covariance-summability} and \ref{ass:mvLSW-regularity} allow the variance constant to be chosen independently of $j$. Hence for all sufficiently large $T$,

    \begin{equation*}
        \max_{p,q} \bigg|\mathbb{E}\Big(\bar{I}_j^{(p,q)}\Big)-\beta_j^{(p,q)}\bigg| \leq \frac{\mu_j}{T},
    \end{equation*}
and 
    \begin{equation*}
        \max_{1\leq p,q \leq P_T} \mathbb{V}\mathrm{ar}\Big(\bar{I}_j^{(p,q)}\Big) \leq \frac{2^j \kappa}{T},
    \end{equation*}
    for finite constants $\mu_j$ and $\kappa$ satisfying the stated independence conditions. The result follows.
    
\end{proof}

We next couple this result with the bias-correction of the time-smoothed periodogram in Corollary \ref{prop:high-dimensional-Ij-consistency}, and obtain a uniform concentration bound for the spectral estimator $\hat{\SS}_j$.

\begin{lemma}\label{lemma1}
    Let $\XX$ be a stationary process as in Proposition~\ref{prop:high-dimensional-Ij-consistency} and assume Assumptions \ref{ass:spectral-boundness}-\ref{ass:mvLSW-regularity} hold. Let $\hat{\SS}_j=\sum_{l=1}^J (C^{-1})_{jl}\bar{\mathbf{I}}_l^{(X)}$ denote its bias-corrected spectral estimator at a fixed scale $j$, where $\bar{\mathbf{I}}_l^{(X)}$ denotes the scale-$l$ raw periodogram smoothed over the entire time domain. 
    Then, for $\nu>2$, there exist finite constants $\tilde{\mu}_j, \tilde{\kappa}_j>0$, independent of $T, P_T, p$ and $q$, such that 
    \begin{equation}\label{eq:corollary1-bound}
        \mathbb{P}\Bigg(\Big\|\hat{\SS}_j-\SS_j\Big\|_\infty > \frac{\tilde{\mu}_j}{\sqrt{T}} + \sqrt{\frac{J\tilde{\kappa}_jP_T^\nu}{T}}\Bigg) \leq P_T^{2-\nu},
    \end{equation}
    where $\|\mathbf{M}\|_\infty=\max_{p,q}|M_{p,q}|$ for a generic matrix $\mathbf{M} \in \mathbb{R}^{P_T \times P_T}$. 
    Hence as $T \rightarrow \infty$ and $JP_T^\nu/T \rightarrow 0$,
    \begin{equation}
        \Big\|\hat{\SS}_j-\SS_j\Big\|_\infty=\mathcal{O}_P\Bigg(\frac{\tilde{\mu}_j}{\sqrt{T}} + \sqrt{\frac{J\tilde{\kappa}_jP_T^\nu}{T}}\Bigg) \,\,\,\ \mbox{ and } \,\,\,\, \Big\|\hat{\SS}_j-\SS_j\Big\|_\infty=o_P(1)
    \end{equation} hold with probability at least $1-P_T^{2-\nu}$.
\end{lemma}

\begin{proof}
    For each fixed scale $j$, akin to the bias-correction result of \cite{sanderson2010nonstationary},
    \begin{equation}\label{eq:bias-bound-S-hat-X}
        \Big|\mathbb{E}\big(\hat{S}_j^{(p,q)}\big) -S_j^{(p,q)}\Big| \leq \frac{\tilde{\mu}_j}{\sqrt{T}}.
    \end{equation}

     By Corollary \ref{prop:high-dimensional-Ij-consistency} and the covariance Cauchy-Schwarz inequality,
    \begin{align}\label{eq:corollary1-variance-1}
        \mathbb{V}\mathrm{ar}\Big(\hat{S}_j^{(p,q)}\Big) 
        & =\sum_{l=1}^J \sum_{m=1}^J (C^{-1})_{jl}(C^{-1})_{jm} \mathbb{C}\mathrm{ov}\Big(\bar{I}_l^{(p,q)}, \bar{I}_m^{(p,q)}\Big) \nonumber \\ 
        & \leq \sum_{l=1}^J\sum_{m=1}^J\big|(C^{-1})_{jl}\big|\big|(C^{-1})_{jm}\big|\sqrt{\mathbb{V}\mathrm{ar}\Big(\bar{I}_l^{(p,q)}\Big)\mathbb{V}\mathrm{ar}\Big(\bar{I}_m^{(p,q)}\Big)} \nonumber \\
        & =  \Bigg[\sum_{l=1}^J\big|(C^{-1})_{jl}\big|\sqrt{\mathbb{V}\mathrm{ar}\Big(\bar{I}_l^{(p,q)}\Big)}\Bigg]^2 \nonumber \\ 
        & \leq \frac{\kappa}{T}\Bigg[\sum_{l=1}^J \big|(C^{-1})\big|_{jl}2^{l/2}\Bigg]^2.
    \end{align}

    \citet[Proposition 4.3.2]{mcgonigle2020wavelet} shows that the rescaled autocorrelation wavelet inner-product operator $\mathbf{B}$ has a bounded inverse on $\ell_2$ where $\mathbf{B}=\mathbf{DCD}$ where $\mathbf{D}=\text{diag}(2^{-1/2},\ldots 2^{-J/2})$. Hence there exists a finite constant $C_B>0$ such that $\|\mathbf{B}^{-1}\mathbf{v}\|_2\leq C_B\|\mathbf{v}\|_2$ for all $\mathbf{v} \in \ell_2$, or equivalently $\|\mathbf{B}^{-1}\|_2 \leq C_B< \infty$. Therefore for a fixed $j$, 
    \begin{equation}\label{eq:mcgonigle-1}
        \Bigg[\sum_{l=1}^J \Big|(C^{-1})_{jl}\Big|2^{l/2}\Bigg]^2=2^{-j}\Bigg[\sum_{l=1}^J \Big|(B^{-1})_{jl}\Big|\Bigg]^2\leq 2^{-j} J \sum_{l=1}^J(B^{-1})_{jl}^2
    \end{equation}
    by the Cauchy-Schwarz inequality. Let $\mathbf{e}_j$ denote the $j$-th standard basis vector. Since $\mathbf{e}_j^\top\mathbf{B}^{-1}$ is the $j$-th row of $\mathbf{B}^{-1}$, $\sum_{l=1}^J (B^{-1})_{jl}^2=\|\mathbf{e}_j^\top\mathbf{B}^{-1}\|_2^2\leq \|\mathbf{e}_j^\top\|_2^2\|\mathbf{B}^{-1}\|_2^2=\|\mathbf{B}^{-1}\|_2^2$. Hence equation~\eqref{eq:mcgonigle-1} becomes 
    \begin{equation*}
        \Bigg[\sum_{l=1}^J \Big|(C^{-1})_{jl}\Big|2^{l/2}\Bigg]^2\leq 2^{-j}J\|\mathbf{B}^{-1}\|_2^2\leq 2^{-j}JC_B^2.
    \end{equation*}
    
    Hence combining with equation~\eqref{eq:corollary1-variance-1} there exists a finite constant $\tilde{\kappa}_j:=\kappa2^{-j}C_B^2$, independent of $T, P_T, p$ and $q$ such that 
    \begin{equation}\label{eq:var-bound-S-hat-X}
        \mathbb{V}\mathrm{ar}\Big(\hat{S}_j^{(p,q)}\Big)\leq \frac{J\tilde{\kappa}_j}{T}.
    \end{equation}

    Therefore for any $\delta>0$, Chebyshev's inequality gives, 
    \begin{equation*}
        \mathbb{P}\bigg(\Big|\hat{S}_j^{(p,q)}-\mathbb{E}\big(\hat{S}_j^{(p,q)}\big)\Big| \leq  \delta \bigg) \geq 1-\frac{J\tilde{\kappa}_j}{T\delta^2}.
    \end{equation*}

    By the triangle inequality and the bias bound of equation~\eqref{eq:bias-bound-S-hat-X}, the events below are related as follows
    $$\bigg\{\Big|\hat{S}_j^{(p,q)}-\mathbb{E}\big(\hat{S}_j^{(p,q)}\big)\Big|\leq \delta\bigg\} \subseteq \bigg\{\Big|\hat{S}_j^{(p,q)}-S_j^{(p,q)}\Big|\leq \delta +\frac{\tilde{\mu}_j}{\sqrt{T}}\bigg\}.$$

    Hence combining the previous two equations, we have 
    \begin{equation*}
        \mathbb{P}\Bigg(\Big|\hat{S}_j^{(p,q)}-S_j^{(p,q)}\Big| \leq  \delta +\frac{\tilde{\mu}_j}{\sqrt{T}} \Bigg) \geq 1-\frac{J\tilde{\kappa}_j}{T\delta^2},
    \end{equation*}
    or equivalently, for any channel pairs $(p,q)$,
    \begin{equation*}
        \mathbb{P}\Bigg(\Big|\hat{S}_j^{(p,q)}-S_j^{(p,q)}\Big| >  \delta +\frac{\tilde{\mu}_j}{\sqrt{T}} \Bigg) \leq \frac{J\tilde{\kappa}_j}{T\delta^2}.
    \end{equation*}

    Taking a union bound over the $P_T^2$ component pairs, 
    \begin{equation*}
        \mathbb{P}\Bigg(\Big\|\hat{\SS}_j-\SS_j\Big\|_\infty > \delta + \frac{\tilde{\mu}_j}{\sqrt{T}}\Bigg) \leq P_T^2\frac{J\tilde{\kappa}_j}{T\delta^2}.
    \end{equation*}

    Setting $\delta:=\sqrt{\frac{J\tilde{\kappa}_jP_T^\nu}{T}}$ for $\nu>2$, we have $ P_T^2\frac{J\tilde{\kappa}_j}{T\delta^2}=P_T^{2-\nu}$, thus
    \begin{equation}\label{eq:lemma1-final}
        \mathbb{P}\Bigg(\Big\|\hat{\SS}_j-\SS_j\Big\|_\infty > \frac{\tilde{\mu}_j}{\sqrt{T}} + \sqrt{\frac{J\tilde{\kappa}_jP_T^\nu}{T}}\Bigg) \leq P_T^{2-\nu}.
    \end{equation}

    Consequently, if $JP_T^\nu/T\rightarrow 0$ as $T \rightarrow \infty$, the threshold in \eqref{eq:lemma1-final} converges to zero with probability at least $1-P_T^{2-\nu}$, hence $\Big\|\hat{\SS}_j-\SS_j\Big\|_\infty =o_P(1)$ with probability at least $1-P_T^{2-\nu}$.

\end{proof}

\subsubsection{Auxiliary results for the penalised estimator}

Having established the required concentration bound for the bias-corrected spectral estimator, we now turn to the auxiliary results needed to analyse the penalised optimisation problem. The following lemmas control the stochastic error entering through the replicated-averaged spectral estimator and establish the bounds required for the log-determinant term. Together, these results provide the ingredients required for the proof of Theorem~\ref{thm1}.

\begin{lemma}\label{lemma2}  
    Let $\XX$ be a stationary process as in Proposition~\ref{prop:high-dimensional-Ij-consistency}, and denote by $\hat{\hat\SS}_j=\frac{1}{R}\sum_{r=1}^R \bar{\mathbf{I}}_j^{(Y^{(r)})}$ the spectral estimator used in optimisation problem \eqref{eq:S-inverse_optimisation}. The replicates $\YY^{(1)},\ldots,\YY^{(R)}$ are conditionally independent given the estimated spectral structure $\mathcal{F}_X=\Big\{\hat{\SS}_j^{(X)}\Big\}_{j=1}^J$, and the number of replicates $R\geq 1$ is fixed. Suppose Assumptions \ref{ass:spectral-boundness}-\ref{ass:mvLSW-regularity} hold. Then for a fixed scale $j$ and $\nu>2$, there exist finite constants $\tilde{\mu}_j, \kappa_j^*>0$, independent of $T,P_T,p,q$ such that the following holds, 
    \begin{equation}\label{eq:lemma2-Pbound}
        \mathbb{P}\bigg(\Big\|\hat{\hat\SS}_j-\SS_j\Big\|_\infty > a_{j,T}\bigg)\leq P_T^{2-\nu},
    \end{equation}
    where $a_{j,T}=\frac{\tilde{\mu}_j}{\sqrt{T}}+\sqrt{\frac{J\kappa_j^*P_T^\nu}{T}}$, and $\|\mathbf{M}\|_\infty=\max_{p,q}|M_{p,q}|$ for a generic matrix $\mathbf{M} \in \mathbb{R}^{P_T \times P_T}$.    
    If $JP_T^\nu/T\rightarrow 0$  as $T \rightarrow \infty$, then $a_{j,T}\rightarrow 0$ and $\Big\|\hat{\hat\SS}_j-\SS_j\Big\|_\infty = o_P(1)$ with probability at least $1-P_T^{2-\nu}$.
\end{lemma}

\begin{proof}
    For each auxiliary replicate $\YY^{(r)}$, conditional on the observed process $\XX$ and therefore on its estimated spectral structure $\mathcal{F}_X=\Big\{\hat{\SS}_j^{(X)}\Big\}_j$, the simulated process has spectral structure $\hat{\SS}_j^{(Y)}=\sum_{l=1}^J (\mathbf{C}^{-1})_{jl}\hat{\SS}_l^{(X)}$. Hence its $\beta$-spectrum at scale $j$ is given by $\hat{\boldsymbol{\beta}}_j^{(Y)}=\sum_{l=1}^JC_{jl}\hat{\SS}_l^{(Y)}=\sum_{l=1}^JC_{jl}\sum_{m=1}^J(\mathbf{C}^{-1})_{l m}\hat{\SS}_m^{(X)}=\hat{\SS}_j^{(X)}$. Conditional on $\mathcal{F}_X$, $\YY^{(r)}$ is stationary, hence its spectral structure does not depend on time. Then following \citet[Proposition 4]{park2014estimating}, 
    \begin{align*}
        \mathbb{E}(I_{j,k}^{(Y^{(r)});(p,q)}|\mathcal{F}_X)&=\sum_{l=1}^J\sum_{n\in \mathbb{Z}}\Big(\sum_t \psi_{l,n+k-t}\psi_{j,k-t}\Big)^2\hat{S}_l^{(Y);(p,q)}, \\ 
        &=\sum_{l=1}^J C_{jl}\hat{S}_l^{(Y);(p,q)}, \mbox{ from the matrix }\mathbf C \mbox{ definition},\\
        &=\sum_{l=1}^J C_{jl}\sum_{l'=1}^J(\mathbf{C}^{-1})_{ll'}\hat{S}_{l'}^{(X);(p,q)} =\hat{S}_j^{(X);(p,q)} \mbox{ for all times }k.
    \end{align*} 
    Thus averaging over the time dimension gives for every replicate $r$ and nodal pair $(p,q)$, 
    \begin{equation*}
        \mathbb{E}\Big(\bar{I}_j^{(Y^{(r)});(p,q)}\big| \mathcal{F}_X\Big)=\frac{1}{T}\sum_{k=0}^{T-1}\mathbb{E}\Big(I_{j,k}^{(Y^{(r)});(p,q)}\big| \mathcal{F}_X\Big)=\hat{S}_j^{(X);(p,q)},
    \end{equation*}
    where we recall that $\bar{I}_j^{(Y^{(r)});(p,q)}=\frac{1}{T}\sum_{k=0}^{T-1} {I}_{j,k}^{(Y^{(r)});(p,q)}$. Further averaging over the $R$ replicates yields the conditional expectation of the final estimator $\hat{\hat{S}}_j^{(p,q)}=\frac{1}{R}\sum_{r=1}^R\bar{I}_j^{(Y^{(r)};(p,q)}$, as
    \begin{equation*}
        \mathbb{E}\Big(\hat{\hat{S}}_j^{(p,q)}\big| \mathcal{F}_X\Big)=\frac{1}{R}\sum_{r=1}^R\mathbb{E}\Big(\bar{I}_j^{(Y^{(r)});(p,q)}\big| \mathcal{F}_X\Big)=\hat{S}_j^{(X);(p,q)}.
    \end{equation*}
   Coupling the tower property, $\mathbb{E}\Big(\hat{\hat{S}}_j^{(p,q)}\Big)=\mathbb{E}\Big(\mathbb{E}\Big(\hat{\hat{S}}_j^{(p,q)}\big| \mathcal{F}_X \Big)\Big)=\mathbb{E}\Big(\hat{S}_j^{(X);(p,q)}\Big)$, with Lemma \ref{lemma1}, we bound the bias of the proposed estimator as follows,
    \begin{equation*}
        \bigg| \mathbb{E}\Big(\hat{\hat{S}}_j^{(p,q)}\Big)-S_j^{(p,q)}\bigg| = \bigg|\mathbb{E}\Big(\hat{\hat{S}}_j^{(X);(p,q)}\Big)-S_j^{(p,q)}\bigg| \leq \frac{\tilde{\mu}_j}{\sqrt{T}}.
    \end{equation*}
Following the variance proof of Proposition~\ref{prop:Ij-consistent}, applied conditionally on $\mathcal{F}_X$, we can write
\begin{align*}
    \mathbb{E}\left(\mathbb{V}\mathrm{ar}\left(\bar{I}_j^{(Y^{(r)});(p,q)}\big| \mathcal{F}_X\right)\right) &=\frac{1}{T^2}\sum_{k=0}^{T-1} \mathbb{E}\left(\mathbb{V}\mathrm{ar}\left(I_{j,k}^{(Y^{(r)});(p,q)} \big| \mathcal{F}_X \right)\right) \\
    &\,\,\,\,\,\,\,\, + \frac{2}{T^2}\sum_{k=0}^{T-1}\sum_{k'=k+1}^{T-1} \mathbb{E}\left(\mathbb{C}\mathrm{ov}\left(I_{j,k}^{(Y^{(r)});(p,q)}, I_{j,k'}^{(Y^{(r)});(p,q)}\big| \mathcal{F}_X\right)\right)\\
    &=\mathcal{O}(2^jT^{-1})
\end{align*}
Hence, 
\begin{equation*}
    \mathbb{E}\left(\mathbb{V}\mathrm{ar}\left(\bar{I}_j^{(Y^{(r)});(p,q)}\big| \mathcal{F}_X\right)\right) \leq \frac{K2^j}{T}, \,\,\,\, \mbox{for some finite constant } K.
\end{equation*}
Averaging over $R$ replicates, 
\begin{equation*}
    \mathbb{E}\left(\mathbb{V}\mathrm{ar}\left(\hat{\hat{S}}_j^{(p,q)}\big| \mathcal{F}_X\right)\right) = \frac{1}{R^2}\sum_{r=1}^R \mathbb{E}\left(\mathbb{V}\mathrm{ar}\left(\bar {I}_j^{(Y^{(r)});(p,q)}\big| \mathcal{F}_X\right)\right) \leq \frac{K2^j}{RT}.
\end{equation*}
Then by the law of total variance, 
\begin{align*}
     \mathbb{V}\mathrm{ar}\left(\hat{\hat{S}}_j^{(p,q)}\right) &=\mathbb{V}\mathrm{ar}\left(\mathbb{E}\left(\hat{\hat{S}}_j^{(p,q)}\big| \mathcal{F}_X\right)\right)+\mathbb{E}\left(\mathbb{V}\mathrm{ar}\left(\hat{\hat{S}}_j^{(p,q)}\big| \mathcal{F}_X\right)\right) \\
     &\leq \frac{J\tilde{\kappa}_j}{T}+ \frac{K2^j}{RT}, \mbox{ where the first term follows from equation~\eqref{eq:var-bound-S-hat-X}.} 
\end{align*}
Denoting $\kappa_j^*=\tilde{\kappa}_j+\frac{K2^j}{R}$, we obtain from the above
\begin{equation*}
    \mathbb{V}\mathrm{ar}\Big(\hat{\hat{S}}_j^{(p,q)}\Big) \leq\frac{J\kappa_j^*}{T}.
\end{equation*}
Therefore, for any $\delta>0$ by Chebyshev's inequality,
\begin{equation*}
    \mathbb{P}\Bigg(\Bigg|\hat{\hat{S}}_j^{(p,q)}-\mathbb{E}\Big(\hat{\hat{S}}_j^{(p,q)}\Big)\Bigg| > \delta \Bigg) \leq  \frac{J\kappa_j^{*}}{T\delta^2},
\end{equation*}
and by the same argument as in Lemma \ref{lemma1} we obtain,
\begin{equation*}
    \mathbb{P}\Bigg(\Bigg|\hat{\hat{S}}_j^{(p,q)}-S_j^{(p,q)}\Bigg| > \frac{\tilde\mu_j}{\sqrt{T}} + \delta  \Bigg) \leq  \frac{J\kappa_j^{*}}{T\delta^2}.
\end{equation*}
Applying a union bound over the $P_T^2$ component pairs and letting $\delta=\sqrt{\frac{J\kappa_j^{*}P_T^\nu}{T}}$ for $\nu>2$, gives
\begin{equation*}
    \mathbb{P}\Bigg(\Big\|\hat{\hat{\SS}}_j-\SS_j\Big\|_\infty > \frac{\tilde\mu_j}{\sqrt{T}}+\sqrt{\frac{J\kappa_j^{*} P_T^\nu}{T}}\Bigg) \leq P_T^{2-\nu},
\end{equation*}
hence defining $a_{j,T}=\frac{\tilde\mu_j}{\sqrt{T}}+\sqrt{\frac{J\kappa_j^{*} P_T^\nu}{T}}$ gives the stated result. 

Additionally, if $JP_T^\nu/T\rightarrow 0$  as $T \rightarrow \infty$, it is immediate that $a_{j,T}\rightarrow 0$ and $\Big\|\hat{\hat{\SS}}_j-\SS_j\Big\|_\infty=o_P(1)$ with probability at least $1-P_T^{2-\nu}$.

\end{proof}

Next, Lemma \ref{lemma3} gives a second-order Taylor expansion for terms of the form $\log \det (\TTheta_0 + \DDelta)$, for some matrix $\DDelta$.

\begin{lemma} \label{lemma3}
    At each scale $j$, denote $g(\tilde{\boldsymbol{\Theta}}_j)=\log\det(\tilde{\boldsymbol{\Theta}}_j)$ and let $\tilde{\boldsymbol{\Theta}}_j=\TTheta_{j}+\boldsymbol{\Delta}_j\succ 0$, with $\DDelta_j^{\top}=\DDelta_j$. Then the Taylor series expansion of $g(\tilde{\boldsymbol{\Theta}}_j)$ at $\TTheta_{j}$ is given by
\begin{equation*}
    g(\tilde{\boldsymbol{\Theta}}_j)=g(\TTheta_{j})+\tr(\TTheta_j^{-1}\DDelta_j)-\frac{1}{2}\text{vec}(\DDelta_j)^{\top}(\TTheta_j^{-1} \otimes \TTheta_j^{-1})\text{vec}(\DDelta_j) + \text{h.o.t.},
\end{equation*}
where the acronym $\text{h.o.t.}$ stands for higher order terms in $\DDelta_j$.
\end{lemma}
\medskip
\begin{proof}
In the notation above,  $g(\tilde{\TTheta}_j)=\log\det(\TTheta_j+\DDelta_j)$. Note that using the invertibility of the spectra/precision matrices, we can write \begin{equation*}
    \TTheta_j+\DDelta_j=\TTheta_j^{1/2}(\mathbf{I}_{P_T}+\TTheta_j^{-1/2}\DDelta_j\TTheta_j^{-1/2})\TTheta_j^{1/2},
\end{equation*}
where $\mathbf{I}_{P_T}$ denotes the identity matrix of size $P_T \times P_T$.
As all matrices are square, using basic properties of determinants we obtain
\begin{eqnarray}
    \det(\TTheta_j+\DDelta_j)&=&\det(\TTheta_j)\det(\mathbf{I}_{P_T}+\TTheta_j^{-1/2}\DDelta_j\TTheta_j^{-1/2}), \mbox { from which we derive}\nonumber\\
    \label{eq:lemma4-logdet}
    \log\det(\TTheta_j+\DDelta_j)&=&\log\det(\TTheta_j)+\log\det(\mathbf{I}_{P_T}+\mathbf{U}),    
    \end{eqnarray}
where to simplify notation we use $\mathbf{U}=\TTheta_j^{-1/2}\DDelta_j\TTheta_j^{-1/2}$. 

Focusing on the second term in the right-hand side of equation~\eqref{eq:lemma4-logdet}, we have $\log\det(\mathbf{I}_{P_T}+\mathbf{U})=\tr(\log(\mathbf{I}_{P_T}+\mathbf{U}))$ \citep{logdettrlog} since $(\mathbf{I}_{P_T}+\mathbf{U})$ is symmetric and positive definite. The symmetry follows immediately from the definition of $\mathbf U$ and the symmetry of the precision matrices. The positive definiteness follows from the assumption $\tilde{\TTheta}_j=
\TTheta_j+\DDelta_j \succ 0$, thus $\mathbf{I}_{P_T}+\mathbf{U}= \TTheta_j^{-1/2} (\TTheta_j+\DDelta_j) \TTheta_j^{-1/2}\succ 0$.

Using a Taylor expansion to write $\log\det(\mathbf{I}_{P_T}+\mathbf{U})=\tr(\mathbf{U})-\frac{1}{2}\tr(\mathbf{U}^2) + \text{h.o.t.}$, and coupled with equation~\eqref{eq:lemma4-logdet}, it yields 
\begin{equation} \label{eq:lemma4-placeholder}
\log\det(\TTheta_j+\DDelta_j)=\log\det(\TTheta_j)+\tr(\mathbf{U})-\frac{1}{2}\tr(\mathbf{U}^2) + \text{h.o.t.}.
\end{equation}
Considering each term separately, we have 
\begin{align}\label{eq:tr}
    \tr(\mathbf{U}) &=\tr(\TTheta_j^{-1/2}\DDelta_j\TTheta_j^{-1/2})=\tr(\TTheta_j^{-1}\DDelta_j), \mbox{ since }\tr(\mathbf{ABC})=\tr(\mathbf{BCA}),\\
    \tr(\mathbf{U}^2)&=\tr(\TTheta_j^{-1/2}\DDelta_j\TTheta_j^{-1}\DDelta_j\TTheta_j^{-1/2}) \nonumber 
    =\tr(\TTheta_j^{-1}\DDelta_j\TTheta_j^{-1}\DDelta_j) \nonumber \\
    &=\text{vec}(\DDelta_j)^{\top}(\TTheta_j^{-1} \otimes\TTheta_j^{-1})\text{vec}(\DDelta_j),\label{eq:trsq}
\end{align}
where for the last equality we used the relationship $\tr(\mathbf{ABCD}) = \text{vec}(\mathbf{D}^{\top})^{\top}(\mathbf{C}^{\top} \otimes \mathbf{A})\text{vec}(\mathbf{B})$ \citep[Chapter 10.2]{abadir2005matrix} and the symmetry of $\DDelta_j$.
Substituting equations~\eqref{eq:tr} and \eqref{eq:trsq} into equation~\eqref{eq:lemma4-placeholder}, we have 
\begin{equation*}
    g(\tilde{\TTheta}_j) = g(\TTheta_j) + \tr(\TTheta_j^{-1}\DDelta_j) - \frac{1}{2} \text{vec}(\DDelta_j)^{\top} (\TTheta_j^{-1}\otimes\TTheta_j^{-1}) \text{vec}(\DDelta_j) + \text{h.o.t.},
\end{equation*}
where we have also used $g(\TTheta_j)=\log\det(\TTheta_j)$.
\end{proof}

\medskip

\noindent Lemma \ref{lemma4} refines the expansion in Lemma \ref{lemma3} to an integral remainder form to aid bounding. 
\begin{lemma} \label{lemma4}
    Let $g(\tilde{\TTheta}_j)=\log\det(\tilde{\TTheta}_{j})$ where $\tilde{\TTheta}_j=\TTheta_j+\DDelta_j$ and let $\TTheta_j$ be the precision matrix and $\boldsymbol{\Delta}_j$ be a symmetric matrix such that $\TTheta_j(v)=\TTheta_j+v\DDelta_j\succ 0$ for any $v \in [0,1]$. The Taylor series expansion of $g(\tilde{\TTheta}_j)$ in integral remainder form is given by 
\begin{equation*}
    g(\tilde{\TTheta}_j) = g(\TTheta_j) + \tr(\TTheta_j^{-1}\DDelta_j)-\text{vec}(\DDelta_j)^{\top} \biggl[\int_0^1(1-v)H(\TTheta_j, \DDelta_j, v) dv\biggr] \text{vec}(\DDelta_j),
\end{equation*}
where $H(\TTheta_j, \DDelta_j,v)= \big( (\TTheta_j+v\DDelta_j)^{-1}\big)^{\top} \otimes (\TTheta_j+v\DDelta_j)^{-1}$ and we use the usual notation, $\text{vec}(\DDelta_j)$ for the operator that stacks the columns of the matrix $\DDelta_j$; $\otimes$ for the Kronecker product, see e.g. \cite{abadir2005matrix}.
\end{lemma}
\medskip
\begin{proof}
Using the integral form of the remainder for the Taylor expansion of     a function $h(\cdot)$, we have 
\begin{equation}\label{eq:taylorint}
h(1)=h(0)+h'(0)+\int_0^1(1-v)h''(v) dv.
\end{equation}
Let $h(v):=g(\TTheta_j+v\DDelta_j)$, and we denote $\TTheta(v)=\TTheta_j+v\DDelta_j$ such that $\mathrm{det}(\TTheta(v)) \neq 0$ for $v \in [0,1]$.
Then $h(0)=\log\det(\TTheta_j)$, $h(1)=\log\det(\TTheta_j+\DDelta_j)$, and
\begin{equation*}
    h'(v) = \frac{d}{dv}g(\TTheta_j+v\Delta_j) 
    = \tr\left(\TTheta^{-1}(v)\frac{d}{dv}\TTheta(v)\right)
    = \tr\left(\TTheta^{-1}(v)\DDelta_j\right),
\end{equation*}
where the second equality follows from Jacobi's identity \citep{magnus_neudecker88} and $\frac{d}{dv}\TTheta(v)=\DDelta_j$. Hence we obtain $h'(0)=\tr(\TTheta_j^{-1}\DDelta_j)$. 

For the second derivative, $h''(v)=\frac{d}{dv}\tr\left(\TTheta^{-1}(v)\DDelta_j\right)$, thus $h''(v)=\tr\bigg(\frac{d}{dv}(\TTheta^{-1}(v))\DDelta_j\bigg)$ and we first find $\frac{d}{dv}\left(\TTheta^{-1}(v)\right)$. 
This follows from $\frac{d}{dv}\left(\TTheta(v)\TTheta^{-1}(v)\right)=\frac{d}{dv}\mathbf{I}_{P_{T}}=\mathbf 0_{P_T}$ which gives $$\frac{d}{dv}\left(\TTheta^{-1}(v)\right)=-\TTheta^{-1}(v)\left(\frac{d}{dv}\TTheta(v)\right)\TTheta^{-1}(v).$$
Hence, 
\begin{align*}
    h''(v) &= -\tr\left( \TTheta^{-1}(v)\bigg(\frac{d}{dv}\TTheta(v)\bigg)\TTheta^{-1}(v)\DDelta_j \right) \\
    & = -\text{vec}(\DDelta_j^{\top})^{\top} ((\TTheta^{-1}(v))^{\top}\otimes \TTheta^{-1}(v))\text{vec}(\DDelta_j)  \\
    & = -\text{vec}(\DDelta_j)^{\top} H(\TTheta_j, \DDelta_j,v)\text{vec}(\DDelta_j),
\end{align*}
where we have used $\frac{d}{dv}\TTheta(v)=\DDelta_j$ in the second equality and $\tr(\mathbf{ABCD})=\text{vec}(\mathbf{D}^{\top})^{\top}(\mathbf{C}^{\top} \otimes \mathbf{A})\text{vec}(\mathbf{B})$. We are now in a position to substitute each component into equation~\eqref{eq:taylorint} and obtain
\begin{align*}
    g(\tilde{\TTheta}_j) &= g(\TTheta_j)+\tr(\TTheta_j^{-1}\DDelta_j) - \int_0^1(1-v)\text{vec}(\DDelta_j)^{\top}H(\TTheta_j,\DDelta_j, v)\text{vec}(\DDelta_j) dv.
\end{align*}

\end{proof}

\subsubsection{Proof of Theorem~\ref{thm1}}
We are now in a position to prove Theorem~\ref{thm1}. Combining Lemmas \ref{lemma2}-\ref{lemma4} allows us to control the change in the penalised objective in a neighbourhood of the true precision matrix and hence establish the required the high probability Frobenius-norm bound.

\begin{proof}
Recall the objective function in the optimisation problem \eqref{eq:S-inverse_optimisation} is
\begin{equation*}
    \mathcal{L}^*(\TTheta_j) = -T[\log\det(\TTheta_j)-\tr(\hat{\hat{\SS}}_j\TTheta_j)] + \lambda_j \| \TTheta_j \|_{1,\text{off}}
\end{equation*}
for all $j$, where we write $\TTheta_j$ to denote $\TTheta_j^{(X)}$ for notational simplicity and $\hat{\hat{\SS}}_j=\frac{1}{R}\sum_{r=1}^R\bar{\mathbf{I}}_j^{(Y^{(r){}})}$. Recall that $\| \cdot \|_{1,\text{off}}$ denotes the $\ell_{1}$-norm of a matrix excluding the diagonal elements.
Dropping the dependence on the scale $j$, we equivalently define \begin{equation*}
    \tilde{\mathcal{L}}(\TTheta) = \frac{\mathcal{L}^*(\TTheta)}{T}=-\log\det(\TTheta) + \tr(\hat{\hat{\SS}}\TTheta) + \lambda' \| \TTheta \|_{1,\text{off}}
\end{equation*}
for any symmetric positive definite matrix $\TTheta$ where $\lambda'=\lambda/T$. Set $\DDelta=\tilde{\TTheta}-\TTheta$, for some symmetric positive-definite matrix $\tilde\TTheta$, and define $\mathcal{Q}(\DDelta):=\tilde{\mathcal{L}}(\TTheta+\DDelta)-\tilde{\mathcal{L}}(\TTheta).$
The estimate $\hat\TTheta$ minimises $\mathcal{L}^*(\TTheta)$, or equivalently $\tilde{\mathcal{L}}(\TTheta)$, and therefore $\hat\DDelta = \hat\TTheta-\TTheta$ minimises $\mathcal{Q}(\DDelta)$.

Following the argument in \cite{tugnait2022sparse}, if we can show that with high probability $$\inf_{\{\DDelta: \|\DDelta\|_F=\tilde{C}r_T\}}\mathcal{Q}(\DDelta)>0,$$ where $r_T:=r_{j,T}=P_Ta_T$ with $a_T:=a_{j,T}$ for fixed $j$ and $\tilde{C}>\frac{1+C_2}{C_1}$ where $C_1$ and $C_2$ are as defined in the statement of the theorem, then the minimiser $\hat\DDelta=\hat\TTheta-\TTheta$ is such that $\|\hat\DDelta\|_F \leq \tilde{C}r_T$. It would therefore follow that $$\|\hat\TTheta-\TTheta\|_F \leq \tilde{C}r_T.$$ 

We derive this argument from the proof of Theorem 1 of \cite{Rothman2008thm1}. We now express $\mathcal{Q}(\DDelta)$ in terms of $\TTheta$ and $\DDelta$, as 
\begin{equation*}
    \mathcal{Q}(\DDelta) = \tr(\hat{\hat{\SS}}\DDelta)-[\log\det(\TTheta+\DDelta)-\log\det(\TTheta)] + \lambda'[ \|\TTheta + \DDelta \|_{1,\text{off}}-\|\TTheta\|_{1,\text{off}} ].
\end{equation*}

Let us split the spectral estimator as $\hat{\hat{\SS}}= \SS + \bigg(\hat{\hat{\SS}}-\SS\Big)$ which gives $\tr(\hat{\hat{\SS}}\DDelta)=\tr(\SS\DDelta)+\tr\Big(\big(\hat{\hat\SS}-\SS\big) \DDelta\Big)$ and leads to the decomposition $\mathcal{Q}(\DDelta) = A_1+A_2+A_3$, where
\begin{eqnarray*}
    A_1 &:=& \tr(\SS\Delta)-\log\det(\TTheta+\DDelta)+\log\det(\TTheta),\\
    A_2 &:=& \tr\Big(\Big(\hat{\hat{\SS}}-\SS\Big)\DDelta\Big),\\
    A_3 &:=& \lambda'(\|\TTheta + \DDelta\|_{1,\text{off}}-\|\TTheta\|_{1,\text{off}}).
\end{eqnarray*}

Our next aim is to lower bound each of the three terms $A_1, A_2$ and $A_3$ in order to show that $\mathcal{Q}(\DDelta)>0$ when $\DDelta$ is such that $\|\DDelta\|_F=\tilde{C}r_T$. 

By Lemma~\ref{lemma4}, $A_1$ can be equivalently written as $A_1 = \text{vec}(\DDelta)^{\top} K(\TTheta,\DDelta) \text{vec}(\DDelta)$, where $K(\TTheta, \DDelta)=\int_0^1(1-v)H(\TTheta, \DDelta,v) dv$, with $H(\TTheta, \DDelta,v)$ defined as in Lemma~\ref{lemma4}.
Using the notation therein, recall that $\TTheta(v)=\TTheta+v\DDelta$ with $v \in [0,1]$ fixed.

Let $\phi_{\min}$ and $\phi_{\max}$ denote the minimal and maximal eigenvalues of $\TTheta$, which are positive and finite by Assumption~\ref{ass:eigenvalues}. Given $\eta = \|\DDelta\|_2$, Weyl's inequality \citep{Horn_Johnson_1985} implies
\begin{equation*}
    \phi_{\max}(\TTheta(v))\leq \phi_{\max}+v\|\DDelta\|_2 \leq \phi_{\max}+\eta.
\end{equation*}
Because $\TTheta(v) = (1-v)\TTheta + v\hat{\TTheta}$ is a convex combination of positive definite matrices, $\TTheta(v)$ is positive definite for all $v \in [0,1]$ and, consequently, $\TTheta^{-1}(v)$ is well-defined for all $v \in [0,1]$. Since $H(\TTheta, \DDelta,v)=\TTheta^{-1}(v) \otimes \TTheta^{-1}(v)$, then its eigenvalues are the pairwise products of the eigenvalues of $\TTheta^{-1}(v)$. Altogether,
\begin{equation*}
    \phi_{\min}(H(\TTheta, \DDelta, v))=\phi_{\min}(\TTheta^{-1}(v))^2=\phi_{\max}(\TTheta(v))^{-2}\geq  (\phi_{\max}+\eta)^{-2}.
\end{equation*}
Using the property $\mathbf{x}^{\top}\mathbf{A}\mathbf{x}\geq \phi_{\min}(\mathbf{A})\|\mathbf{x}\|_2^2$ for any positive definite matrix $\mathbf{A}$, we obtain
\begin{align}\label{eq:A1}
    A_1 &=\int_0^1(1-v)\text{vec}(\DDelta)^{\top}(H(\TTheta, \DDelta, v))\text{vec}(\DDelta) dv \nonumber \\
    &\geq\|\text{vec}(\DDelta)\|^2_2 \bigg(\int_0^1(1-v)\phi_{\min}(H(\TTheta, \DDelta, v)) dv\bigg) \nonumber\\
   & \geq \|\DDelta\|_F^2(\phi_{\max}+\eta)^{-2}\int_0^1(1-v)dv \nonumber\\
   &=C_1\|\DDelta\|_F^2, \, \text{ with } C_1=\frac{(\phi_{\max}+\eta)^{-2}}{2}.
\end{align}

We next consider the term $A_2$, which can be written equivalently as $A_2=\sum_{p,q}(\hat{\hat{\SS}}-\SS)_{p,q}\DDelta_{q,p}$. Lemma \ref{lemma2} gives $$\Big\| \hat{\hat{\SS}}-\SS\Big\|_\infty =\max_{p,q}\left|\left(\hat{\hat{S}}-S\right)_{p,q}\right|\leq a_{T}$$ with probability at least $1-P_T^{2-\nu}$. 

Therefore, 
\begin{align*}
    |A_2|\leq \Big\| \hat{\hat{\SS}}-\SS\Big\|_\infty\sum_{p,q}|\Delta_{q,p}|\leq a_{T}\|\DDelta\|_1.
\end{align*}

By the Cauchy-Schwarz inequality, 
\begin{equation*}
    \|\DDelta\|_1 = \sum_{p,q} |\Delta_{p,q}| \leq \sqrt{P_T^2} \bigg(\sum_{p,q}\Delta_{p,q}^2\bigg)^{1/2} = P_T \|\DDelta\|_F.
\end{equation*}

Hence, $|A_2|\leq P_Ta_{T}\|\DDelta\|_F$ and we obtain the lower bound, 
\begin{equation}\label{eq:A2}
    A_2 \geq -P_{T}a_T\|\DDelta\|_F=-r_T\|\DDelta\|_F.
\end{equation}

Finally, we consider $A_3$. Using the edge set notation $\mathcal{E}$ and its complement $\mathcal{E}^c$ from Definition \ref{def1}, we split the matrix $\DDelta$ into matrices $\DDelta_{\mathcal{E}}$ and $\DDelta_{\mathcal{E}^c}$, where  $\left(\DDelta_{\mathcal{E}}\right)_{p,q}=\left(\DDelta\right)_{p,q}$ if $(p,q) \in \mathcal{E}$ and $0$ otherwise, and similarly for the edge complement set. Since the penalty excludes the diagonal, $A_3$ can be written as

\begin{equation*}
     A_3 = \lambda'\sum_{p,q \in \mathcal{E}} \left[|(\TTheta)_{p,q} + (\DDelta)_{p,q} | - |(\TTheta)_{p,q} | \right]
    + \lambda' \sum_{p,q \in \mathcal{E}^c} \left[|(\TTheta)_{p,q} + (\DDelta)_{p,q} | - |(\TTheta)_{p,q} | \right].
\end{equation*}
Since $(\TTheta)_{p,q}=0$ if $(p,q) \in \mathcal{E}^c$, we have
\begin{align*}
   A_3 & = \lambda'\sum_{p,q \in \mathcal{E}} \left[ |(\TTheta)_{p,q} + (\DDelta)_{p,q} | - |(\TTheta)_{p,q} | \right] + \lambda'\| \DDelta_{\mathcal{E}^c}\|_{1, \text{off}}\\
    & = \lambda' \left( \|\TTheta +\DDelta_{\mathcal{E}} \|_{1,\text{off}} - \|\TTheta \|_{1,\text{off}} \right) + \lambda' \|\DDelta_{\mathcal{E}^c} \|_{1,\text{off}} \\
    & \geq -\lambda' \|\DDelta_{\mathcal{E}}\|_{1,\text{off}} + \lambda' \|\DDelta_{\mathcal{E}^c}\|_{1,\text{off}}, 
\end{align*}
where the third step is an application of the triangle inequality.
Hence, $A_3 \geq \lambda'(\| \DDelta_{\mathcal{E}^c} \|_{1,\text{off}} -\| \DDelta_{\mathcal{E}} \|_{1,\text{off}} ) \geq  - \lambda'\| \DDelta_{\mathcal{E}} \|_{1,\text{off}}$, and, similarly to $A_2$, we can write
\begin{equation}\label{eq:A3}
    A_3 \geq -\lambda'\sqrt{E}\|\DDelta\|_F,
\end{equation}
since $|\mathcal{E}|=E$ by Definition \ref{def1}.

Putting together all three lower bounds in equations~\eqref{eq:A1},~\eqref{eq:A2} and~\eqref{eq:A3}, we obtain
\begin{align*}
    \mathcal{Q}(\DDelta) &\geq C_1\|\DDelta\|_F^2-r_T\|\DDelta\|_F-\lambda'\sqrt{E}\|\DDelta\|_F \\
    &=r_T(C_1\tilde{C}^2r_T-\tilde{C}r_T-\lambda'\sqrt{E}\tilde{C}) \\
    &\geq r_T(C_1\tilde{C}^2r_T - \tilde{C}r_T - C_2a_T\sqrt{E}\tilde{C}) \\
    &=r_T\big(C_1\tilde{C}^2r_T-\tilde{C}r_T-C_2\frac{r_T}{P_T}\sqrt{E}\tilde{C}\big)
\end{align*}
for some $C_2>0$ such that $\lambda'\leq C_2a_T$. We recall that the first equality holds since $\DDelta$ is such that $\|\DDelta\|_F=\tilde{C}r_T$. 

Since $\mathcal{E}=\{(p,q):(\TTheta_{j})_{p,q}\neq 0,\, p\neq q\}$ there are at most $P_T(P_T-1)$ non-zero off-diagonal entries by Definition \ref{def1}, so that $E\leq P_T(P_T-1)$ and $\frac{\sqrt{E}}{P_T}\leq 1$. Hence we obtain, 

\begin{equation}\label{eq:Q-Delta}
    \mathcal{Q}(\DDelta) \geq r_T^2(C_1\tilde{C}^2-(1+C_2)\tilde{C}).
\end{equation}
For the event $\bigg\{\Big\|\hat{\hat\SS}-\SS\Big\|_\infty\leq a_T\bigg\}$, the right-hand side of equation~\eqref{eq:Q-Delta} is strictly positive whenever $\tilde{C}>\frac{1+C_2}{C_1}$. Therefore, by Lemma \ref{lemma2}, with probability at least $1-P_T^{2-\nu}$, no minimiser of $\tilde{\mathcal{L}}(\TTheta)$ (and therefore of $\mathcal{L}^*(\TTheta)$) can lie outside the Frobenius ball of radius $\tilde{C}r_T$ around $\TTheta$, which implies that $\Big\|\hat\TTheta-\TTheta\Big\|_F=\mathcal{O}_P(\tilde{C}r_T)$ with probability at least $1-P_T^{2-\nu}$. Since $r_T=o(1)$, it follows that  $\|\hat\TTheta-\TTheta\|_F=o_P(1)$ with probability at least $1-P_T^{2-\nu}$.

\end{proof}

\subsection{Proof of Theorems \ref{thm2} and \ref{thm3}}\label{app:proof-of-graph-recovery}

This appendix contains the proofs of Theorem~\ref{thm2} and \ref{thm3}. Both results build on the concentration bound for the spectral estimator established in Lemma \ref{lemma2}. The proof of Theorem~\ref{thm2} uses this bound together with Assumptions \ref{ass:eigenvalues}--\ref{ass:high-dim-regularity} and the primal-dual witness argument of \cite{Ravikumar2011consistency} to establish a high-probability element-wise norm bound of the penalised precision matrix estimator. Theorem~\ref{thm3} then combines this result with the minimum signal condition in Assumption~\ref{ass:theta-min} to establish exact recovery of the estimated graph support with high probability.

\subsubsection{Proof of Theorem~\ref{thm2}}\label{app:proof-thm2}

We first establish the high-probability element-wise bound using the high-probability event supplied by Lemma \ref{lemma2} and the primal-dual witness construction of \cite{Ravikumar2011consistency}.

\begin{proof}
For notational simplicity we write $\lambda_j'=\lambda_j/T$ and let $\hat{\hat{\SS}}_j$ denote the spectral estimator defined in equation~\eqref{eq:S-hat-hat}. We fix a scale $j$, and drop the explicit dependence on $j$ in what follows.

Define the event $\mathcal{A}_T:=\left\{\left\|\hat{\hat{\SS}}-\SS\right\|_\infty \leq a_{T}\right\}$. By Lemma \ref{lemma2}, under Assumptions \ref{ass:spectral-boundness}-\ref{ass:mvLSW-regularity}, $\mathbb{P}(\mathcal{A}_T)\geq 1-P_T^{2-\nu}$, for $\nu>2$. By the choice of regularisation parameter $\lambda'=\frac{8}{\alpha}a_T$, we have $a_T=\alpha\lambda'/8$. 

After dividing the graphical lasso objective of equation~\eqref{eq:S-inverse_optimisation} by $T$, the estimator can be written as, 
\begin{equation*}
    \hat{\TTheta}=\arg\min_{\TTheta \succ 0} \left\{-\log\det(\TTheta)+\text{tr}\left(\hat{\hat{\SS}}\TTheta\right)+\lambda'\|\TTheta\|_{1,\text{off}}\right\}.
\end{equation*}

Following the primal-dual witness argument of \cite{Ravikumar2011consistency}, define the corresponding restricted estimator,
\begin{equation*}
    \tilde{\TTheta}={\arg\min}_{\TTheta \succ 0, \TTheta_{\mathcal{S}^c}=0} \left\{-\log\det(\TTheta)+\text{tr}\left(\hat{\hat{\SS}}\TTheta\right)+\lambda'\|\TTheta\|_{1,\text{off}}\right\},
\end{equation*}
where $\mathcal{S}$ denotes the augmented support of $\TTheta$ (as in Assumption~\ref{ass:mutual-incoherence}). Thus the restricted problem differs from the original graphical lasso problem only by requiring entries outside the true augmented support to be zero. Define $\DDelta=\tilde{\TTheta}-\TTheta$, for some $\tilde{\TTheta}$. 

Following \citet[Lemma 6]{Ravikumar2011consistency}, define $\rho_T =2\kappa_{\Gamma,T}\left(\left\|\hat{\hat{\SS}}-\SS\right\|_\infty +\lambda'\right)$. On $\mathcal{A}_T$, we have $\rho_T \leq2\kappa_{\Gamma,T}(a_T +\frac{8}{\alpha}a_T)$ and hence 
\begin{equation}\label{eq:rT-1}
    \rho_T \leq 2\kappa_{\Gamma,T}\bigg(1+\frac{8}{\alpha}\bigg)a_T.
\end{equation}
Assumption~\ref{ass:high-dim-regularity} states that $\text{deg}_T\kappa_{S,T}\kappa_{\Gamma,T}(1+8/\alpha)a_T \rightarrow 0$ and $\text{deg}_T\kappa_{S,T}^3\kappa_{\Gamma,T}^2(1+8/\alpha)a_T \rightarrow 0$. Hence from equation~\eqref{eq:rT-1} we have that $3\text{deg}_T\kappa_{S,T}\rho_T \leq 6\text{deg}_T\kappa_{S,T}\kappa_{\Gamma,T}(1+8/\alpha)a_T \rightarrow 0$. Thus for sufficiently large $T$, it holds that $3\text{deg}_T\kappa_{S,T}\rho_T \leq 1$ and
\begin{equation*}
    \rho_T \leq \frac{1}{3\text{deg}_T\kappa_{S,T}}.
\end{equation*}
Similarly from equation~\eqref{eq:rT-1}, $3\text{deg}_T\kappa_{S,T}^3\kappa_{\Gamma,T}\rho_T \leq 6\text{deg}_T\kappa_{S,T}^3\kappa_{\Gamma,T}^2(1+8/\alpha)a_T \rightarrow 0$. Hence for sufficiently large $T$, it holds that $3\text{deg}_T\kappa_{S,T}^3\kappa_{\Gamma,T}\rho_T  \leq 1$ and hence
\begin{equation*}
    \rho_T \leq  \frac{1}{3\text{deg}_T\kappa_{S,T}^3\kappa_{\Gamma,T}}.
\end{equation*}

Combining the preceding bounds on $\rho_T$ gives 
\begin{equation}\label{eq:rT-min}
    \rho_T \leq \min\Bigg\{\frac{1}{3\text{deg}_T\kappa_{S,T}}, \frac{1}{3\text{deg}_T\kappa_{S,T}^3\kappa_{\Gamma,T}}\Bigg\}.
\end{equation}

Thus the conditions of \citet[Lemma 6]{Ravikumar2011consistency} are satisfied and therefore $\|\DDelta\|_\infty \leq \rho_T$ and consequently on $\mathcal{A}_T$, 
\begin{equation}\label{eq:Delta-1}
    \|\DDelta\|_\infty \leq \rho_T\leq2\kappa_{\Gamma,T}\Big(1+{8}/{\alpha}\Big)a_T,
\end{equation}
from equation~\eqref{eq:rT-1}. 

We next control the remainder term arising from the expansion of the inverse matrix as in \cite{Ravikumar2011consistency}. Let us define $\mathcal{R}(\DDelta)=(\TTheta+\DDelta)^{-1}-\TTheta^{-1}+\TTheta^{-1}\DDelta\TTheta^{-1}$. 
Equations \eqref{eq:rT-min} and \eqref{eq:Delta-1} ensure that $\|\DDelta\|_\infty \leq 1/3\text{deg}_T\kappa_{S,T}$. Hence by \citet[Lemma 5]{Ravikumar2011consistency}, $\|\mathcal{R}(\DDelta)\|_\infty \leq \frac{3}{2}\text{deg}_T\kappa_{S,T}^3\|\DDelta\|_\infty^2$, and applying equation~\eqref{eq:Delta-1} we obtain 
\begin{equation*}
    \|\mathcal{R}(\DDelta)\|_\infty \leq 6\text{deg}_T\kappa_{S,T}^3\kappa_{\Gamma,T}^2(1+8/\alpha)^2a_T^2.
\end{equation*}
Therefore for all sufficiently large $T$, 
\begin{equation*}
    \|\mathcal{R}(\DDelta)\|_\infty \leq a_T =\alpha\lambda'/8.
\end{equation*}
Furthermore, by definition of $\mathcal{A}_T$,
\begin{equation*}
    \left\|\hat{\hat{\SS}}-\SS\right\|_\infty \leq a_T =\alpha\lambda'/8.
\end{equation*}
Combining the bounds on $\mathcal{R}(\DDelta)$ and $\hat{\hat{\SS}}-\SS$, we obtain $\max\left\{\left\|\hat{\hat{\SS}}-\SS\right\|_\infty , \|\mathcal{R}(\DDelta)\|_\infty\right\} \leq \alpha\lambda'/8$. Under the mutual incoherence of Assumption~\ref{ass:mutual-incoherence}, \citet[Lemma 4]{Ravikumar2011consistency} establishes strict dual feasibility. Hence the restricted estimator $\tilde{\TTheta}$ satisfies the optimality conditions of the unrestricted graphical lasso problem and, by uniqueness of a solution to the graphical lasso, we obtain $\tilde{\TTheta}=\hat{\TTheta}$. It follows from equation~\eqref{eq:Delta-1} that for all sufficiently large $T$, on $\mathcal{A}_T$, 
\begin{equation}\label{eq:Theta-1}
    \Big\|\hat{\TTheta}-\TTheta\Big\|_\infty \leq 2\kappa_{\Gamma,T}(1+8/\alpha)a_T.
\end{equation}
By Assumption~\ref{ass:high-dim-regularity}, together with the uniform eigenvalue bounds of Assumption~\ref{ass:eigenvalues}, the right hand side of equation~\eqref{eq:Theta-1} converges to $0$. Hence for any $\epsilon>0$, there exists $T_\epsilon<\infty$ such that for all $T \geq T_\epsilon$, $2\kappa_{\Gamma,T}(1+8/\alpha)a_T \leq \epsilon$ and therefore $\mathbb{P}\Big(\Big\|\hat{\TTheta}-\TTheta\Big\|_\infty>\epsilon\Big)\leq \mathbb{P}(\mathcal{A}_T^c) \leq P_T^{2-\nu}$. Thus $\Big\|\hat{\TTheta}-\TTheta\Big\|_\infty =o_P(1)$ with probability at least $1-P_T^{2-\nu}$.

\end{proof}

\subsubsection{Proof of Theorem~\ref{thm3}}\label{app:proof-of-thm3}

We now use the element-wise bound obtained in the proof of Theorem~\ref{thm2} to  establish exact recovery of the graph support with high probability following the model selection consistency argument of \citet[Theorem 2]{Ravikumar2011consistency}. The primal-dual witness argument excludes false-positive edges, while Assumption~\ref{ass:theta-min} ensures that the estimation error is sufficiently small to prevent true non-zero entries from being estimated as zero. 

\begin{proof}
    For notational simplicity we drop the dependence on $j$. As in the proof of Theorem~\ref{thm2}, we define $\mathcal{A}_T:=\left\{\left\|\hat{\hat{\SS}}-\SS\right\|_\infty\leq a_T\right\}$. By Lemma \ref{lemma2}, $\mathbb{P}(\mathcal{A}_T\big)\geq 1-P_T^{2-\nu}$.

    The primal-dual argument used in the proof of Theorem~\ref{thm2} and \cite{Ravikumar2011consistency}, establishes that for all sufficiently large $T$, on $\mathcal{A}_T$, $\tilde{\TTheta}=\hat{\TTheta}$, where $\tilde{\TTheta}$ is the restricted estimator satisfying $\tilde{\TTheta}_{\mathcal{S}^c}=0$ and  $\mathcal{S}$ is the augmented support defined in Assumption~\ref{ass:mutual-incoherence}. It follows that $\hat{\TTheta}_{\mathcal{S}^c}=0$, meaning that all the true zeros are correctly identified by $\hat{\TTheta}$. Since $\mathcal{E}$ (as defined in Definition \ref{def1}) is the off-diagonal support of $\TTheta$, this implies that $\hat{\mathcal{E}}\subseteq \mathcal{E}$ on $\mathcal{A}_T$. Hence $\mathbb{P}\Big(\hat{\mathcal{E}}\subseteq \mathcal{E}\Big) \geq \mathbb{P}(\mathcal{A}_T)\geq 1-P_T^{2-\nu}$.

    If $\mathcal{E} = \emptyset$, then $\hat{\mathcal{E}}\subseteq \mathcal{E}$ immediately gives $\hat{\mathcal{E}}=\mathcal{E}=\emptyset$, and the result follows.

    It remains to show the reverse inclusion for $\mathcal{E} \neq \emptyset$. Let $\theta_{\min, T}=\min_{(p,q) \in \mathcal{E}}|\Theta_{p,q}|$ as defined in Assumption~\ref{ass:theta-min}. From the proof of Theorem~\ref{thm2} on $\mathcal{A}_T$, $\Big\|\hat{\TTheta}-\TTheta\Big\|_\infty \leq 2\kappa_{\Gamma,T}(1+8/\alpha)a_T$. By Assumption~\ref{ass:theta-min}, for all sufficiently large $T$, $2\kappa_{\Gamma,T}(1+8/\alpha)a_T<\theta_{\min,T}$. Therefore for any $(p,q) \in \mathcal{E}$, 
    $$\Big|\hat{\Theta}_{p,q}\Big|\geq |\Theta_{p,q}|-\Big|(\hat{\TTheta}-\TTheta)_{p,q}\Big| \geq \theta_{\min,T}-\Big\|\hat{\TTheta}-\TTheta\Big\|_\infty > \theta_{\min,T} -\theta_{\min,T}=0.$$ Thus every true edge remains non-zero in the estimated precision matrix, so $\mathcal{E}\subseteq \hat{\mathcal{E}}$. 

    Combining the two inclusions gives $\hat{\mathcal{E}}=\mathcal{E}$ on $\mathcal{A}_T$ for all sufficiently large $T$. Consequently, $\mathbb{P}\Big(\hat{\mathcal{E}}=\mathcal{E}\Big) \geq \mathbb{P}(\mathcal{A}_T)\geq 1-P_T^{2-\nu}$.
\end{proof}

\section{Simulation Study Results}
\begin{figure}[H]
\centering
\begin{subfigure}[t]{0.22\linewidth}
    \centering
    \includegraphics[width=\linewidth]{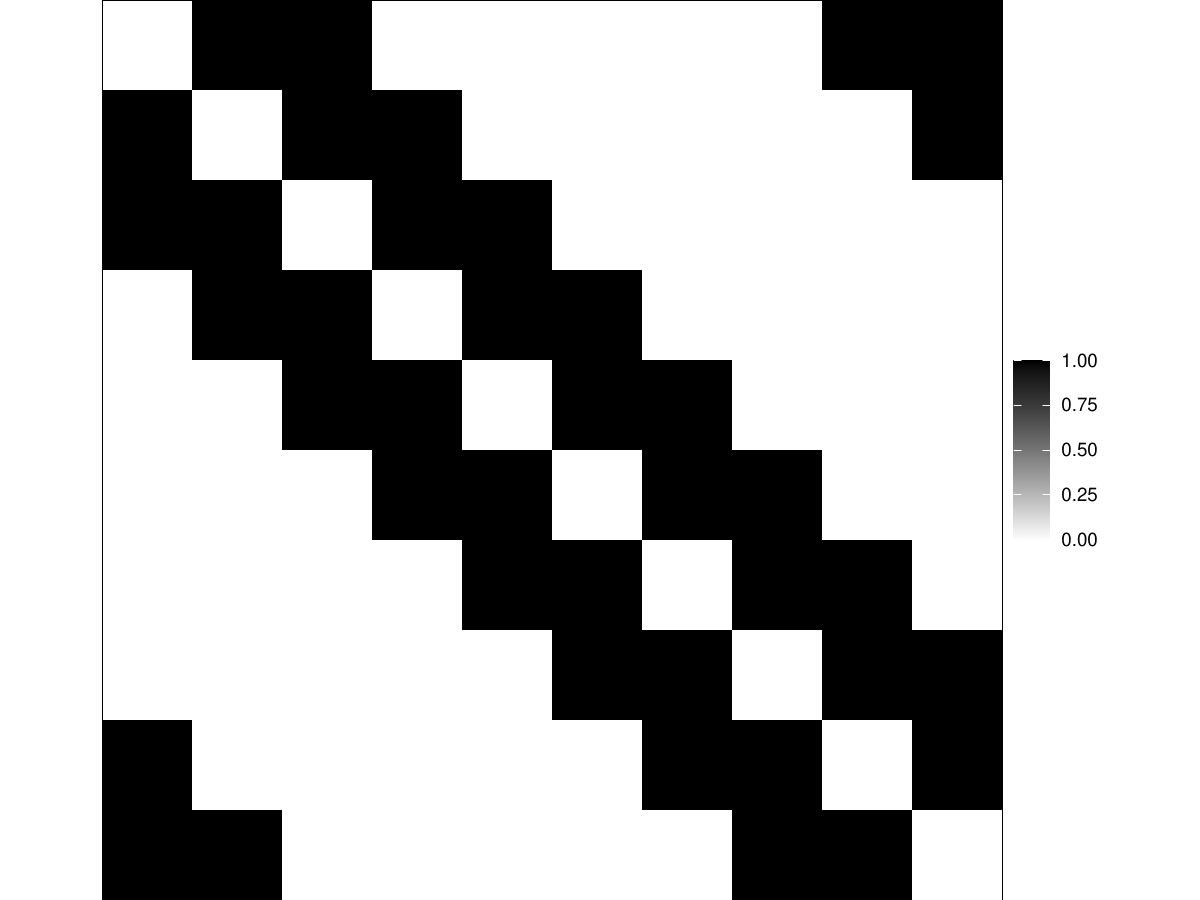}
    \caption{True ($\beta_{1,1}=0.85$)}
\end{subfigure}
\hspace{10mm}
\begin{subfigure}[t]{0.22\linewidth}
    \centering
    \includegraphics[width=\linewidth]{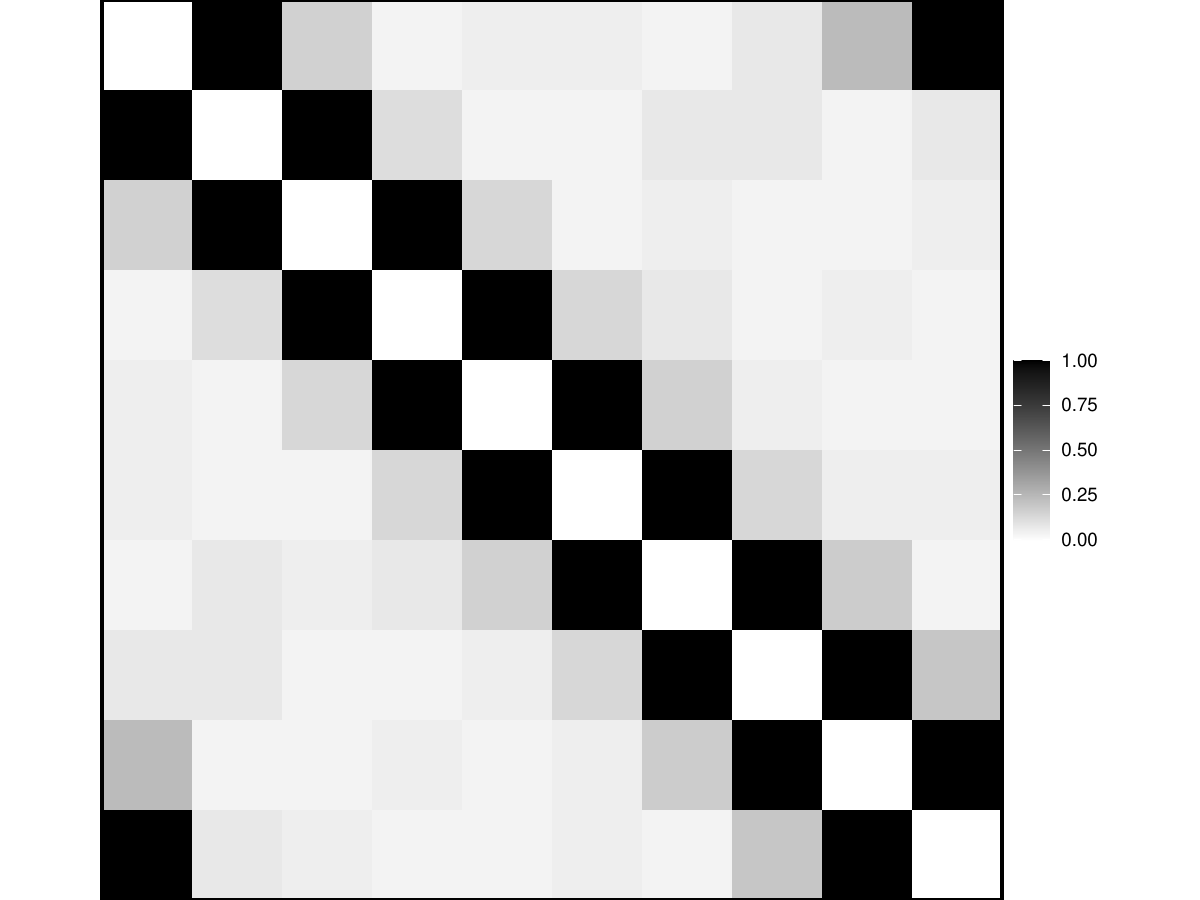}
    \caption{Fourier-$\ell_2$}
\end{subfigure}
\hspace{10mm}
\begin{subfigure}[t]{0.22\linewidth}
    \centering
    \includegraphics[width=\linewidth]{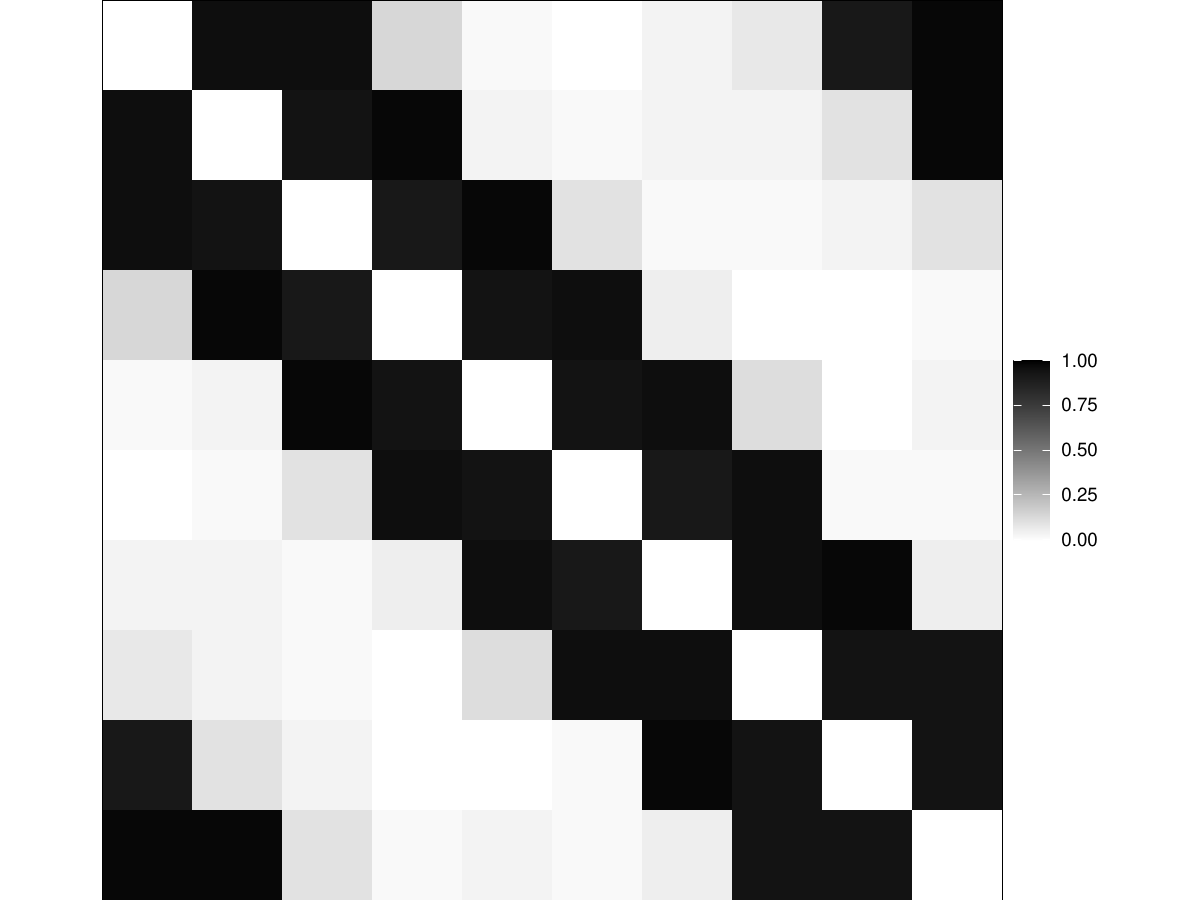}
    \caption{Wavelet}
\end{subfigure}

\vspace{4mm}

\begin{subfigure}[t]{0.22\linewidth}
    \centering
    \includegraphics[width=\linewidth]{Ring/10D_Ring_True.pdf}
    \caption{True ($\beta_{1,1}=0.55$)}
\end{subfigure}
\hspace{10mm}
\begin{subfigure}[t]{0.22\linewidth}
    \centering
    \includegraphics[width=\linewidth]{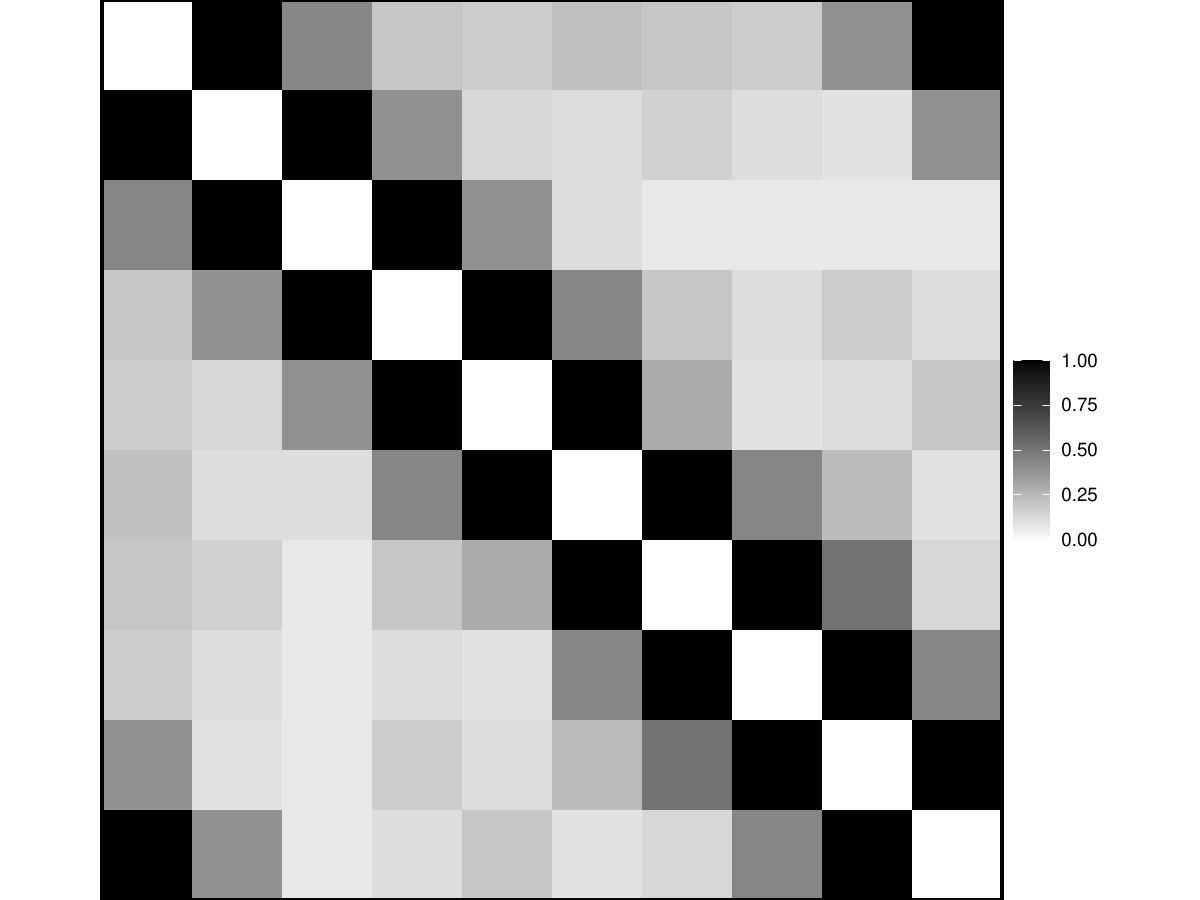}
    \caption{Fourier-$\ell_2$}
\end{subfigure}
\hspace{10mm}
\begin{subfigure}[t]{0.22\linewidth}
    \centering
    \includegraphics[width=\linewidth]{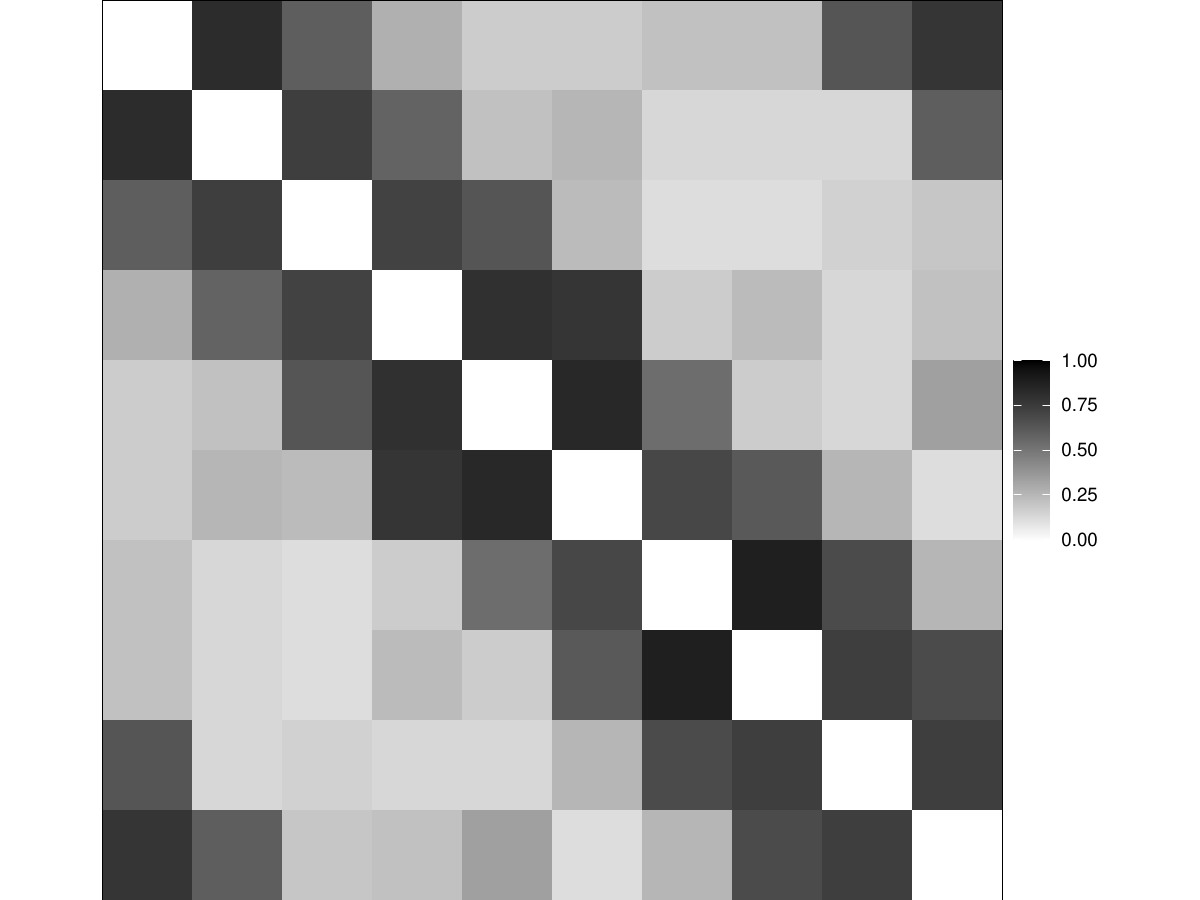}
    \caption{Wavelet}
\end{subfigure}

\caption{Adjacency matrices for 10-dimensional GNAR(1,[1]) models with $\beta_{1,1}=0.65$ (top) or with $\beta_{1,1}=0.35$ (bottom) over Ring graph (Truth, left column) versus averaged estimates using existing and proposed methods, middle and right columns respectively.}
\label{fig:10D-Ring-beta65-beta35}
\end{figure}

\begin{figure}[H]
    \centering

    \begin{subfigure}[t]{0.22\linewidth}
        \centering
        \includegraphics[width=\linewidth]{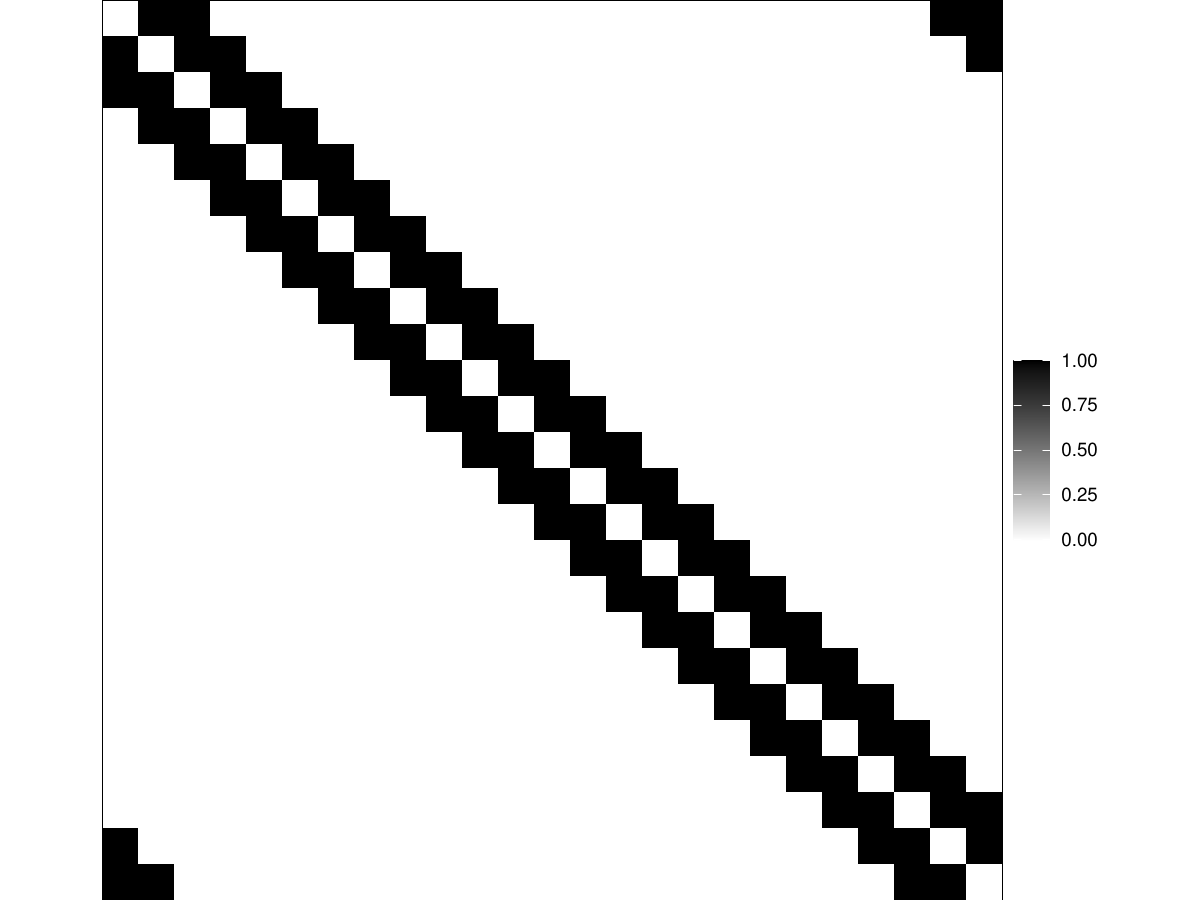}
        \caption{True ($\beta_{1,1}=0.85$)}
    \end{subfigure}
    \hspace{10mm}
    \begin{subfigure}[t]{0.22\linewidth}
        \centering
        \includegraphics[width=\linewidth]{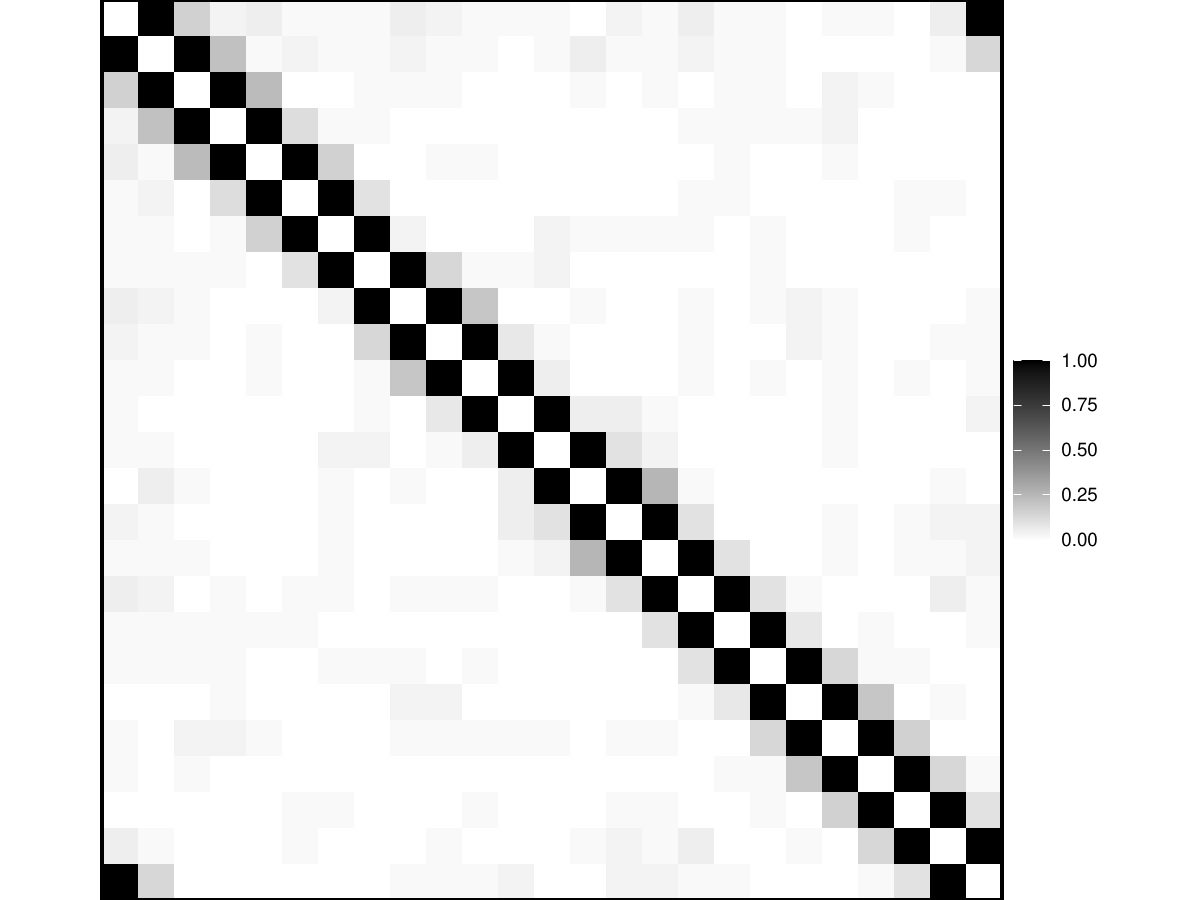}
        \caption{Fourier-$\ell_2$}
    \end{subfigure}
    \hspace{10mm}
    \begin{subfigure}[t]{0.22\linewidth}
        \centering
        \includegraphics[width=\linewidth]{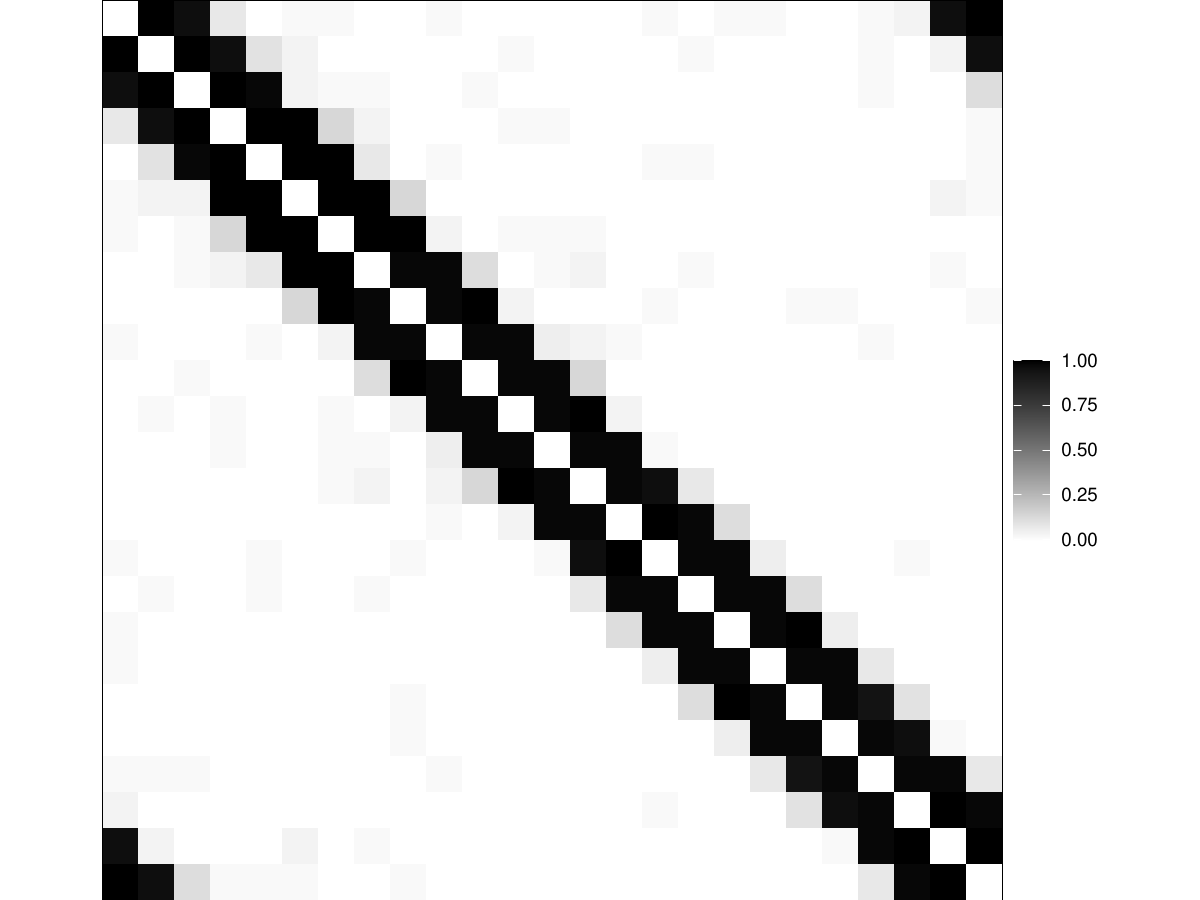}
        \caption{Wavelet}
    \end{subfigure}

    \vspace{4mm}

    \begin{subfigure}[t]{0.22\linewidth}
        \centering
        \includegraphics[width=\linewidth]{Ring/25D_Ring_True.pdf}
        \caption{True ($\beta_{1,1}=0.55$)}
    \end{subfigure}
    \hspace{10mm}
    \begin{subfigure}[t]{0.22\linewidth}
        \centering
        \includegraphics[width=\linewidth]{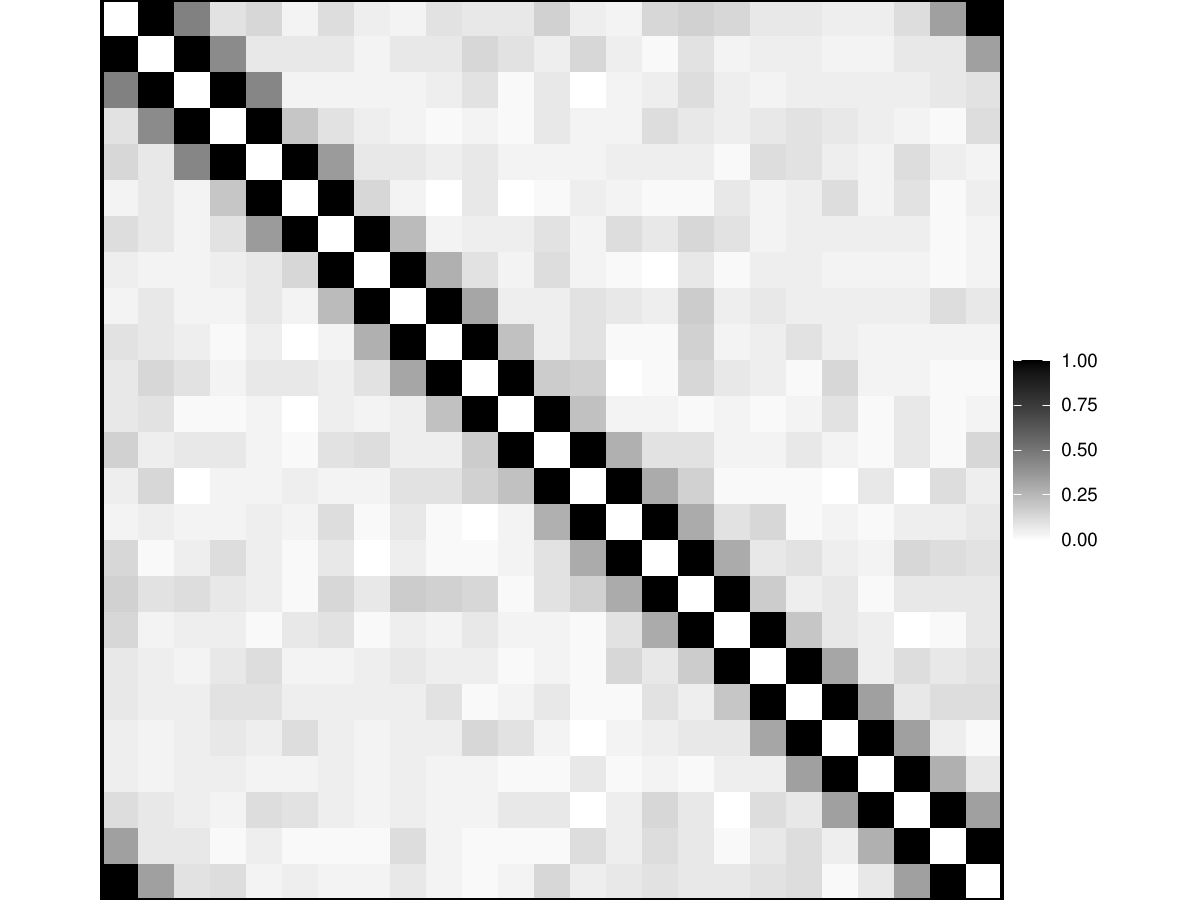}
        \caption{Fourier-$\ell_2$}
    \end{subfigure}
    \hspace{10mm}
    \begin{subfigure}[t]{0.22\linewidth}
        \centering
        \includegraphics[width=\linewidth]{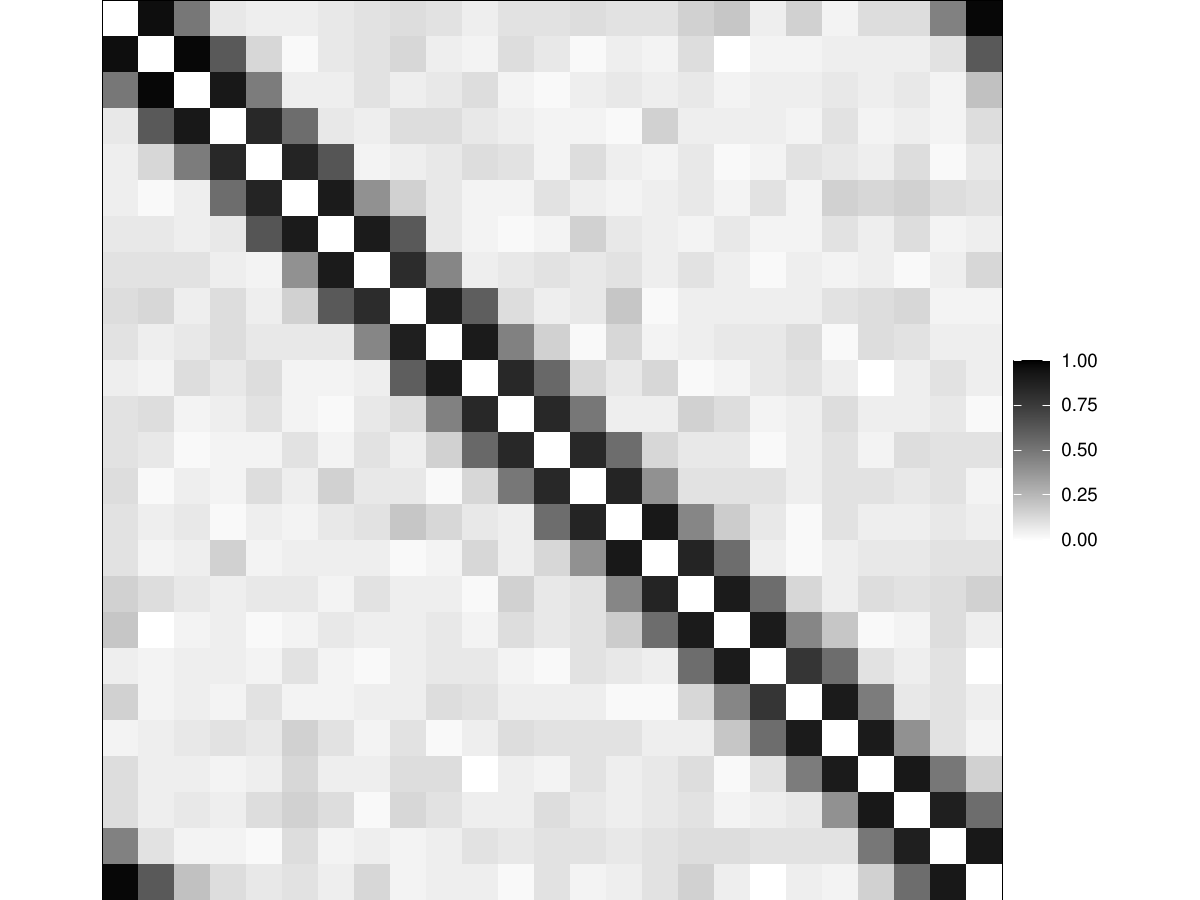}
        \caption{Wavelet}
    \end{subfigure}

    \caption{Adjacency matrices for 25-dimension GNAR(1,[1]) models with $\beta_{1,1}=0.65$ (top) or with $\beta_{1,1}=0.35$ (bottom) over Ring graph (Truth, left column) versus averaged estimates using existing and proposed methods, middle and right columns respectively.}
    \label{fig:25D-Ring-beta65-beta35}
\end{figure}

\begin{figure}[H]
\centering

\begin{subfigure}[t]{0.22\linewidth}
    \centering
    \includegraphics[width=\linewidth]{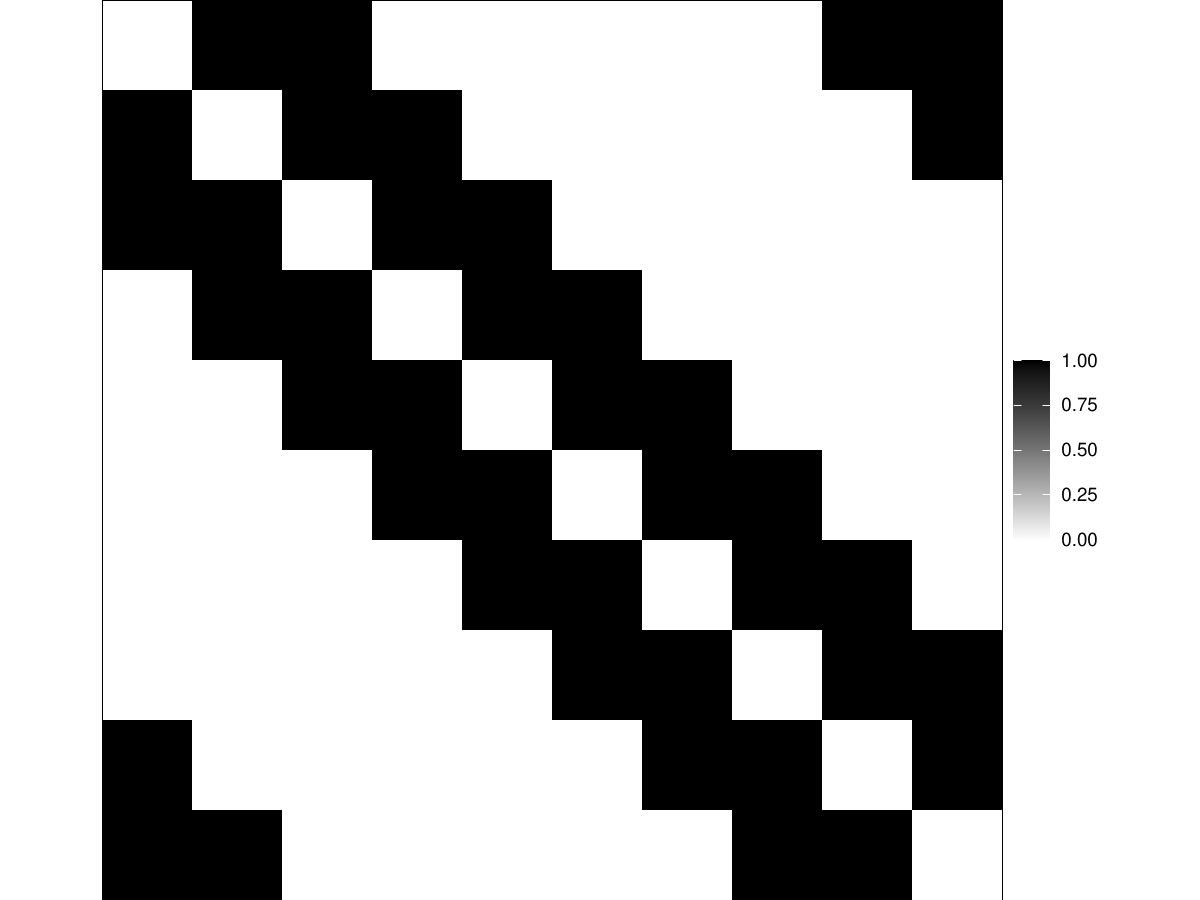}
    \caption{True ($\beta_i\in I_{1}$)}
\end{subfigure}
\hspace{10mm}
\begin{subfigure}[t]{0.22\linewidth}
    \centering
    \includegraphics[width=\linewidth]{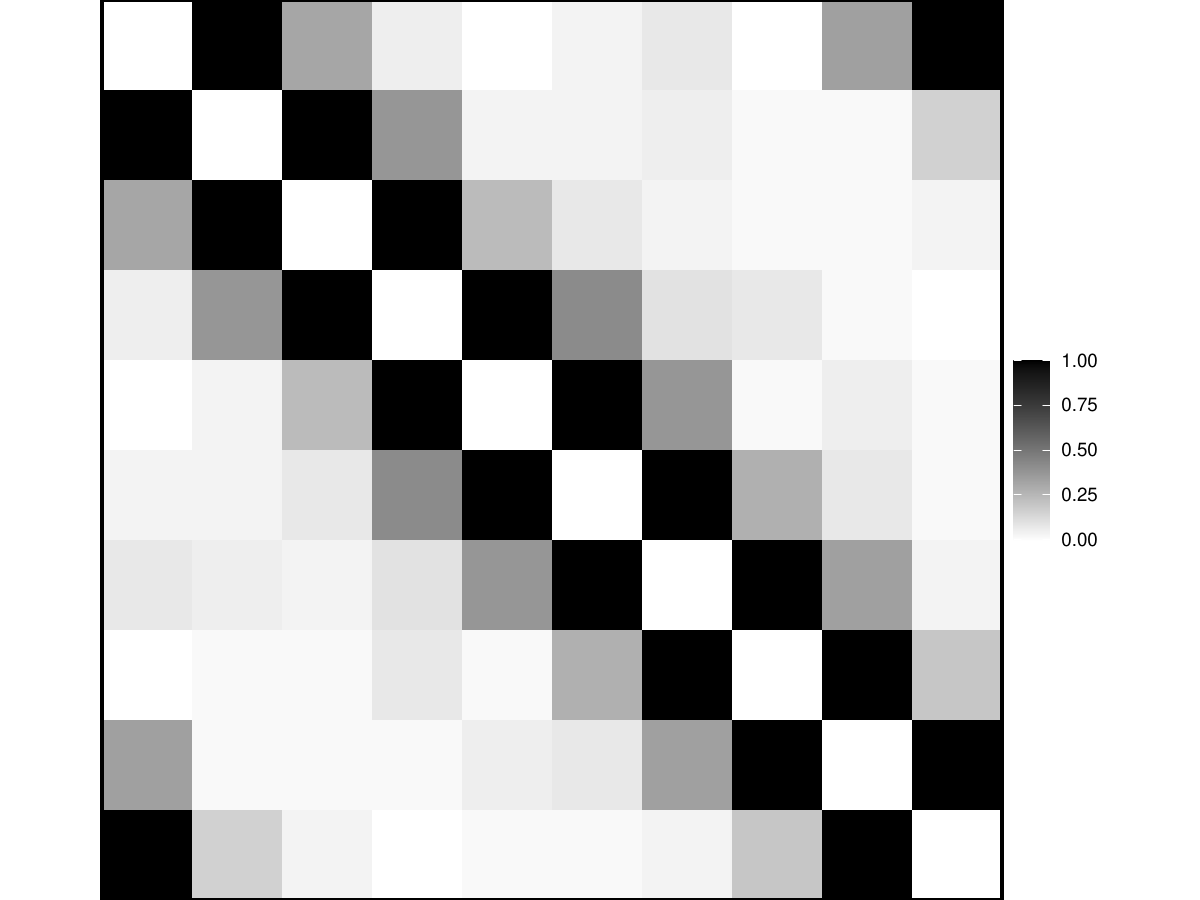}
    \caption{Fourier-$\ell_2$}
\end{subfigure}
\hspace{10mm}
\begin{subfigure}[t]{0.22\linewidth}
    \centering
    \includegraphics[width=\linewidth]{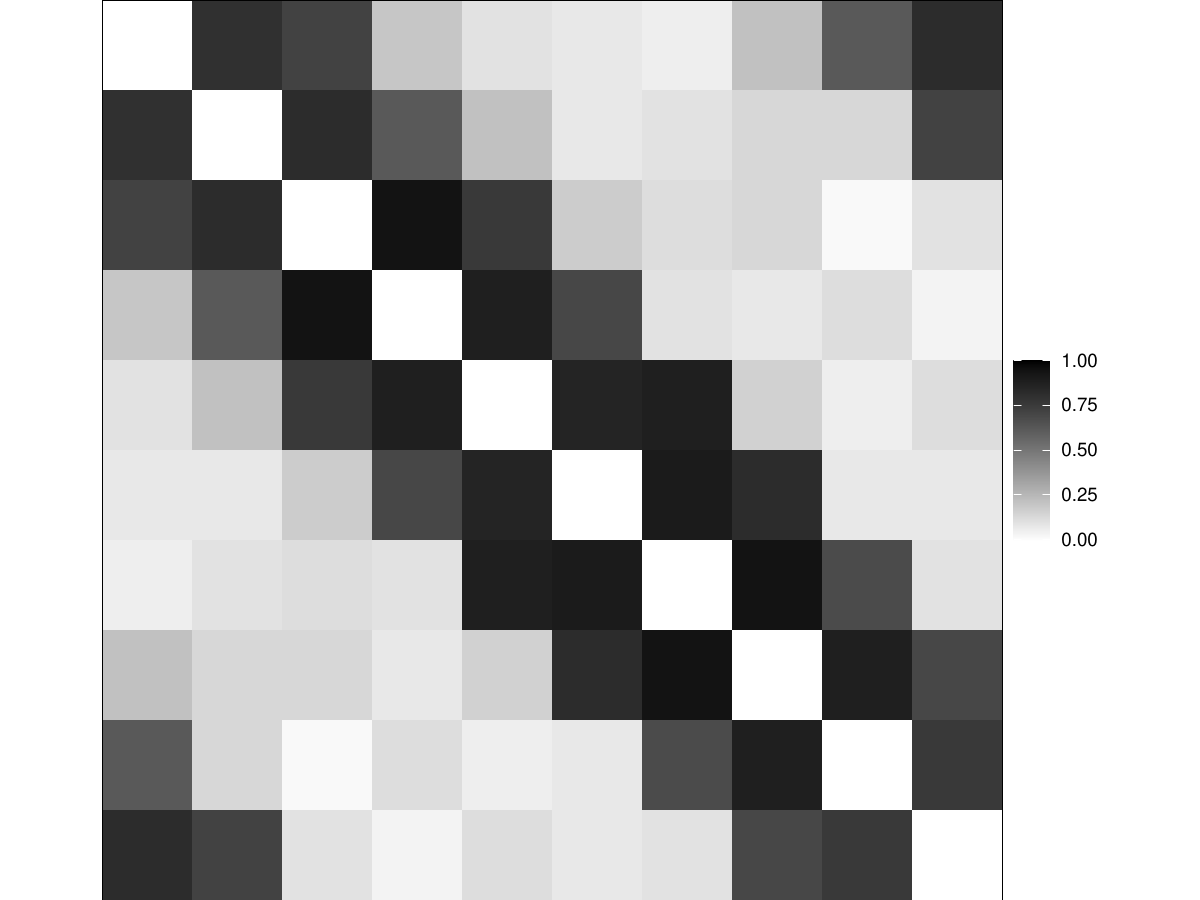}
    \caption{Wavelet}
\end{subfigure}

\vspace{4mm}

\begin{subfigure}[t]{0.22\linewidth}
    \centering
    \includegraphics[width=\linewidth]{VAR/VAR_05_True.pdf}
    \caption{True ($\beta_i\in I_{2}$)}
\end{subfigure}
\hspace{10mm}
\begin{subfigure}[t]{0.22\linewidth}
    \centering
    \includegraphics[width=\linewidth]{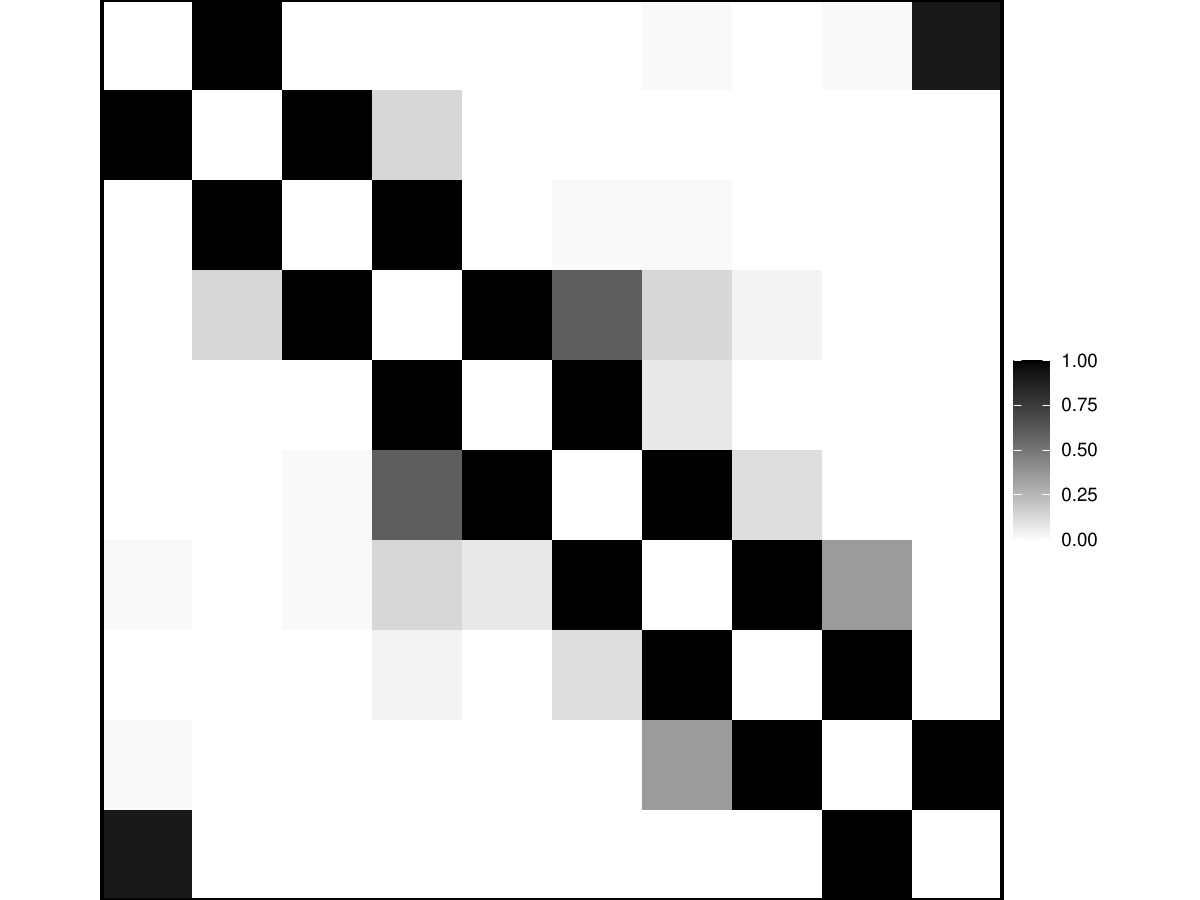}
    \caption{Fourier-$\ell_2$}
\end{subfigure}
\hspace{10mm}
\begin{subfigure}[t]{0.22\linewidth}
    \centering
    \includegraphics[width=\linewidth]{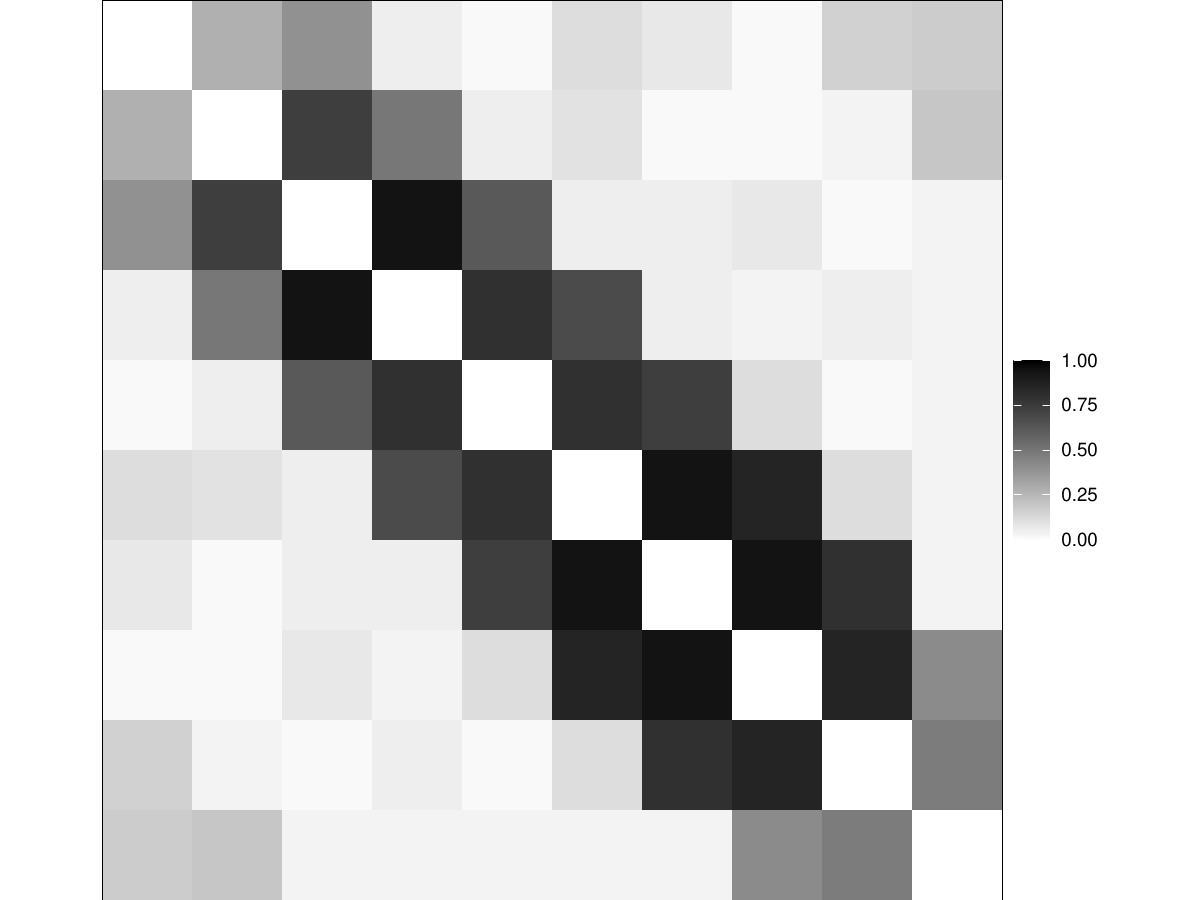}
    \caption{Wavelet}
\end{subfigure}

\caption{Adjacency matrices for 10-dimensional VAR(1) models with coefficient matrix $\mathbf{A}$ in \eqref{eq:VAR Model1} with $\beta_i\in I_{1}$ (top) or $\beta_i\in I_{2}$ (bottom), $i=1,\ldots,10$. Truth (left column) versus averaged estimates using existing and proposed methods (middle and right columns, respectively).}
\label{fig:VAR-I1-I2}
\end{figure}

\begin{figure}[H]
\centering
\begin{subfigure}[t]{0.22\linewidth}
    \includegraphics[width=\linewidth]{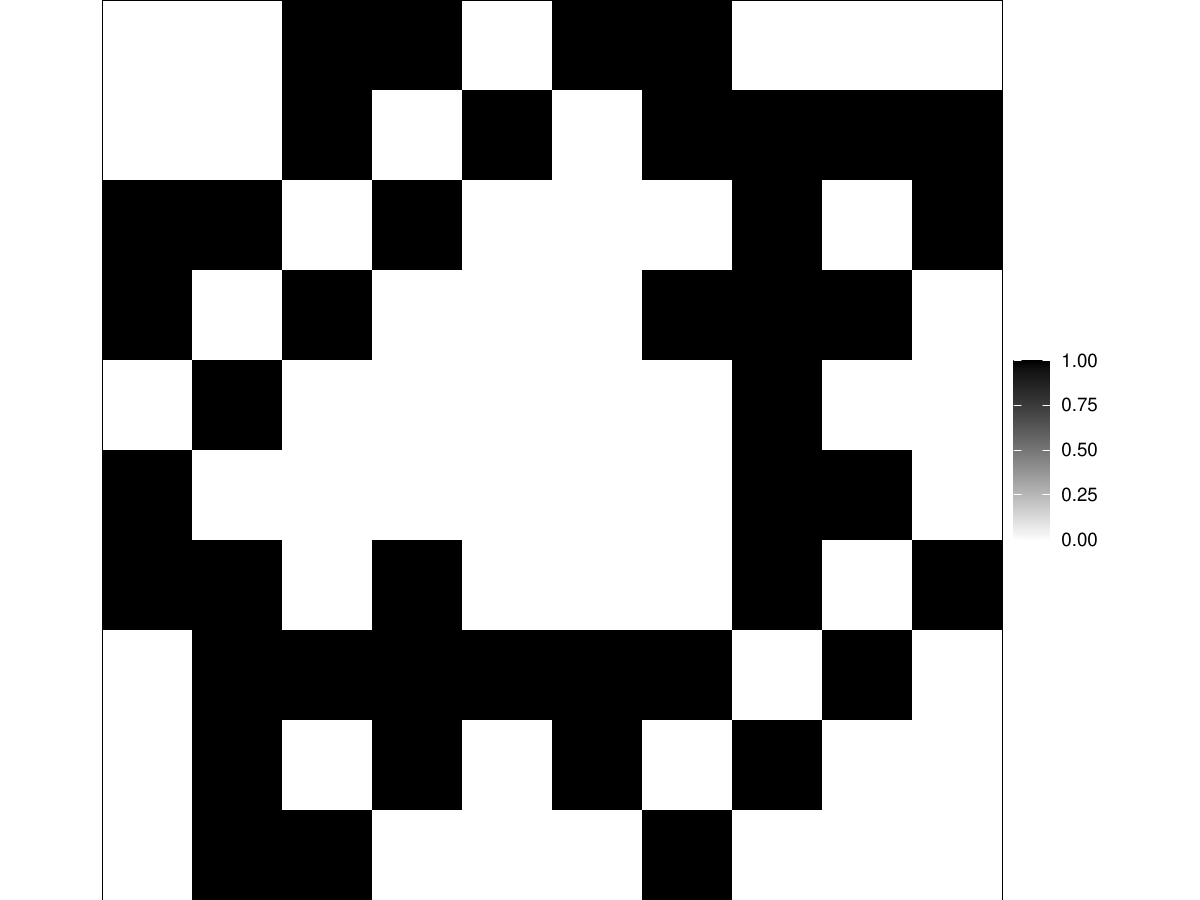}
    \caption{True}
\end{subfigure}
\hspace{10mm}
\begin{subfigure}[t]{0.22\linewidth}
    \includegraphics[width=\linewidth]{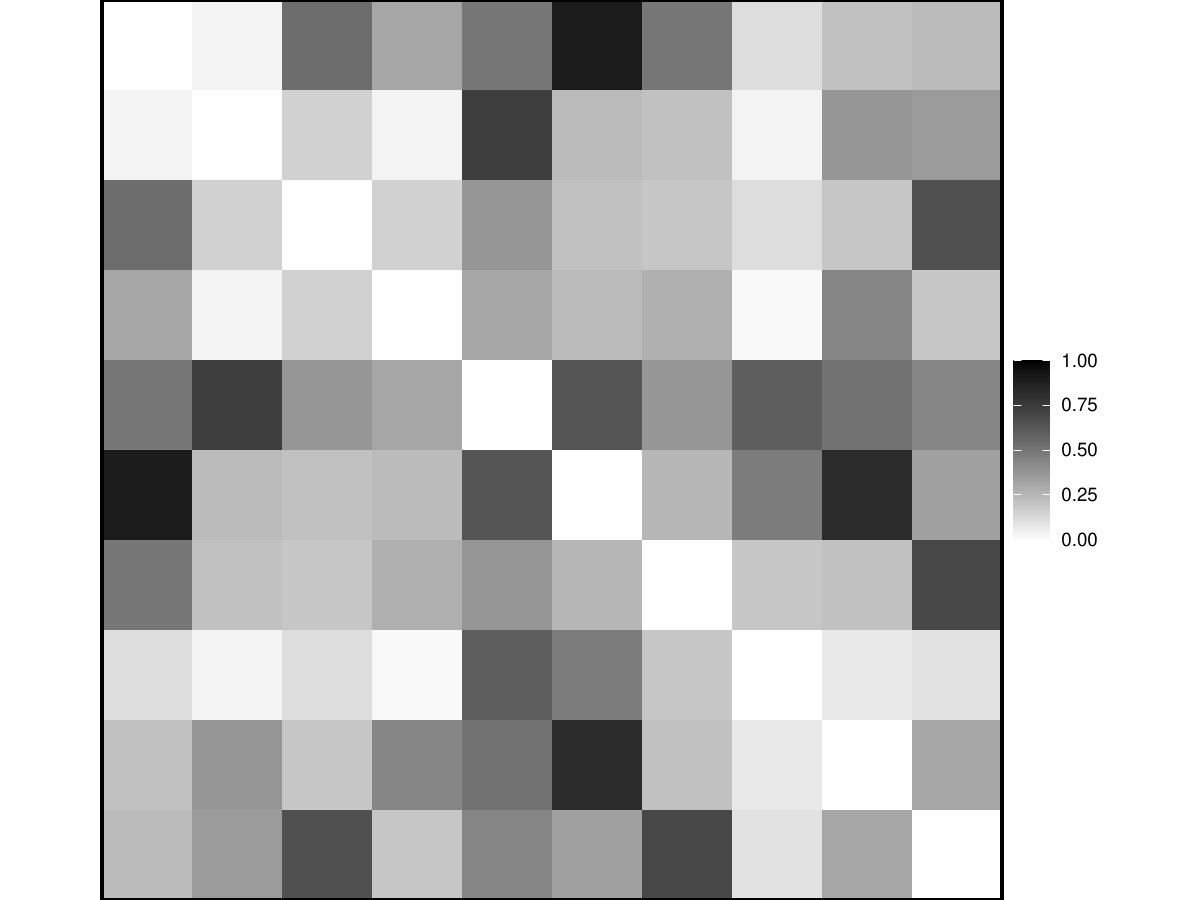}
    \caption{Fourier-$\ell_2$}
\end{subfigure}
\hspace{10mm}
\begin{subfigure}[t]{0.22\linewidth}
    \includegraphics[width=\linewidth]{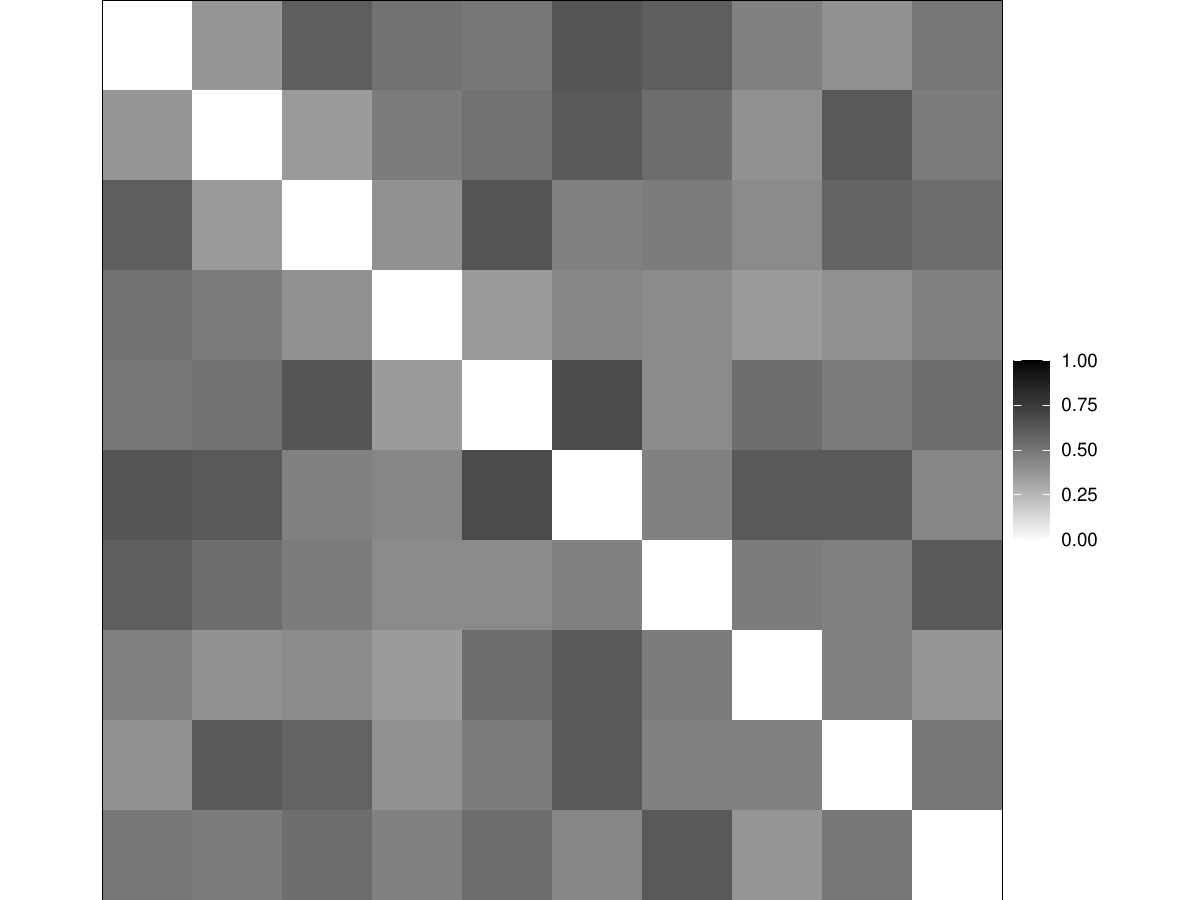}
    \caption{Wavelet}
\end{subfigure}
\caption{Adjacency matrices for 10-dimensional VAR(1) model with diagonal coefficient matrix $\mathbf{A}$ and noise precision matrix $\Sigma_{\vveps}^{-1}$. Truth (left) versus averaged estimates using existing and proposed methods (middle and right, respectively).}
\label{fig:Fiecas19}
\end{figure}

\begin{figure}[H]
\centering
\begin{subfigure}[t]{0.22\linewidth}
    \includegraphics[width=\linewidth]{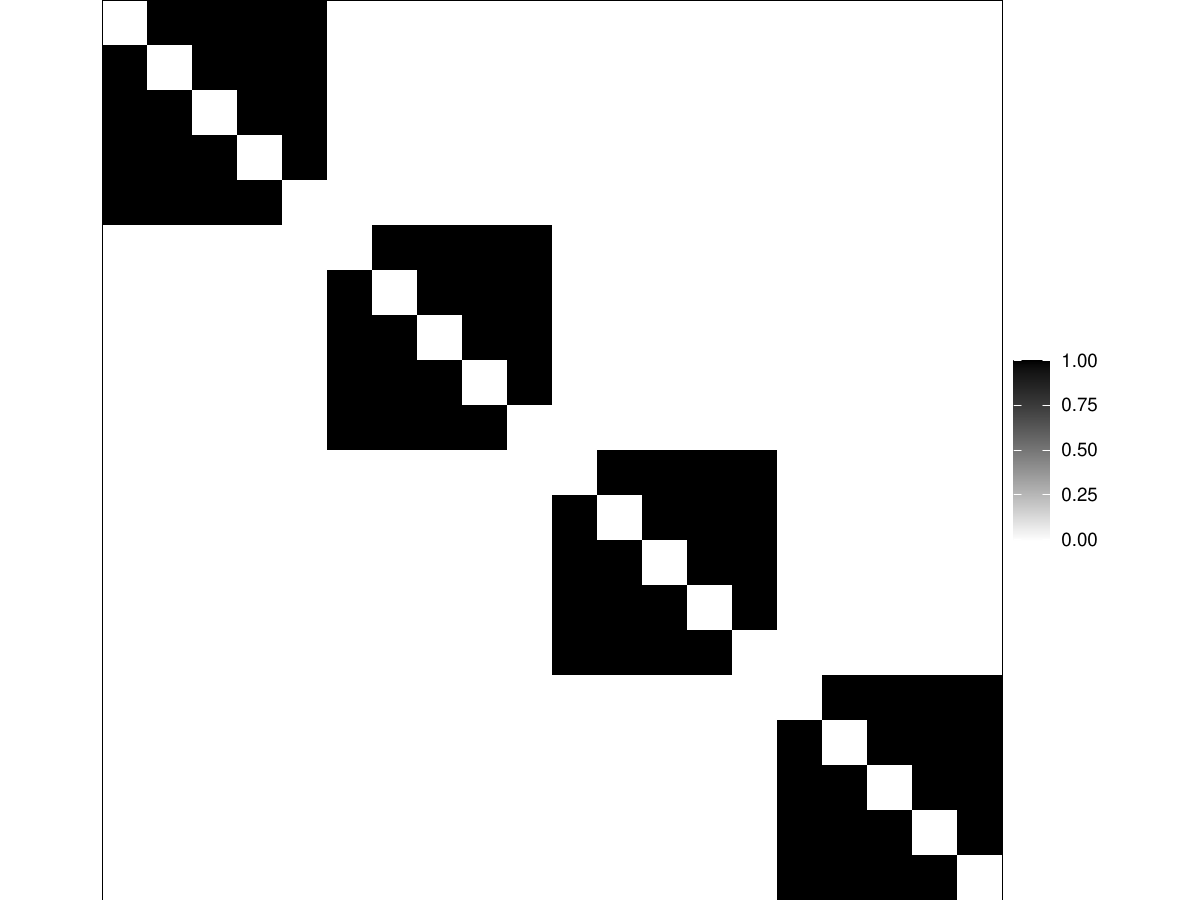}
    \caption{True}
\end{subfigure}
\hspace{10mm}
\begin{subfigure}[t]{0.22\linewidth}
    \includegraphics[width=\linewidth]{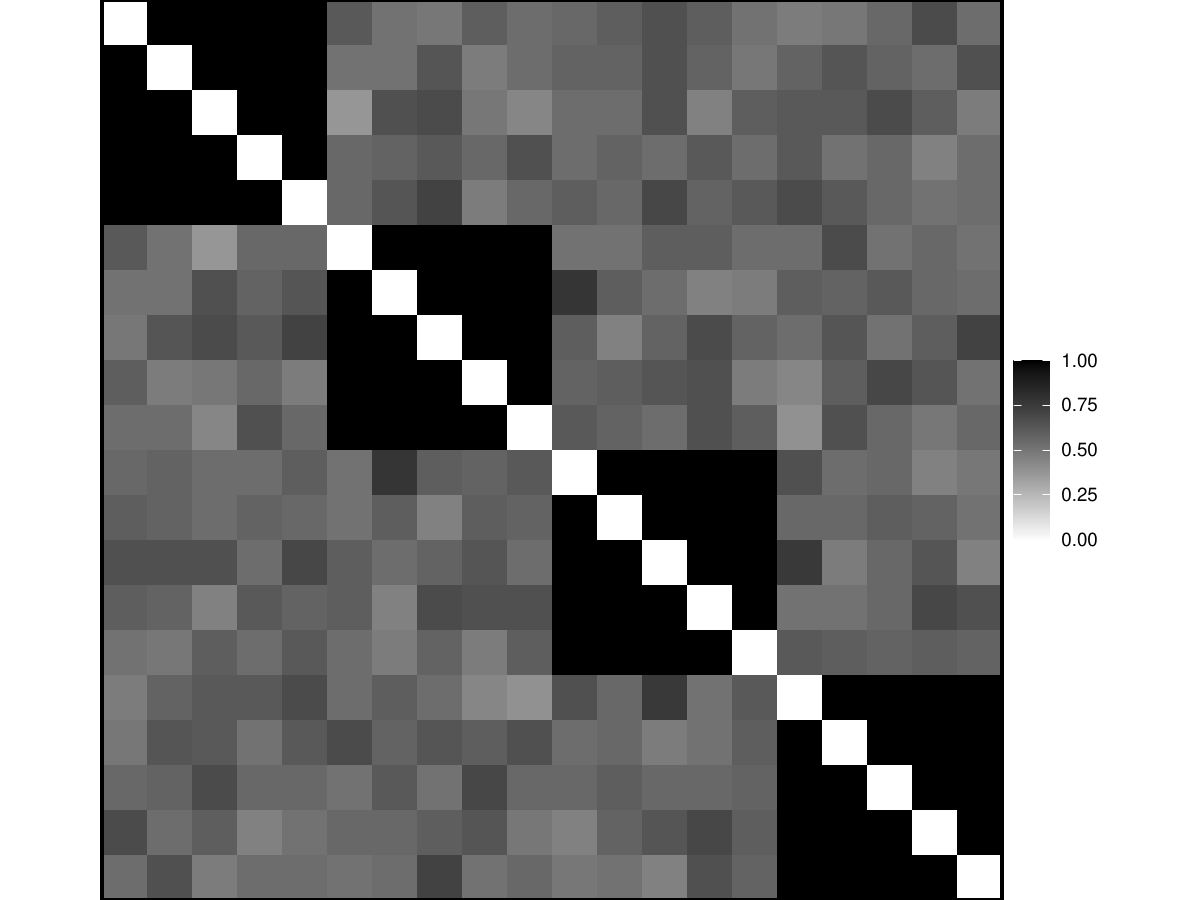}
    \caption{Fourier-$\ell_1$}
\end{subfigure}
\hspace{10mm}
\begin{subfigure}[t]{0.22\linewidth}
    \includegraphics[width=\linewidth]{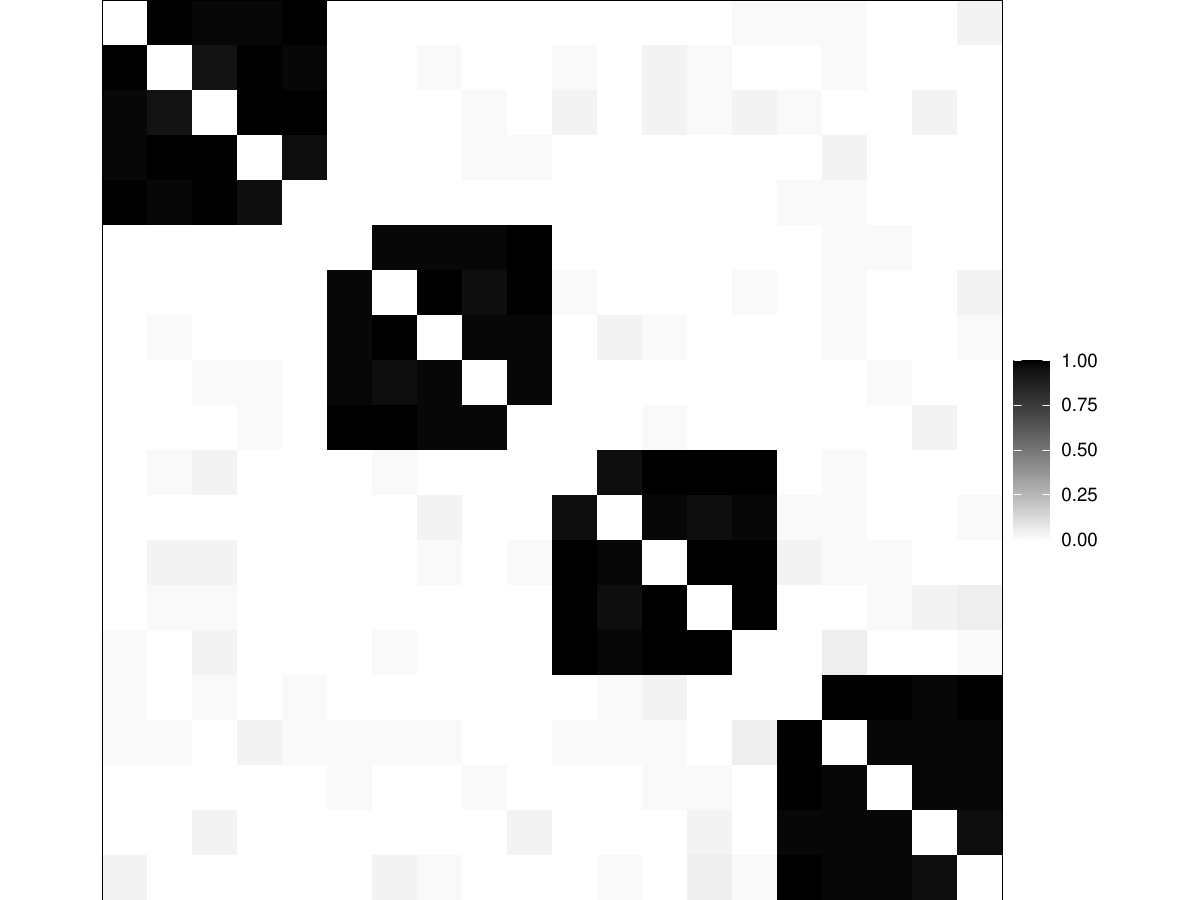}
    \caption{Wavelet}
\end{subfigure}

\caption{Adjacency matrices for 20-dimensional VARMA(2,2) model in~\eqref{eq:VARMA Model} with block diagonal structure. Truth (left) versus averaged estimates using existing and proposed methods (middle and right, respectively).}
\label{fig:Deb24}
\end{figure}

This appendix provies supplementary tables and figures for the simulation study discussed in Section~\ref{sec:Simulation Study}. Figures \ref{fig:10D-Ring-beta65-beta35}--\ref{fig:Deb24} compare the true adjacency alongside the wavelet estimate and the best-performing Fourier-based alternative (Fourier-$\ell_1$ or Fourier-$\ell_2$). The heatmaps indicate the frequency with which each edge is selected across simulations, with black denoting edges always selected and white denoting edges never selected on a grayscale.


\subsection{Sparsity of Simulated Models}\label{app:Sparsity-tbl}
For each model, estimated sparsity in Table~\ref{tbl:Sparsity} is computed as the ratio of connections in the \texttt{WavTSglasso} estimated adjacency matrix to the total number of possible connections, and the values are averaged across simulations.

\begin{table}[H]
\centering
\small
\begin{tabular}{ll|ll|c|c}
\toprule
\multicolumn{2}{c|}{\multirow{2}{*}{\textbf{Graph}}} & \multicolumn{2}{c|}{\multirow{2}{*}{\textbf{Model}}} & \textbf{True} & \textbf{Estimated} \\
& & & & \textbf{Sparsity,} $s$ & \textbf{Sparsity,} $\hat{s}$ \\
\midrule
    \multirow{2}{*}{10-dim ER} & $\rho=0.1$ & \multirow{2}{*}{GNAR(1,[1])} & \multirow{2}{*}{$\beta_{1,1}=0.85$} & 14.8\% & 13.6\% \\
    & $\rho=0.25$ & & & 50.6\% & 39.8\% \\
    \hline
    \multirow{2}{*}{10-dim ER} & $\rho=0.1$ & \multirow{2}{*}{GNAR(2,[1,1])} & \multirow{2}{*}{$\beta_{1,1} = \beta_{2,1} = 0.4$} & 14.8\% & 16.1\% \\
    & $\rho=0.25$ & & & 50.6\% & 40.0\% \\
    \hline
    \multicolumn{2}{c|}{25-dim ER} & GNAR(1,[2]) & $\rho=0.05$ & 21.3\% & 15.9\% \\
    \hline
    \multicolumn{2}{c|}{\multirow{2}{*}{10-dim Ring}} & \multirow{2}{*}{GNAR(1,[1])} & $\beta_{1,1}=0.85$ & 40.0\% & 40.7\% \\
    &&& $\beta_{1,1}=0.55$ & 40.0\% & 38.3\% \\
    \hline
    \multicolumn{2}{c|}{\multirow{2}{*}{25-dim Ring}} & \multirow{2}{*}{GNAR(1,[1])} & $\beta_{1,1}=0.85$ & 16.0\% & 16.8\% \\
    &&& $\beta_{1,1}=0.55$ & 16.0\% & 17.8\% \\
    \hline
    \multicolumn{2}{c|}{\multirow{2}{*}{10-dim}} & \multirow{2}{*}{VAR(1)} & $\beta_{i} \in I_{1}$ & 40.0\% & 37.3\% \\
    &&& $\beta_{i} \in I_{2}$ & 40.0\% & 27.4\% \\
    \hline
    \multicolumn{2}{c|}{10-dim} & VAR(1) & Random $\Sigma_{\vveps}^{-1}$ & 44.0\% & 44.6\% \\
    \hline
    \multicolumn{2}{c|}{20-dim} & VARMA(2,2) & Block-diagonal & 20.0\% & 20.3\% \\
    \bottomrule
\end{tabular}
\caption{True and estimated sparsity levels of the graphs for the simulated models in Section~\ref{sec:Simulation Study}. Estimates are obtained using \texttt{WavTSglasso}.}\label{tbl:Sparsity}
\end{table}

\subsection{Additional Simulation Study Results} \label{app:add_sims} 
\begin{figure}[!ht]
\centering
\begin{subfigure}[H]{0.31\linewidth}
    \includegraphics[width=\linewidth]{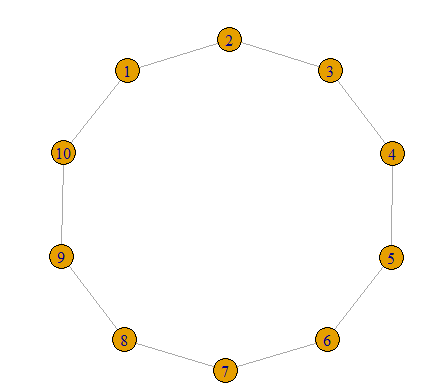}
    \caption{$P=10$.}
\end{subfigure}\bigskip
\hspace{20mm}
\begin{subfigure}[H]{0.31\linewidth}
    \includegraphics[width=\linewidth]{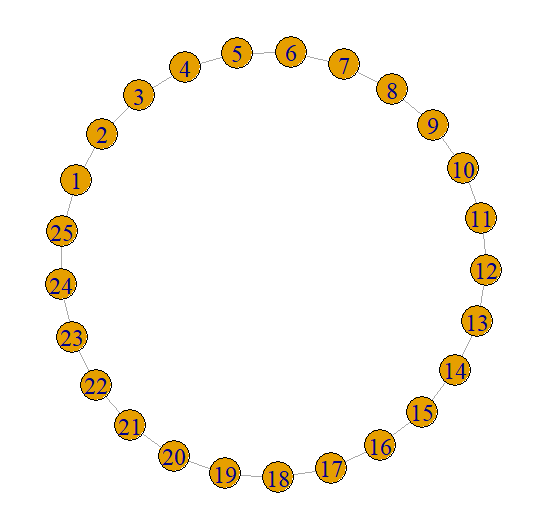}
    \caption{$P=25$.}
\end{subfigure}
\caption{10- and 25-dimensional Ring graphs.}
\label{fig:Ring Networks}
\end{figure}

Tables~\ref{tbl:ER Model Boot T=256}--\ref{tbl:10D Ring Results} report the the True Positive Rates (TPR), False Positive Rates (FPR), True Discovery Rates (TDR) and $F_1$-scores ($F_1$) for the various models presented in Section~\ref{sec:Simulation Study}. Values in bold identify the best method overall, considering these four aggregated measures. Figure~\ref{fig:Ring Networks} provides a visual illustration of the Ring graphs. 

\subsubsection*{Bootstrapping}

\begin{table}[H]
\centering
\setlength{\tabcolsep}{4pt}
\begin{tabular}{l*{68}{c}}
\toprule
& \multicolumn{4}{c}{Sparse ER} & \multicolumn{4}{c}{Dense ER} \\
\cmidrule(lr){2-5}\cmidrule(lr){6-9}
& TPR & FPR & TDR & $F1_1$ & TPR & FPR & TDR & $F_1$ \\
\midrule
$R=1$
& 0.8619 & 0.0602 & 0.8052 & 0.8326
& 0.5400 & 0.1696 & 0.7934 & 0.6426\\
\midrule

$R=10$
& 0.8910 & 0.0350 & 0.8785 & 0.8847
& 0.5756 & 0.1346 & 0.8490 & 0.6861\\
\midrule

$R=50$
& 0.8982 & 0.0360 & 0.8801 & 0.8891
& 0.5848 & 0.1336 & 0.8539 & 0.6942 \\

\bottomrule
\end{tabular}
\caption{10-dimensional GNAR(1,[1]) model over Erd\H{o}s-Rényi graphs ($\rho=0.1,\,0.25$) for $T=256$.}
\label{tbl:ER Model Boot T=256}
\end{table}

\vspace{4mm} 

\begin{table}[H]
\centering
\setlength{\tabcolsep}{4pt}
\centering
\begin{tabular}{l*{8}{c}}
\toprule
& \multicolumn{4}{c}{Sparse ER} & \multicolumn{4}{c}{Dense ER} \\
\cmidrule(lr){2-5}\cmidrule(lr){6-9}
& TPR & FPR & TDR & $F_1$ & TPR & FPR & TDR & $F_1$ \\
\midrule
$R=1$
& 0.9098 & 0.0115 & 0.9650 & 0.9366
& 0.6044 & 0.0936 & 0.9275 & 0.7319 \\
\midrule

$R=10$
& 0.9112 & 0.0117 & 0.9633 & 0.9365
& 0.6597 & 0.0972 & 0.9267 & 0.7707\\
\midrule

$R=50$
& 0.9064 & 0.0122 & 0.9642 & 0.9344
& 0.6664 & 0.1015 & 0.9236 & 0.7742 \\

\bottomrule
\end{tabular}
\caption{10-dimensional GNAR(1,[1]) model over Erd\H{o}s-Rényi graphs ($\rho=0.1,\,0.25$) for $T=2048$.}
\label{tbl:ER Model Boot T=2048}
\end{table}

\subsubsection*{GNAR Models on Erd\H{o}s-Rényi Graphs}

\begin{table}[H]

\centering
\begin{tabular}{l*{4}{c}}
\toprule
& TPR & FPR & TDR & $F_1$\\
\midrule
W - $\ell_1$ 
&{0.6269}& {0.0396} & {0.8221} & 0.7114 \\
\midrule

F - $\ell_1$ (glasso) 
& 0.7490 & 0.0429 & 0.8235 & 0.7845 \\
\midrule

F - $\ell_1$ (ADMM) 
& \textbf{0.7765} &\textbf{0.0299} &\textbf{0.8750} & \textbf{0.8228}\\
\midrule

F - $\ell_2$ (ADMM) 
&{0.7260} &{0.0099} &{0.9541}  & 0.8246\\
\bottomrule
\end{tabular}
\caption{25-dimensional GNAR(1,[2]) models over Erd\H{o}s-Rényi graphs for $T=1024$.}
\label{tbl:ER Model Results GNAR(1,[2])}
\end{table}

\subsubsection*{GNAR Models on Ring Graphs}
\begin{table}[H]
\begin{center}
\setlength{\tabcolsep}{4pt}
\begin{tabular}{l*{8}{c}}
\toprule
& \multicolumn{4}{c}{$\beta=0.85$} & \multicolumn{4}{c}{$\beta=0.55$} \\
\cmidrule(lr){2-5}\cmidrule(lr){6-9}
& TPR & FPR & TDR & $F_1$ & TPR & FPR & TDR & $F_1$ \\
\midrule

W - $\ell_1$ 
& \textbf{0.9510} & \textbf{0.0447} & \textbf{0.9368} & \textbf{0.9438}
& \textbf{0.7410} & \textbf{0.1433} & \textbf{0.7911} & \textbf{0.7652}\\
\midrule

F - $\ell_1$ (glasso) 
& 0.7280 & 0.2793 & 0.6432 & 0.6830
& 0.7240 & 0.1753 & 0.7485 & 0.7360\\
\midrule

F - $\ell_1$ (ADMM) 
& 0.8040 & 0.2833 & 0.6703 & 0.7311
& 0.3630 & 0 & 1 & 0.5326 \\
\midrule

F - $\ell_2$ (ADMM) 
& 0.5780 & 0.0440 & 0.9539 & 0.7198 
& \textbf{0.7090} & \textbf{0.1193} & \textbf{0.8246} & \textbf{0.7624}  \\

\end{tabular}
\caption{10-dimensional GNAR(1,[1]) models over Ring graph for $T=1024$.}
\label{tbl:10D Ring Results}
\end{center}
\end{table}

\subsection{GNAR Models on Ring Graphs} \label{app:Ring Models}

\subsection{VAR and VARMA Models}\label{app:VAR Models}

\section{Underlying Graph Discovery Results}\label{app:graph-discovery}
In this section we show plots of the true and estimated adjacency matrices using the clustering method as described in Section~\ref{sec:graph-discovery}. Each heatmap is shown in a grayscale, with shading proportional to the frequency at which an edge is selected: black denotes edges always selected, and white denotes edges never selected. 

\begin{figure}[H]
\centering
\begin{subfigure}[t]{0.22\linewidth}
    \includegraphics[width=\linewidth]{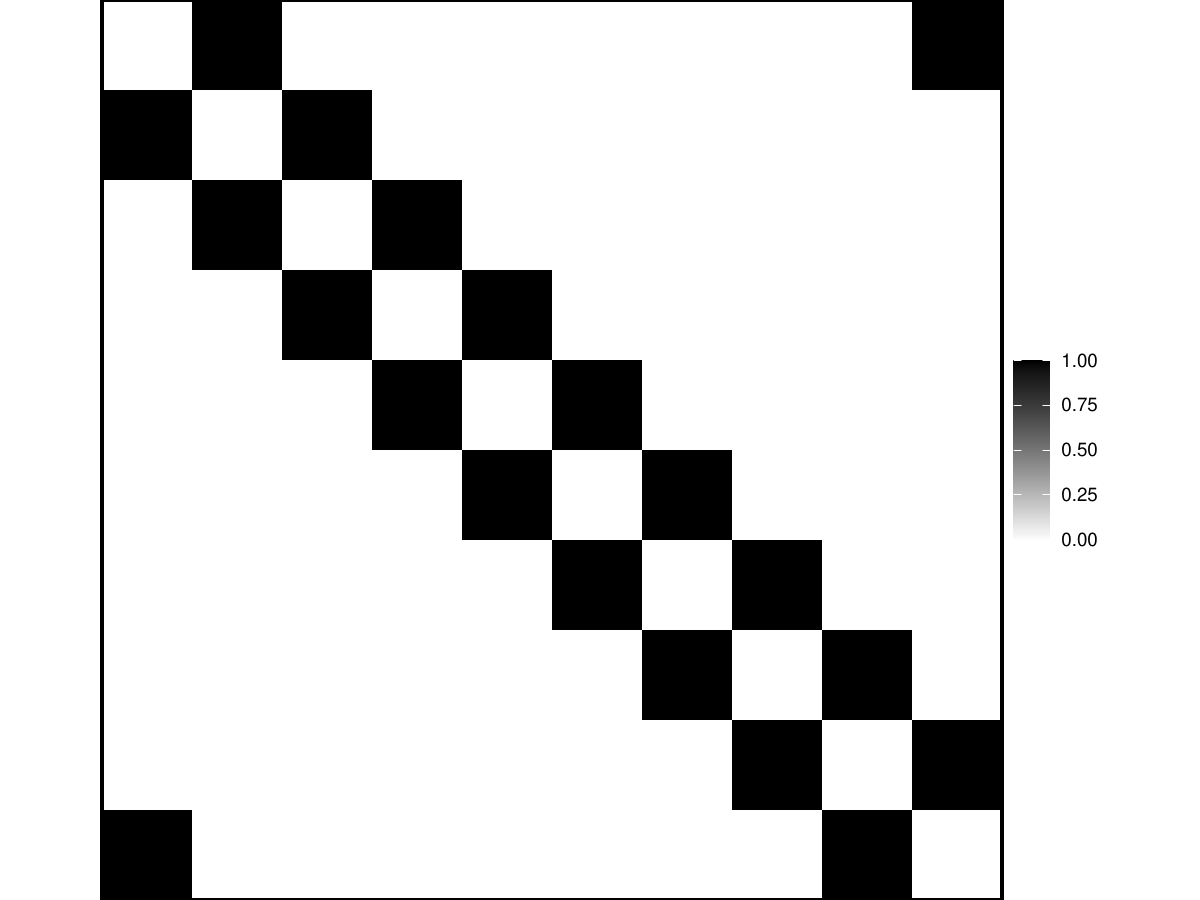}
    \caption{True, 10-dim}
\end{subfigure}
\hspace{10mm}
\begin{subfigure}[t]{0.22\linewidth}
    \includegraphics[width=\linewidth]{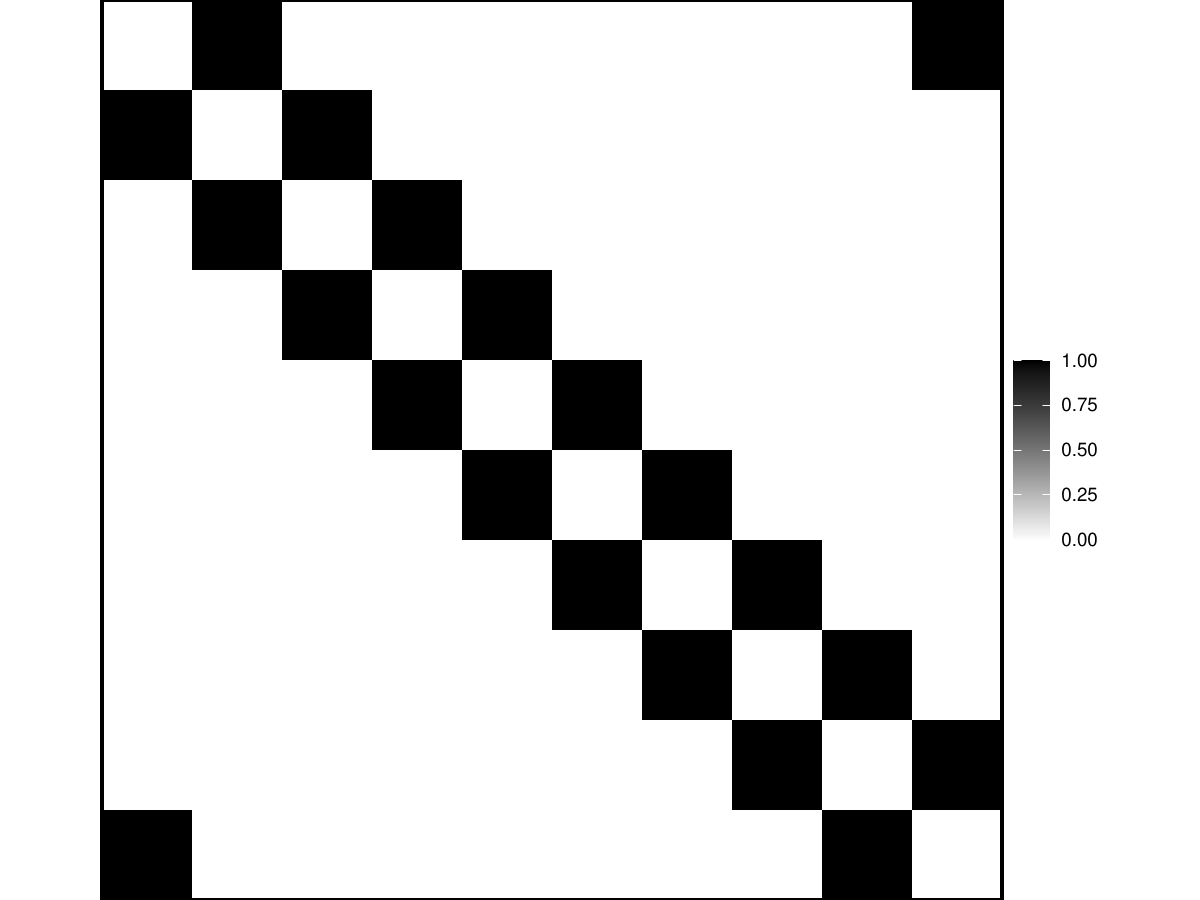}
    \caption{Estimated ($\beta_{1,1}=0.85$)}
\end{subfigure}
\hspace{10mm}
\begin{subfigure}[t]{0.22\linewidth}
\includegraphics[width=\linewidth]{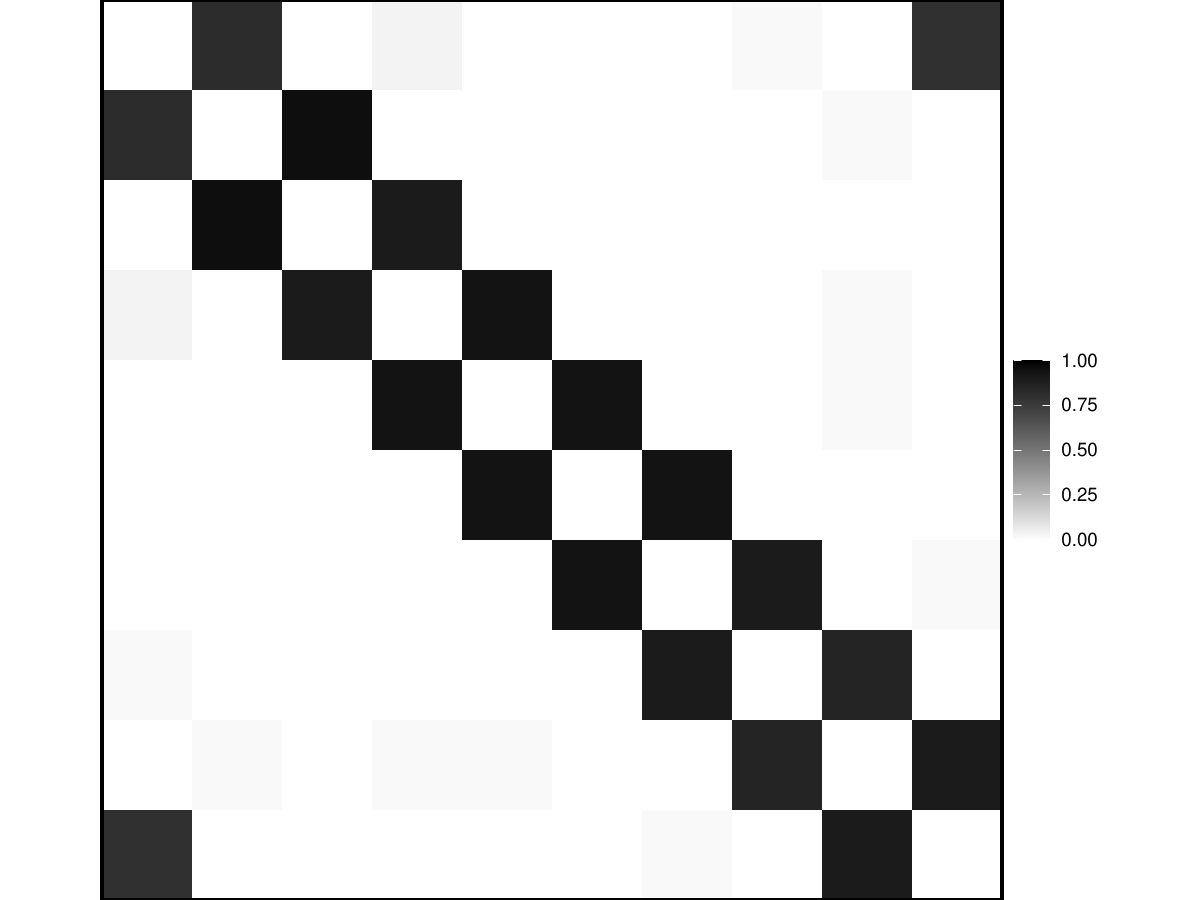}
\caption{Estimated ($\beta_{1,1}=0.55$)}
\end{subfigure}

\hspace{4mm}

\begin{subfigure}[t]{0.22\linewidth}
    \includegraphics[width=\linewidth]{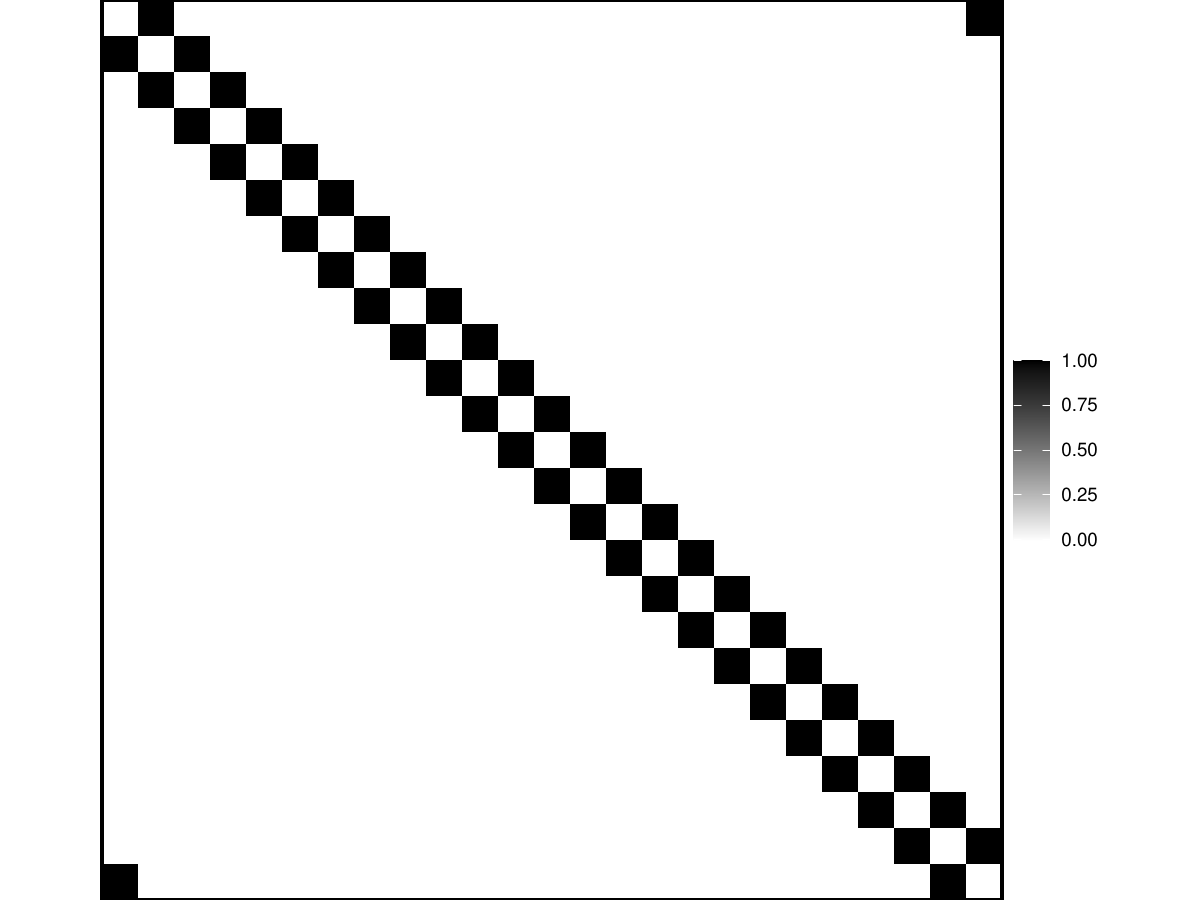}
    \caption{True, 25-dim}
\end{subfigure}
\hspace{10mm}
\begin{subfigure}[t]{0.22\linewidth}
    \includegraphics[width=\linewidth]{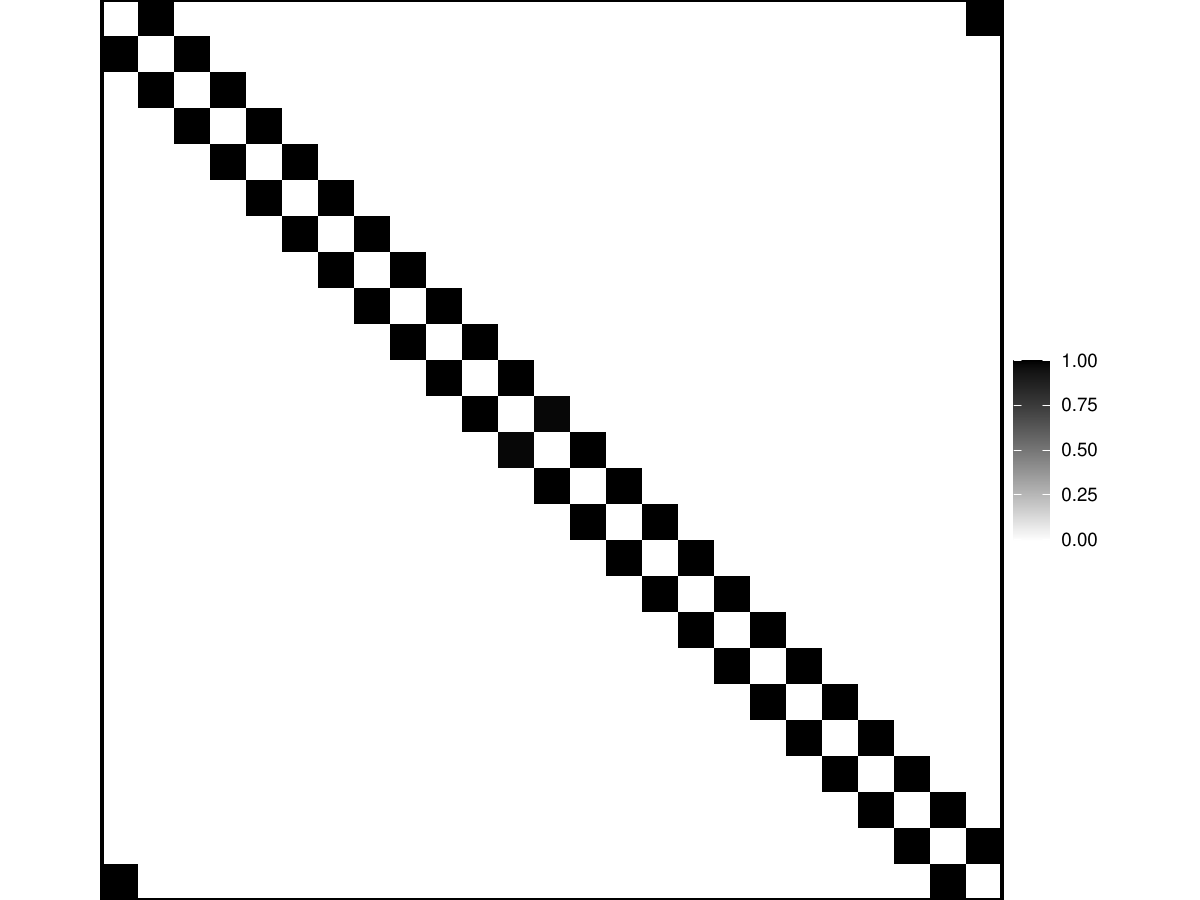}
    \caption{Estimated ($\beta_{1,1}=0.85$)}
\end{subfigure}
\hspace{10mm}
\begin{subfigure}[t]{0.22\linewidth}
    \includegraphics[width=\linewidth]{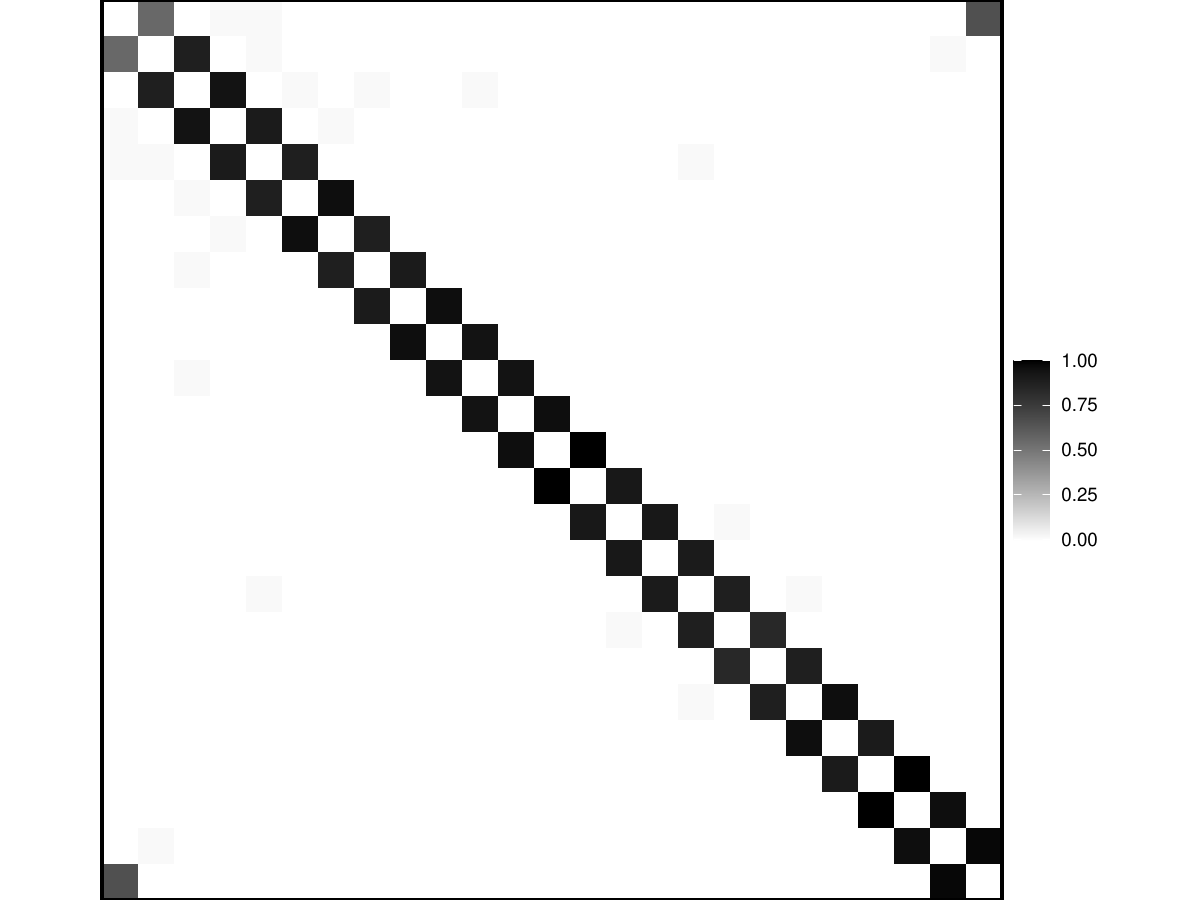}
    \caption{Estimated ($\beta_{1,1}=0.55$)}
\end{subfigure}
\caption{Estimated adjacency matrices of 10-dimensional (top row) and 25-dimensional (bottom row) Ring graphs, based on GNAR(1,[1]) model with $\beta_{1,1}=0.85$ (centre column) or with $\beta_{1,1}=0.55$ (right column).}
\label{fig:10D-beta65-beta35-Ring-clustering}
\end{figure}

\section{COVID-19 Data Information}\label{app:covid}
\begin{table}[H]
    \centering
    \begin{tabular}{p{10cm}|c|c}
       \textbf{NHS England Region} & \textbf{Nodes} & \textbf{Number of Trusts}  \\
       \midrule
        NHS England Cumbria and North East  & 1-4 & 4 \\
        NHS England Yorkshire and Humber & 5-11 & 7\\
        NHS England Lancashire & 12-14 & 3\\
        NHS England Greater Manchester & 15-19 & 5 \\
        NHS England Cheshire and Merseyside & 20-25 & 6  \\
        NHS England North Midlands & 26-30 & 5\\
        NHS England West Midlands & 31-40  & 10 \\
        NHS England Central Midlands & 41-42 & 2\\
        NHS England East & 43-50 & 8 \\
        NHS England London & 51-65 & 15 \\
        NHS England South West (South West North) & 66-68 & 3\\
        NHS England South East (Hampshire, Isle of Wight, Thames Valley) & 69-72 & 4\\
        NHS England South East (Kent, Surrey and Sussex) & 73-81 & 9 \\
        NHS England South West (South West South) & 82-84 & 3 \\
        
    \end{tabular}
    \caption{14 different NHS England Trust regions and corresponding nodes for the $P=84$ Trusts retained following preprocessing steps.}
\label{tbl:covid-tbl}
\end{table}
\end{appendix}

\end{document}